\documentclass[a4paper,12pt]{article}

\usepackage[margin=2.5cm]{geometry}
\usepackage{graphicx}
\usepackage{amssymb}
\usepackage{multirow}
\usepackage{amsmath}
\usepackage{caption}
\usepackage{comment}
\usepackage{subcaption}
\usepackage[colorlinks=true,bookmarks=true]{hyperref}
\usepackage{float}
\usepackage{mathrsfs}
\usepackage[nosort]{cite}

\hypersetup{citecolor=blue}
\newcommand{\dif}{\mathrm{d}}

\allowdisplaybreaks

\makeatletter
\renewcommand\section{\@startsection {section}{1}{\z@}%
                                   {-3.5ex \@plus -1ex \@minus -.2ex}%
                                   {2.3ex \@plus.2ex}%
                                   {\setcounter{equation}{0}\large\bfseries}}
\renewcommand\subsection{\@startsection{subsection}{2}{\z@}%
                                     {-3.25ex\@plus -1ex \@minus -.2ex}%
                                     {1.5ex \@plus .2ex}%
                                     {\normalfont\bfseries}}
\renewcommand\subsubsection{\@startsection{subsubsection}{3}{\z@}%
                                     {-3.25ex\@plus -1ex \@minus -.2ex}%
                                     {1.5ex \@plus .2ex}%
                                     {\normalfont\it}}
\makeatother

\begin{document}

\thispagestyle{empty}
\begin{flushright}

\end{flushright}
\vbox{}

\vspace{2cm}

\begin{center}
{\LARGE{Gravitational multi-soliton solutions on flat space
        }}\\[16mm]
{{Yu Chen}}
\\[6mm]
{\it Department of Physics,
National University of Singapore, 
Singapore 119260}\\[15mm]

\end{center}
\vspace{1cm}

\centerline{\bf Abstract}
\bigskip
\noindent
It is well known that, for even $n$, the $n$-soliton solution on the Minkowski seed, constructed using the inverse-scattering method (ISM) of Belinski and Zakharov (BZ), is the multi-Kerr-NUT solution. We show that, for odd $n$, the natural seed to use is the Euclidean space with two manifest translational symmetries, and the $n$-soliton solution is the accelerating multi-Kerr-NUT solution. We thus define the $n$-soliton solution on flat space for any positive integer $n$. It admits both Lorentzian and Euclidean sections. In the latter section, we find that a number, say $m$, of solitons can be eliminated in a non-trivial way by appropriately fixing their corresponding so-called BZ parameters. The resulting solutions, which may split into separate classes, are collectively denoted as $[n-m]$-soliton solutions on flat space. We then carry out a systematic study of the $n$- and $[n-m]$-soliton solutions on flat space. This includes, in particular, an explicit presentation of their ISM construction, an analysis of their local geometries, and a classification of all separate classes of solutions they form. We also show how even-soliton solutions on the seeds of the collinearly centred Gibbons--Hawking and Taub-NUT arise from these solutions.


\newpage

{\hypersetup{linkcolor=black}
\tableofcontents
}

\section{Introduction}

\label{sec_introduction}

The Einstein equations of general relativity are notoriously difficult to solve due to their highly non-linear nature. A number of solution-generating techniques have been developed to find new solutions from old known ones \cite{Stephani:2003tm}. One of the most powerful solution-generating techniques is the inverse-scattering method (ISM), developed by Belinski and Zakharov (BZ) in 1978 \cite{Belinsky:1971nt,Belinsky:1979mh}, applicable to solutions which admit two commuting Killing vectors. Using this method, soliton solutions can be constructed by a purely algebraic procedure on a given solution known as a seed. The ISM has been applied to generate various new solutions in different physical contexts, such as cosmological models, cylindrically symmetric solutions, gravitational plane waves and their collisions, as well as stationary and axisymmetric solutions. The reader is referred to \cite{Belinski:2001ph} for a recent comprehensive review. The last class is most relevant to black hole solutions and includes the Kerr solution in particular. In this paper, we are interested in the application of the ISM to generate this class of solutions.

To construct stationary and axisymmetric solutions using the ISM, one often uses the Minkowski metric in the following form with a manifest axial symmetry as the seed:
\begin{align} \label{metric_seed_even_Lorentzian}\dif s_0^2{}^{\text{(even)}}=-\dif t^2+\rho^2\dif\phi^2+\left(\dif \rho^2+\dif z^2\right).
\end{align}
Using the ISM, $n$ solitons can be placed on this seed; one also says that an $n$-soliton transformation can be performed on the seed. For the generated solution to have the Lorentzian signature, $n$ is required to be even. In this case, the final solution is known to describe a number $\frac{n}{2}$ of Kerr-NUT black holes in superposition \cite{Israel:1964mba,Neugebauer:1980mba}.

When $n$ is odd, the solution of $n$ solitons placed on the above seed has the Euclidean signature and contains a curvature singularity at $\rho\rightarrow 0$, and no known physical relevance has been known about it. In fact, an odd-soliton solution is known to be a \emph{finite} perturbation on its seed: we cannot recover a solution from it that is arbitrarily close to the seed by tuning the so-called BZ parameters \cite{Belinski:2001ph}. This finite perturbation changes the signature of the metric, among other things. Hence, for an odd-soliton solution to be Lorentzian, the seed must be Euclidean \cite{Verdaguer:1982si}. In \cite{Verdaguer:1982si}, Verdaguer studied in detail the 1-soliton solution on the Euclidean version of (\ref{metric_seed_even_Lorentzian}), i.e., Euclidean space with one translational and one axial symmetry. He found that although its signature is Lorentzian, the 1-soliton solution still possesses a curvature singularity at $\rho\rightarrow 0$. To have well-behaved stationary and axisymmetric odd-soliton solutions, we should look for new Euclidean seeds. Since an odd-soliton transformation is a finite perturbation, we should not insist that the seeds have an axial symmetry.

In this paper, we are interested in applying the ISM on the simplest seed, i.e., the flat space. At this point, we recall that the Minkowski space (\ref{metric_seed_even_Lorentzian}), or a part of it, has different sets of two commuting Killing vectors; thus, it can be cast in different coordinates so that the ISM can be applied. In fact, in 1980, Jantzen \cite{Jantzen:1980mba} classified all 2-parameter Abelian isometries of the Minkowski space and identified the corresponding coordinate systems. Subsequently he applied the ISM on the Minkowski space in so-called boost-rotation coordinates \cite{Jantzen:1983mba}. For the static 2-soliton solution, by computing its Petrov type, he concluded that it is the C-metric, which is known to describe the accelerating Schwarzschild black hole \cite{Kinnersley:1970zw}. He further conjectured that even-soliton solutions on his seed are the accelerating multi-Kerr-NUT solution. His seed has explicit dependence on the coordinate $z$ and is obviously not the simplest to use, and multi-soliton solutions on it are difficult to analyse. The $z$-dependence suggests that this seed itself can be viewed as a soliton solution on a simpler seed. The accelerating multi-Kerr-NUT solution can then be viewed as a multi-soliton solution on this simpler seed. We remark that the ISM has also been applied on another flat-space seed, namely, Milne's model in the cosmological context \cite{Ibanez:1986iv}, although no cosmological singularity has been found in the 2-soliton solution.

One major purpose of this paper is to demonstrate that for odd $n$, a more natural flat-space seed is the following metric of the Euclidean space:
\begin{align} \label{metric_seed_odd_Lorentzian} \dif {s}_0^2{}^{\text{(odd)}}=\dif t^2+\dif\phi^2+\left(\dif \rho^2+\dif z^2\right).
\end{align}
Three unusual properties of the seed (\ref{metric_seed_odd_Lorentzian}) are observed: Firstly, it is Euclidean; secondly, it looks too trivial; thirdly, its $g$-matrix, defined as the $2\times 2$ matrix $g_{ab}$ for $a,b=t,\phi$, does not obey the condition (\ref{rho}). We will show that these properties are not obstacles to apply the ISM on this seed; in fact, upon placing an odd number of solitons on the seed, non-trivial well-behaved Lorentzian solutions are generated.

The seeds (\ref{metric_seed_even_Lorentzian}) and (\ref{metric_seed_odd_Lorentzian}) have another common origin. The seed (\ref{metric_seed_even_Lorentzian}) is obviously a special case of the Levi-Civita metric \cite{Levi-Civita:1919mba,Kasner:1921zz}
\begin{align}
\label{LeviC}
\dif s^2=\epsilon_1\rho^{1+d}\dif t^2+\epsilon_2\rho^{1-d}\dif \phi^2+\rho^{(d^2-1)/2}\left(\dif \rho^2+\dif z^2\right),
\end{align}
with the constant $d=\pm 1$ and appropriately chosen signs $\epsilon_{1,2}$. If we instead set $d=0$ and divide the $g$-matrix of (\ref{LeviC}) by a factor $\rho$, the $g$-matrix of the seed (\ref{metric_seed_odd_Lorentzian}) is recovered. It is known that two seeds whose $g$-matrices differ by an overall factor are equivalent (see, e.g., \cite{Belinsky:1971nt,Pomeransky:2005sj}). The exact meaning of this will be explained in the next section. So setting $d=0$ in (\ref{LeviC}), we effectively recover the seed (\ref{metric_seed_odd_Lorentzian}).

Now, we know that the flat-space seeds (\ref{metric_seed_even_Lorentzian}) and (\ref{metric_seed_odd_Lorentzian}) are special cases of the Levi-Civita seed with $d=\pm1 ,0$. The natural question to ask at this point is what happens if $d$ is taken as a general parameter. Multi-soliton solutions on the general Levi-Civita seed have been extensively studied in the literature (see, e.g., \cite{Letelier:1985mba,Belinski:2001ph}). These solutions are stationary and axisymmetric. However, it is known that they possess a curvature singularity at their axes or Killing horizons, located at $\rho\rightarrow 0$. When $\rho$ is appropriately analytically continued to a time-like variable, the singularity at $\rho\rightarrow 0$ is usually interpreted as a cosmological singularity. The seed (\ref{LeviC}) in this case is usually referred to as the Kasner seed, and multi-soliton solutions based on this seed have been thoroughly investigated in the literature (see, e.g., \cite{Belinski:2001ph}).

In this paper, we are interested in stationary and axisymmetric solutions that are \emph{locally regular} in the following sense. We demand that the axes or Killing horizons of these solutions possess no curvature singularities. These axes or Killing horizons can be made regular at least locally. So the points at $\rho=0$ can be made as \emph{part of} the space-time manifold, at least locally. The two special cases with $d=\pm 1$ for even $n$ and $d=0$ for odd $n$ are the only ones among (general non-diagonal) multi-soliton solutions on the Levi-Civita seed that possess the above demanded properties, as far as the author could find.\footnote{A few diagonal soliton solutions (including so-called generalised soliton solutions \cite{Belinski:2001ph}) constructed on the Levi-Civita seed (\ref{LeviC}), or the Kasner seed, in various physical contexts, were known in the literature not to develop a curvature singularity at $\rho\rightarrow 0$ in the special case $d=0$. The reader is referred to the review \cite{Belinski:2001ph} and references therein on this point.} In this view, using flat-space seeds (or their equivalence classes) is really a necessity; it is a consequence of regularity considerations.

We want to point out that, in the seed (\ref{metric_seed_odd_Lorentzian}), or (\ref{LeviC}) with $d=0$, neither $\partial_t$ nor $\partial_\phi$ generates an axial symmetry. They instead generate cylindrical or translational symmetries. However, after applying the ISM with an odd number of solitons, the generated solution does have an axial symmetry, at least locally. It is thus clearly a finite perturbation of the seed (\ref{metric_seed_odd_Lorentzian}). We further show that it is locally regular stationary and axisymmetric and describes the accelerating generalisation of the superposition of a number $\lfloor\frac{n}{2}\rfloor$ of Kerr-NUT black holes.

We thus define the $n$-soliton solution on flat space, for any positive integer $n$: for even $n$, the seed (\ref{metric_seed_even_Lorentzian}) is used, and for odd $n$, the seed (\ref{metric_seed_odd_Lorentzian}) is used. The $n$-soliton solution on flat space is locally regular, and it describes the multi-Kerr-NUT and its accelerating generalisation. We show that there is a transition limit, from which we can recover the odd-soliton solutions from the even-soliton solutions, and vice versa. In this limit, one soliton is sent to infinity and disappears from the solution completely: the $n$-soliton solution on flat space is reduced to the $(n-1)$-soliton solution on flat space. So, in this limit, we can recover the accelerating multi-Kerr-NUT solution from the multi-Kerr-NUT solution and vice versa.

Interestingly, the $n$-soliton solution on flat space, for any $n$, has a Euclidean section, obtainable by appropriate analytic continuation. Euclidean sections of Lorentzian solutions have all-plus signature and form a particular class of Euclidean solutions. The latter are also of great interest in gravitational physics. Globally regular Euclidean solutions have been known as gravitational instantons, a well-known example being the Eguchi--Hanson instanton \cite{Eguchi:1978xp}. In this paper, however, we are not concerned with the issue of global regularity, but instead focus on locally regular solutions. Hence by abuse of notation, we call Euclidean solutions (gravitational-)instanton solutions. Since they are vacuum solutions to the Einstein equations, they play an important role in Euclidean quantum gravity \cite{Gibbons:1994cg}, and, in particular, they are stationary-phase points in the Euclidean path integral.

A large class of instanton solutions is the multi-centred Gibbons--Hawk\-ing solution \cite{Gibbons:1979zt}, with the Eguchi--Hanson instanton as a special case with two centres and equal charges. A closely related solution is the multi-centred Taub-NUT solution \cite{Newman:1963yy,Hawking:1976jb}. These two classes of multi-centred solutions have self-dual Riemann tensors and were constructed by using a particular, so-called Gibbons--Hawking ansatz \cite{Hawking:1976jb,Gibbons:1979zt}. In the case of non-self-dual Riemann tensors, the Gibbons--Hawking ansatz is no longer useful, and it becomes much more difficult to find instanton solutions.

The Euclidean sections of stationary and axisymmetric Lorentzian solutions are gravita\-tional-instanton solutions with two commuting Killing vectors. Upon analytic continuation, the Killing horizons in the Lorentzian section now also become axes. The two Killing vectors thus generate two axial symmetries, at least locally. But recall that for instanton solutions with two axial symmetries, we can, however, directly apply the ISM to generate them rather than obtain them from Lorentzian ones by analytic continuation. 

Similarly as motivated in the Lorentzian solutions, to construct locally regular instanton solutions with two axial symmetries, we should use flat-space seeds. We demand all axes of these solutions, including their meeting points, are regular at least locally. Not surprisingly, the seeds we need to use turn out to be the analytic continuation of the seeds (\ref{metric_seed_even_Lorentzian}) and (\ref{metric_seed_odd_Lorentzian}), obtained by defining $t=i\tau$. Hence for even $n$, the seed is
\begin{align}
\label{metric_seed_even_Euclidean}
 \dif {s}_{ 0}^2{}^{\text{(even)}}=\dif \tau^2+\rho^2\dif\phi^2+\left(\dif \rho^2+\dif z^2\right),
\end{align}
and for odd $n$, the seed is
\begin{align}
\label{metric_seed_odd_Euclidean}
\dif {s}_{ 0}^2{}^{\text{(odd)}}=-\dif \tau^2+\dif\phi^2+\left(\dif \rho^2+\dif z^2\right).
\end{align}
Note that (\ref{metric_seed_even_Euclidean}) is Euclidean, while (\ref{metric_seed_odd_Euclidean}) is Lorentzian. For both seeds, the generated soliton solutions have the Euclidean signature as required.

We find that the $n$-soliton solution on the seeds (\ref{metric_seed_even_Euclidean}) and (\ref{metric_seed_odd_Euclidean}) is in fact equivalent to the Euclidean section of the $n$-soliton solution on the seeds (\ref{metric_seed_even_Lorentzian}) and (\ref{metric_seed_odd_Lorentzian}). Thus locally regular instanton solutions with two axial symmetries, constructed using the ISM, are no more than the Euclidean sections of the multi-Kerr-NUT and accelerating multi-Kerr-NUT solutions. Hence, up to analytic continuation, the signatures of the seeds are not really significant for the $n$-soliton solution on flat space. The generated solution can always be analytically continued to have a demanded Lorentzian or Euclidean signature. This seems to be a trivial fact at first sight, but it should be taken as a remarkable property of the soliton solutions of the ISM.

But there are more locally regular instanton solutions with two axial symmetries. In the special case when all the ``nuts'' \cite{Gibbons:1979xm} are collinearly centred, the multi-centred Gibbons--Hawking and Taub-NUT solutions also possess two axial symmetries. Indeed, it has been known that in this case these two classes of solutions can be cast into so-called Weyl--Papapetrou coordinates \cite{Chen:2010zu}. However, their possible relations to soliton solutions have not been studied so far. Another major purpose of this paper is to show how the multi-collinearly-centred Gibbons--Hawking and Taub-NUT solutions arise from the (Euclidean) $n$-soliton solution on flat space. In what follows, we will exclusively refer to their collinearly centred subclasses when discussing the Gibbons--Hawking and Taub-NUT solutions.

The key idea is that in the Euclidean regime, solitons not only can be added, but also can be eliminated in a non-trivial way. For the $n$-soliton solution on flat space, a soliton represents the meeting point of two adjacent axes, as we will show. The BZ parameters, one for each soliton, then control the generators of the axes. By fixing its corresponding BZ parameter in such a way that the two adjacent axes join up into a single and larger one, the soliton is then eliminated. The resulting metric can be entirely written in terms of the rest solitons, but yet it does not degenerate to the $(n-1)$-soliton solution on flat space as in the case of the transition limit. A number, say $m$, of solitons can be eliminated simultaneously, and, depending on the different choices of fixing their BZ parameters, the resulting solutions may split into separate classes. These solutions are collectively denoted as $[n-m]$-soliton solutions on flat space. The multi-centred Gibbons--Hawking and Taub-NUT solutions arise from these soliton solutions.

This paper carries out a systematic study of multi-soliton solutions on flat space, namely, the $n$-soliton and $[n-m]$-soliton solutions on flat space. We focus on the Euclidean section of the $n$-soliton solution on flat space, from which its Lorentzian section can be easily recovered. So we use the seeds (\ref{metric_seed_even_Euclidean}) and (\ref{metric_seed_odd_Euclidean}). We perform a detailed analysis of the local geometries of the $n$- and $[n-m]$-soliton solutions and show how a classification of all separate classes of solutions they form is achieved. We also show that not only the multi-centred Gibbons--Hawking and Taub-NUT solutions but also even-soliton solutions on these solutions are included in multi-soliton solutions on flat space.

The paper is organised as follows. We review the ISM of gravitational field and its soliton solutions in Sec.~\ref{sec_ISM}. In Sec.~\ref{sec_n-soliton}, we explicitly construct the Euclidean $n$-soliton solution on flat space for both even and odd $n$. Some of its basic properties are also studied, including its transition limit, Lorentzian section, and its static limit and asymptotic structure. In Sec.~\ref{sec_n-m-soliton}, we show how its BZ parameter should be tuned to eliminate a soliton and then define and study $[n-m]$-soliton solutions on flat space. To do so, we introduce the so-called rod-structure formalism. It is calculated and used to study the local geometries of multi-soliton solutions on flat space. In Sec.~\ref{sec_examples}, we study in detail several examples of the $n$- and $[n-m]$-soliton solutions on flat space for $n\le 6$ and $n-m\leq 3$ and show how various known solutions are recovered. In Sec.~\ref{sec_classification}, we carry out a classification of multi-soliton solutions on flat space and show how even-soliton solutions on multi-centred Gibbons--Hawking and Taub-NUT solutions are included. A discussion of potential applications and generalisations of multi-soliton solutions on flat space then follows in Sec.~\ref{sec_discussion}. The paper contains an appendix, in which we define prolate spheroidal and C-metric-like coordinates that will be frequently used in Sec.~\ref{sec_examples}.

\section{Inverse-scattering method}

\label{sec_ISM}

In this section, we first review the integration scheme of the ISM and the soliton solutions of the gravitational field, developed by Belinski and Zakharov \cite{Belinsky:1971nt,Belinsky:1979mh}. The presentation here closely follows the review \cite{Belinski:2001ph}. We then generalise the definition of a seed on which the ISM can be applied. The generalised seeds include the seeds (\ref{metric_seed_odd_Lorentzian}) and (\ref{metric_seed_odd_Euclidean}), in particular.

\subsection{Integration scheme of the inverse-scattering method}

Stationary and axisymmetric solutions admit two commuting Killing vectors, denoted by $\partial_t$ and $\partial_\phi$, respectively. These two Killing vectors together generate the time flow and an axial symmetry, at least locally. Under suitable conditions, these two Killing vectors are 2-surface orthogonal (see, e.g., \cite{Wald:1984rg}). The solutions considered in this paper satisfy these conditions, so they admit the following representation in the Weyl--Papapetrou coordinates:
\begin{align}
\label{Weyl_Papapetrou_form}
\dif s^2=g_{ab}\,\dif x^a\dif x^b+f\left(\dif\rho^2+\dif z^2\right),
\end{align}
for $a,b=1,2$ and $x^1=t,x^2=\phi$, subject to a supplementary condition
\begin{align}
\label{rho}
|\det g|=\rho^2\,.
\end{align}
Following \cite{Belinski:2001ph}, we denote by $g$ the $2\times 2$ matrix with elements consisting of metric components $g_{ab}$. For Lorentzian solutions, $\det g$ is negative, so we actually have $\det g=-\rho^2$. If instanton solutions with two axial symmetries generated by $\partial_\tau\equiv\partial_{x^1}$ and $\partial_\phi\equiv\partial_{x^2}$ are considered, we have $\det g=\rho^2$ instead. The matrix $g$ and the function $f$ depend only on the two coordinates $\rho$ and $z$.

In the Weyl--Papapetrou coordinates (\ref{Weyl_Papapetrou_form}) and (\ref{rho}), the Einstein equations in vacuum decouple into two groups of equations. The first group is for the matrix $g$, which can be written as
\begin{align}
\label{G}
(\rho g_{,\rho}g^{-1}){}_{,\rho}+(\rho g_{,z}g^{-1})_{,z}=0\,.
\end{align}
The second group of equations is for $f$ and can be written as
\begin{align}
\label{nu}
(\ln f)_{,\rho}&=-\frac{1}{\rho}+\frac{1}{4\rho}{\text{Tr}}\left(U^2-V^2\right),\nonumber\\
(\ln f)_{,z}&=\frac{1}{2\rho}{\text{Tr}}\left(UV\right),
\end{align}
if we define two $2\times 2$ matrices $U$ and $V$ as follows:
\begin{align}
\label{UV1}
U=\rho g_{,\rho}g^{-1}\,,\quad V=\rho g_{,z}g^{-1}\,.
\end{align}
Notice that the integrability of $f$ in Eq.~(\ref{nu}) is guaranteed by (\ref{G}). Hence, we can always solve the vacuum Einstein equations by first solving (\ref{rho}) and (\ref{G}) for $g$ and subsequently solving (\ref{nu}) for $f$. Solutions $(g,f)$ to Eqs.~(\ref{rho}), (\ref{G}) and (\ref{nu}) then give solutions to the vacuum Einstein equations.

In the ISM of Belinski and Zakharov, Eq.~(\ref{G}) is considered first, and its solutions are constructed. Equation (\ref{G}) can be viewed as the compatibility condition for the so-called ``L--A pair'' of linear differential equations for a generating matrix $\psi(\rho,z,\lambda)$:
\begin{align}
\label{Lax_pair}
D_{\rho}\psi=\frac{\rho U+\lambda V}{\lambda^2+\rho^2}\,\psi\,,\quad D_{z}\psi=\frac{\rho V-\lambda U}{\lambda^2+\rho^2}\,\psi\,,
\end{align}
where the differential operators $D_{\rho}$ and $D_z$ are defined as
\begin{align}
D_{\rho}=\partial_{\rho}+\frac{2\lambda\rho}{\lambda^2+\rho^2}\,\partial_{\lambda}\,,\quad D_{z}=\partial_{z}-\frac{2\lambda^2}{\lambda^2+\rho^2}\,\partial_{\lambda}\,.
\end{align}
Here $\lambda$ is a complex ``spectral parameter'' independent of the coordinates $\rho$ and $z$. It can be checked that the system (\ref{Lax_pair}) is compatible, in terms of $g$, if and only if (\ref{G}) is satisfied. The value of the generating matrix $\psi$ at $\lambda=0$ can be identified as the matrix $g$, i.e., $g=\psi(\rho,z,\lambda=0)$.

The L--A pair of differential equations (\ref{Lax_pair}) allows us to generate new solutions starting from known ones following a purely algebraic procedure. These known solutions are called ``seed solutions'', or simply ``seeds'', in the literature. From a seed $g_0$, we can compute the corresponding matrices $U_0$ and $V_0$ from (\ref{UV1}) and then solve the generating matrix $\psi_0$ from (\ref{Lax_pair}). In the ISM, one seeks a new generating matrix $\psi$ by dressing $\psi_0$ in the following form:
\begin{align}
\psi=\chi\psi_0\,,\label{dressing}
\end{align}
where $\chi$ is a $2\times 2$ matrix. The equations for the dressing matrix $\chi$, obtained by substituting (\ref{dressing}) into (\ref{Lax_pair}), are then
\begin{align} D_{\rho}\chi&=\frac{\rho U+\lambda V}{\lambda^2+\rho^2}\chi-\chi\frac{\rho U_0+\lambda V_0}{\lambda^2+\rho^2}\,,\nonumber\\ D_{z}\chi&=\frac{\rho V-\lambda U}{\lambda^2+\rho^2}\chi-\chi\frac{\rho V_0-\lambda U_0}{\lambda^2+\rho^2}\,. 
\label{chi}
\end{align}
The task is now to look for solutions of the matrix $\chi$. We are primarily interested in the soliton solutions whose $\chi$ admits the following form:
\begin{align}
\chi=I+\sum_{k=1}^{n}\frac{R_k}{\lambda-\mu_k}\,,\label{chi_soliton}
\end{align}
with $n$ simple poles (or pole trajectories) for the parameter $\lambda$. Here $I$ stands for the identity matrix. A pole labelled by $k$ is located at $\lambda=\mu_k$ and has a residue matrix $R_k$, both of which depend only on the coordinates $\rho$ and $z$. Additional conditions must also be imposed on the matrix $\chi$ to ensure that the derived matrix $g$ is real and symmetric. Discussions of these conditions can be found  in \cite{Belinsky:1971nt,Belinsky:1979mh}.

Note that the right-hand sides of Eq.~(\ref{chi}) for $\chi$ do not contain poles of order higher than one, so will be the case for the left-hand sides. This requirement gives the equations for $\mu_k$ as follows:
\begin{align}
\mu_{k,\rho}=\frac{2\rho \mu_k}{\mu_k^2+\rho^2}\,,\quad \mu_{k,z}=\frac{-2\mu_k^2}{\mu_k^2+\rho^2}\,,
\end{align}
whose solutions are
\begin{align}
\mu_k^{\pm}=\pm \sqrt{\rho^2+(z-z_k)^2}-\left(z-z_k\right),
\end{align}
where $z_k$ is an integration constant. These solutions also solve the quadratic algebraic equation
\begin{align}
\mu_k^2+2(z-z_k)\mu_k-\rho^2=0\,.
\end{align}
Note that $\mu_k^-=-\frac{\rho^2}{\mu_k^+}$. We refer to $\mu_k^{\pm}$ as a soliton and an anti-soliton, respectively. So each pole location of the matrix $\chi$ as in (\ref{chi_soliton}) corresponds to a soliton or an anti-soliton, and the number of poles is the number of solitons and anti-solitons. We call the integration constant $z_k$ as the location of the soliton or anti-soliton $\mu_k$. In what follows, unless otherwise specified, solitons and anti-solitons will be collectively called solitons.

The matrices $R_k$ are also completely determined by Eq.~(\ref{chi}) and the condition that the derived matrix $g$ is symmetric. We refer the reader to \cite{Belinsky:1971nt,Belinsky:1979mh} for more details. Here we just quote the results. The matrices $R_k$ are degenerate and, for each fixed $k$, can be written as
\begin{align} (R_k)_{ab}=n^{(k)}_am^{(k)}_b\,,
\end{align}
where $m^{(k)}$ and $n^{(k)}$ are vectors. The vectors $m^{(k)}$ are given by
\begin{align}
m^{(k)}_a=m_{0b}^{(k)}[\psi_0^{-1}(\rho,z,\lambda=\mu_k)]_{ba}\,,
\label{m_vectors}
\end{align}
where $m_0^{(k)}$ are some constant vectors known as BZ vectors, with components $m_{0a}^{(k)}$ known as BZ parameters. The generating matrix $\psi_0$ for the seed is used in (\ref{m_vectors}). So for each soliton $\mu_k$, in addition to its location $z_k$, there are two BZ parameters associated to it. To give the vectors $n^{(k)}$, we need to first define a symmetric $n\times n$ matrix $\Gamma$ with elements $\Gamma_{kl}$ and its inverse matrix $D_{kl}$ as
\begin{align}
\label{Gamma_matrix}
\Gamma_{kl}=\frac{[m^{(k)}]^T g_0 m^{(l)}}{\rho^2+\mu_k\mu_l}\,,\quad \sum_{p=1}^{n} D_{kp}\Gamma_{pl}=\delta_l^k\,.
\end{align}
The vectors $n^{(k)}$ are then given by
\begin{align}
n_a^{(k)}=\sum_{l=1}^n\mu_l^{-1}D_{lk}L_a^{(l)}\,,
\end{align}
where the vectors $L^{(k)}$ are defined as
\begin{align}
\quad L^{(k)}_a=m^{(k)}_cg_{0ca}\,.
\end{align}

Once $\chi$ is constructed, one can calculate the new matrix $g$. As mentioned, $g$ can be obtained from the generating matrix $\psi$ by setting $\lambda=0$, i.e.,
\begin{align}
g=\psi(\lambda=0)=\chi(\lambda=0)g_0=\left(I-\sum_{k=1}^n R_k\mu_k^{-1}\right)g_0\,.
\end{align}
The components of the matrix $g$ can be written as
\begin{align}
g_{ab}=(g_0)_{ab}-\sum_{k,l=1}^n \frac{D_{kl}}{\mu_k\mu_l}L^{(k)}_aL^{(l)}_b\,.
\label{unphysical_g}
\end{align}
The above construction of the matrix $g$ gives a new solution to (\ref{G}). Compared with the seed matrix $g_0$, $n$ solitons have been added to the new matrix $g$, each with one location parameter and two BZ parameters.

We should recall that to find solutions to the vacuum Einstein equations, $g$ should also satisfy (\ref{rho}). Belinski and Zakharov \cite{Belinsky:1971nt,Belinsky:1979mh} showed that once a solution to (\ref{G}) is known, a solution to both (\ref{rho}) and (\ref{G}) can be immediately obtained by a simple normalisation. For any matrix $g$ satisfying (\ref{G}), the rescaled matrix $hg$ satisfies (\ref{G}) if and only if the function $\ln |h|$ solves the following equation:
\begin{align}
\varphi_{,\rho\rho}+\rho^{-1}\varphi_{,\rho}+\varphi_{,zz}=0\,,
\label{lap}
\end{align}
for $\varphi$. Moreover, we can deduce from Eq.~(\ref{G}) that $\ln |\det g|$ solves (\ref{lap}). We can then directly show that the normalised matrix $g^{\text{(ph)}}\equiv \frac{\rho}{\sqrt{|\det{g}|}}g$ solves (\ref{G}) and (\ref{rho})  simultaneously. The determinant of the matrix $g$ (\ref{unphysical_g}) is calculated to be
\begin{align}
\det g=(-1)^n\rho^{2n}\left(\prod_{k=1}^{n}\mu_k^{-2}\right)\det g_0\,.
\end{align}
So the normalised matrix $g^{\text{(ph)}}$ can be written as
\begin{align}
g^{\text{(ph)}}=\frac{\rho^{1-n}}{\sqrt{|\det g_0|}}\left(\prod_{k=1}^{n}\mu_k\right)g\,.
\label{physical_g}
\end{align}

The last part of the ISM construction is to find the solution $f^{\text{(ph)}}$ to (\ref{nu}), with $g$ replaced by $g^{\text{(ph)}}$ constructed as above. To do so, one first solves $f$ to (\ref{nu}) with $g$ given by (\ref{unphysical_g}). Then, $f^{\text{(ph)}}$ can be obtained from $f$ again by a normalisation. The explicit form of $f^{\text{(ph)}}$ can be found in  \cite{Belinsky:1971nt,Belinsky:1979mh}. Now, the pair $(g^{\text{(ph)}}, f^{\text{(ph)}})$ are finally determined, and they satisfy Eqs.~(\ref{rho}), (\ref{G}) and (\ref{nu}). Together they give rise to a new solution to the vacuum Einstein equations. 

Obviously, the normalisation (\ref{physical_g}) does not change the sign of the determinant of the matrix $g$. Hence, for even $n$, $\det g^{\text{(ph)}}$ has the same sign as $\det g_0$; for odd $n$, their signs are opposite. It is thus clear that, to generate Lorentzian solutions, one can add an even number of solitons on Lorentzian seeds or an odd number of solitons on Euclidean seeds. This observation is due to Verdaguer \cite{Verdaguer:1982si}, and it explains the different signatures of the seeds (\ref{metric_seed_even_Lorentzian}) and (\ref{metric_seed_odd_Lorentzian}).

We observe that in the above construction, the vectors $m^{(k)}$ can be rescaled by arbitrary factors that can depend on $\rho$ and $z$ and also on $k$, without changing the final solution \cite{Belinsky:1971nt,Belinsky:1979mh}. Hence, although each soliton has two BZ parameters, only their ratio is non-trivial, and we can set any one of these two BZ parameters to unity. In this paper, we fix $m^{(k)}_{02}=1$, so $m^{(k)}_0=(m^{(k)}_{01},1)$. Here, $m^{(k)}_{01}$ will be referred to as the BZ parameter of the soliton $\mu_k$. Thus, for each soliton added, there are two parameters introduced into the new solution: its location $z_k$ and its BZ parameter $m^{(k)}_{01}$. Note that, in particular, there is a well-defined limit of the new solution when $m^{(k)}_{01}\rightarrow \infty$ for any $k$. By the above-mentioned rescaling property of the vectors $m^{(k)}$, this limit is equivalent to the solution constructed using a BZ vector $(1,0)$ for the soliton $\mu_k$.

When $z_k$ and $m^{(k)}_{01}$ are real, the soliton solution we obtain is real. This is the case we will focus on, which is known to include various interesting stationary and axisymmetric solutions as well as instanton solutions with two axial symmetries. However, we point out that if we have pairs of solitons added, with each pair possessing complex-conjugate locations and BZ parameters, the resulting soliton solution is also real. This case will not be studied in this paper; the interested reader is referred to \cite{Belinski:2001ph} for a review.

\subsection{Generalised seeds}

In Belinski and Zakharov's original definition \cite{Belinsky:1971nt,Belinsky:1979mh}, a seed is a particular solution of the vacuum Einstein equations which is also assumed to satisfy the condition (\ref{rho}). Thus it consists of the information of a pair $(g_0,f_0)$ which satisfy all of (\ref{rho}), (\ref{G}) and (\ref{nu}). The Levi-Civita metric (\ref{LeviC}) is such a seed that has been extensively used in the literature. We now slightly generalise this definition of a seed to incorporate more general classes of metrics on which the ISM can be applied.

We see from the previous subsection that the only necessary input information for the construction of soliton solutions of the ISM is a $2\times 2$ matrix $g_0$ which satisfies (\ref{G}). The entire integration procedure can be carried throughout whether (\ref{rho}) holds or not. In particular, the integrability condition of Eq.~(\ref{nu}) for $f$ and (\ref{Lax_pair}) for $\psi$ does not involve (\ref{rho}). We thus define a generalised seed as a $2\times 2$ matrix $g_0$ that satisfies (\ref{G}), or a metric in the form (\ref{Weyl_Papapetrou_form}) that has such a $g$-matrix. Given a generalised seed, one can construct soliton solutions by first constructing the new $g$-matrix (\ref{unphysical_g}) and normalising it to $g^{\text{(ph)}}$, which satisfies (\ref{rho}). The corresponding $f^{\text{(ph)}}$ can then be solved, which, together with $g^{\text{(ph)}}$, gives rise to the soliton solution on the generalised seed.

This definition of generalised seeds then encompasses, in addition to traditional seeds, the seeds (\ref{metric_seed_odd_Lorentzian}) and (\ref{metric_seed_odd_Euclidean}), which have an $f_0$ that does not satisfy (\ref{nu}). According to this definition, the $g$-matrix (\ref{unphysical_g}) obtained after a soliton transformation on some generalised seed is again a generalised seed. So soliton-transformations can be applied successively on generalised seeds \cite{Belinsky:1971nt,Belinsky:1979mh}. Intermediate seeds in the successive construction obviously may violate (\ref{rho}). It is clear that generalised seeds themselves may not necessarily correspond to solutions to the vacuum Einstein equations. This fact, however, does not prevent us from using them to apply the ISM. We use generalised seeds in the rest of this paper.

It is not difficult to see from the ISM integration scheme that two generalised seeds $g_{A}$ and $g_B$ lead to the same final solution if they are related by an overall factor, i.e., $g_B=hg_A$ for some $h$. This is because an overall scaling of a seed commutes with soliton transformations \cite{Belinsky:1971nt,Belinsky:1979mh,Pomeransky:2005sj}. Note that, as we have shown, $\ln |h|$ should solve (\ref{lap}). We say that these two seeds are equivalent. So, after the normalisation procedure is performed, the final solutions that two equivalent seeds lead to are the same (up to coordinate transformations and redefinitions of parameters). It is then sufficient to consider equivalence classes of generalised seeds to apply the ISM. Within each equivalence class, there is one seed which satisfies Eqs.~(\ref{rho}), (\ref{G}) and (\ref{nu}), and it is a solution to the vacuum Einstein equations. This can be seen from the fact that any given generalised seed $g_A$ can be normalised to an equivalent one ($g_A^{\text{(ph)}}, f_A^{\text{(ph)}}$). So the equivalence classes of generalised seeds are in one-to-one correspondence to traditional seeds. The real advantage of using generalised seeds lies in practice: one is now free to choose an as-simple-as-possible representative in each equivalence class of seeds to carry out the ISM.

As an example, the seed (\ref{metric_seed_odd_Lorentzian}) and the seed (\ref{LeviC}) with $d=0$ are equivalent, and we will choose the former to carry out the ISM for simplicity. As another example, one can now immediately find the general diagonal seed that has no $z$-dependence. In this case, Eq.~(\ref{G}) can be directly solved, and the general solution is
\begin{align}
\label{general_g0}
{g}_{ 0}=\left[ \begin {array}{cc} \epsilon_1\rho^{s_1}&0\\0&\epsilon_2\rho^{s_2}\end {array} \right].
\end{align}
Here, $\epsilon_{1,2}$ are possible sign choices, and $s_{1,2}$ are arbitrary integration constants. Notice that there is now no need to impose the constraint $s_1+s_2=2$. The equivalence classes of the general diagonal seed (\ref{general_g0}) are in one-to-one correspondence to the Levi-Civita seed (\ref{LeviC}): the seed (\ref{general_g0}) is equivalent to (\ref{LeviC}) with $2d=s_1-s_2$ (with sign choices of $\epsilon_{1,2}$ appropriately taken care of). A further simplification of (\ref{general_g0}) can be made by using the normalisation freedom to set $s_1=0$, so the general diagonal seed is indexed by $s_2$, which can be assumed to be positive. The ISM construction is obviously easier to be applied on the seed (\ref{general_g0}) with $s_1=0$ than to be applied on (\ref{LeviC}). Flat-space seeds (\ref{metric_seed_even_Lorentzian}), (\ref{metric_seed_odd_Lorentzian}), (\ref{metric_seed_even_Euclidean}) and (\ref{metric_seed_odd_Euclidean}) correspond to the cases $s_2=0,2$. As mentioned, they are the only values of $s_2$ that give rise to locally regular stationary and axisymmetric solutions or locally regular instanton solutions with two axial symmetries, if an odd and even number of solitons are added, respectively.

One last comment we want to give is that Belinski and Zakharov \cite{Belinsky:1979mh} showed that the $n$-soliton transformation (\ref{unphysical_g}) on a seed is equivalent to the transformation of adding the $n$ solitons successively. Symbolically, we write
\begin{align}
n\sim n_1+n_2+...+n_p\,,
\label{decomposition_solitons}
\end{align}
with $\sum\limits_{i=1}^p n_i=n$. The right-hand side represents the operation of adding these $n$ solitons in $p$ steps, with $n_i$ solitons added in the $i$-th step. As an example of (\ref{decomposition_solitons}), we have $5\sim 1+4\sim 3+2$ on any given seed. In the ISM construction of the next section, we add all solitons at one time. In Sec.~\ref{sec_soliton_calculus}, we use the equivalence equation (\ref{decomposition_solitons}) to deduce relations of various multi-soliton solutions on flat space without using the explicit forms of these solutions.

\section{The $n$-soliton solution on flat space}

\label{sec_n-soliton}

In this section, we first review how even-soliton solutions are constructed on the seed (\ref{metric_seed_even_Euclidean}) and then use the proposed seed (\ref{metric_seed_odd_Euclidean}) to construct odd-soliton solutions. The $n$-soliton solution on flat space for any positive integer $n$ is then defined, and it is shown to reduce to the $(n-1)$-soliton solution on flat space in the transition limit. It has the Euclidean signature, and we show how a natural Lorentzian section of it can be obtained. The static limit and asymptotic structure of the $n$-soliton solution on flat space are then discussed.

\subsection{Even-soliton solutions on flat space}

\label{sec_even}

The construction of even-soliton solutions on the seed (\ref{metric_seed_even_Lorentzian}) was first carried out in \cite{Belinsky:1971nt}. Here we use the seed (\ref{metric_seed_even_Euclidean}). Its $g$-matrix and generating matrix are trivial to write down
\begin{align}
\label{seed2}
{g}_{ 0}^{\text{(even)}}=\left[ \begin {array}{cc} 1&0\\0&\rho^2\end {array} \right],\qquad {\psi}_{ 0}^{\text{(even)}}=\left[ \begin {array}{cc} 1&0\\0&\rho^2-2\lambda z-\lambda^2\end {array} \right].
\end{align}
Choose the BZ vectors $m_{0}^{(k)}$ as follows:
\begin{align}
\label{BZ1}
m_{0}^{(k)\text{(even)}}=\left(-\frac{C_k}{2z_k},1\right),
\end{align}
where $z_k$ is the location of the soliton $\mu_k$, and $C_k$ is its BZ parameter by slight abuse of notation. Using the identity $\rho^2-2z\mu_k-\mu_k^2=-2z_k\mu_k$, the $m$-vectors $m^{(k)}$ are calculated to be
\begin{align}
m^{(k)\text{(even)}}=\left(C_k,\mu_k^{-1}\right),
\end{align}
up to irrelevant constant factors. The $L$-vectors $L^{(k)}$ are then
\begin{align}
L^{(k)\text{(even)}}=\left(C_k,\rho^2\mu_k^{-1}\right),
\label{L_even}
\end{align}
and the components of the $\Gamma$-matrix are given by
\begin{align}
\Gamma_{kl}^{\text{(even)}}=\frac{\rho^2+C_kC_l\mu_k\mu_l}{\mu_k\mu_lR_{kl}}\,,\quad R_{kl}\equiv \rho^2+\mu_k\mu_l\,.
\label{Gamma_even}
\end{align}
The unphysical $g$-matrix (\ref{unphysical_g}) is then determined unambiguously by the above specifications (\ref{L_even}) and (\ref{Gamma_even}).

The physical value of the $g$-matrix is given by (\ref{physical_g}), and the corresponding function $f^{\text{(ph)}}$ can be found after some algebra; they can be written as follows:
\begin{align}
\label{normalisation_even} g^{\text{(ph)}}&=h_1^{\text{(even)}} g\,,\quad h_1^{\text{(even)}}\equiv(i\rho)^{-n}\prod_{k=1}^n\mu_k\,,\nonumber\\
 f^{\text{(ph)}}&=h_2^{\text{(even)}}\det \Gamma^{\text{(even)}}\,,\quad h_2^{\text{(even)}}\equiv k_0\rho^{-n^2/2}\left(\prod_{k=1}^n\mu_k\right)^{3-n}\left[\prod_{n\ge k>l\ge 1}R_{kl}^{2}\right],
\end{align}
where $k_0$ is an arbitrary integration constant. One can check that the metric now satisfies $\det g^{\text{(ph)}}=\rho^2$, as required by a physical solution. This completes the construction of even-soliton solutions on flat space. We define the 0-soliton solution on flat space to be the seed (\ref{metric_seed_even_Euclidean}) itself.

\subsection{Odd-soliton solutions on flat space}

\label{sec_odd}

The seed (\ref{metric_seed_odd_Euclidean}) is used for odd-soliton solutions. Its $g$-matrix and generating matrix are also trivial to write down
\begin{align}
\label{seed4}
{g}_{0}^{\text{(odd)}}=\left[ \begin {array}{cc} -1&0\\0&1\end {array} \right],\qquad {\psi}_{ 0}^{\text{(odd)}}=\left[ \begin {array}{cc} -1&0\\0&{1}\end {array} \right].
\end{align}
The BZ vectors $m_{0}^{(k)}$ and the $m$-vectors $m^{(k)}$ are given as follows:
\begin{align}
\label{BZ2}
m_{0}^{(k)\text{(odd)}}=\left(C_k,1\right),\quad m^{(k)\text{(odd)}}=\left(-C_k,1\right),
\end{align}
where the $C_k$'s are the BZ parameters. The $L$-vectors $L^{(k)}$ are then
\begin{align}
L^{(k)\text{(odd)}}=\left(C_k,1\right),
\end{align}
and the components of the $\Gamma$-matrix are given by
\begin{align}
\Gamma_{kl}^{\text{(odd)}}=\frac{1-C_kC_l}{R_{kl}}\,.
\end{align}
The unphysical $g$-matrix is then given by (\ref{unphysical_g}) unambiguously. The physical value of the $g$-matrix and its corresponding function $f^{\text{(ph)}}$ are found to be
\begin{align}
\label{normalisation_odd}
g^{\text{(ph)}}&=h_1^{\text{(odd)}} g\,,\quad h_1^{\text{(odd)}}\equiv -(i\rho)^{1-n}\left(\prod_{k=1}^n\mu_k\right),\nonumber\\
 f^{\text{(ph)}}&=h_2^{\text{(odd)}}\det \Gamma^{\text{(odd)}}\,,\quad h_2^{\text{(odd)}}\equiv k_0\rho^{-(n-1)^2/2}\left(\prod_{k=1}^n\mu_k\right)^{2-n}\left(\prod_{n\ge k>l\ge1}R_{kl}^{2}\right),
\end{align}
where $k_0$ is an arbitrary integration constant. This completes the construction of odd-soliton solutions on flat space.

From now on, for simplicity, we omit the label ``(ph)''. Only the physical values of the $g$-matrix and $f$ of a solution will be used in the rest of the paper.

\subsection{The $n$-soliton solution on flat space and transition limit}

\label{sec_transition_limit}

We define the $n$-soliton solution on flat space for any positive integer $n$, as the solution (\ref{normalisation_even}) when $n$ is even and the solution (\ref{normalisation_odd}) when $n$ is odd. In the rest of this section, we study some of its properties. Detailed study of a few concrete examples of the $n$-soliton solution on flat space can be found in Sec.~\ref{sec_examples}.

The metric components of the $n$-soliton solution on flat space are algebraic expressions of the three types of quantities
\begin{align}
\rho^2\,,\quad \mu_k\quad\text{and}\quad C_k\,,
\end{align}
for $k=1,2,...,n$, and we recall that $C_k$ is the BZ parameter of the soliton $\mu_k$. In particular, the quantities $f$ and $g_{ab}f$ are all quadratic polynomials in terms of each BZ parameter $C_k$. The constant $k_0$ in the function $f$ is not an independent parameter, so one can set it to an arbitrary non-zero constant without loss of generality. The coordinate $z$ and the locations of the solitons $z_k$ only appear in the metric components through the solitons $\mu_k$. It is then clear that the metric of the $n$-soliton solution on flat space remains invariant under the following operations:
\begin{align}
z\rightarrow z+z_0\,,\quad z_k\rightarrow z_k+z_0\,\,\,\text{for all $1\leq k\leq n$},
\label{symmetry_translational}
\end{align}
where $z_0$ is an arbitrary constant. So, for even $n$, the $n$-soliton solution is a $(2n-1)$-parameter class: there are $n-1$ independent soliton-location parameters, and $n$ independent BZ parameters. For odd $n$, we show that the $n$-soliton solution possesses a one-parameter redundancy among the BZ parameters $C_k$, so it is a $(2n-2)$-parameter class.

Operations such as (\ref{symmetry_translational}) which leave the metric of a solution invariant are called symmetries. One can show that the $n$-soliton solution on flat space possesses a few more symmetries, in addition to the translational one (\ref{symmetry_translational}). They can be summarised as follows.
\begin{enumerate}
  \item Inversion symmetry:
  
  For any $1\leq k\leq n$, the metric is invariant under the simultaneous substitutions:
  \begin{align}
  \mu_k\rightarrow -\frac{\rho^2}{\mu_k}\,,\quad C_k\rightarrow \frac{1}{C_k}\,,\quad k_0\rightarrow C_k^2k_0\,.
  \label{symmetry_inversion}
  \end{align}
  If $\mu_k=\mu_k^{\pm}$, then $-\frac{\rho^2}{\mu_k}=\mu_{k}^{\mp}$. (\ref{symmetry_inversion}) is a symmetry between a soliton and its anti-partner.
  \item Swapping symmetry:
  
  For any pair $1\leq k,l\leq n$, the metric is invariant under the simultaneous substitutions:
  \begin{align}
 \mu_k\leftrightarrow \mu_l\,,\quad  C_k\leftrightarrow  C_l\,.
 \label{symmetry_swapping}
  \end{align}
  This is a symmetry between two solitons.
  \item Reflection symmetry:
  
  The metric is invariant under the simultaneous substitutions:
  \begin{align}
  \tau\rightarrow -\tau\,,\quad C_k\rightarrow -C_k\,\,\,\text{for all $1\leq k\leq n$}.
  \label{symmetry_reflection}
  \end{align}
  This is a symmetry involving all solitons.
\end{enumerate}
These symmetries have interesting implications, as we show in this and subsequent sections.

The inversion symmetry tells us that changing a soliton to its corresponding anti-soliton does not change the metric of the $n$-soliton solution on flat space, as long as one changes its BZ parameter $C_k$ to $C_k^{-1}$ and appropriately redefines the integration constant $k_0$. This means that, for the $n$-soliton solution on flat space, it is sufficient to consider only true solitons $\mu_k^+$ (and not anti-solitons $\mu_k^{-}$) without loss of generality. From this point onwards, unless otherwise specified, the choice $\mu_k=\mu_k^+$ is made for concreteness and clarity of presentation.

The swapping symmetry implies that, in our formulation of the $n$-soliton solution on flat space, all solitons are treated on an equal footing. So without loss of generality, one can impose
\begin{align}
\label{ordering_solitons}
z_1<z_2<...<z_n\,.
\end{align}
This ordering of the locations of solitons also gives an ordering of the solitons themselves, i.e., $\mu_1<\mu_2<...<\mu_n$, for real $\rho$.

We further discuss the implications of the various symmetries in subsequent sections. It will be clear that the above symmetry property is just one among many properties of both the even- and odd-soliton solutions on flat space. It is thus natural to treat these two classes of solutions as one, namely, the $n$-soliton solution on flat space. We now turn to another property that justifies this unified treatment. There should exist a limit in which one can reduce the $n$-soliton solution on flat space to the $(n-1)$-soliton solution on flat space, for any $n$. This limit will be called the transition limit. To find it, we first observe that a soliton $\mu_k$ trivially becomes a constant in the limit $z_k\rightarrow +\infty$. Indeed, focusing on regions where $\rho$ and $z$ are finite, we find that
\begin{align}
\mu_k\simeq 2z_k\,,\,\, \text{if}\,\, z_k\rightarrow +\infty\,.
\end{align}
Bearing in mind the ordering (\ref{ordering_solitons}), we should set $z_n$ to infinity, to remove the effect of a soliton in the transition limit. Since $z_n$ enters the metric only through the soliton $\mu_n$, the transition limit should be taken by treating $\mu_n$ as a constant and setting $\mu_n\rightarrow +\infty$. The coordinates $\rho$, $z$, and other solitons $\mu_k$ for $1\le k\leq n-1$ should be kept fixed in the limit.

More specifically, one can show that, starting from the $n$-soliton solution on flat space for odd $n$, if we set the BZ parameter $C_n$ as
\begin{align}
C_{n}\rightarrow 0\,,
\label{odd-to-even_BZ}
\end{align}
rescale the metric by making the substitutions
\begin{align}
\tau\rightarrow ({\mu_{n}})^{-\frac{1}{2}}\tau\,,\quad \phi\rightarrow({\mu_{n}})^{\frac{1}{2}}\phi \,,\quad 
f\rightarrow(\mu_{n})^{2-n}f\,, \label{odd-to-even}
\end{align}
and take the limit $\mu_n\rightarrow +\infty$, the $(n-1)$-soliton solution on flat space is recovered exactly. Notice that in the limit $\mu_n\rightarrow +\infty$, the substitutions in (\ref{odd-to-even}) are just constant rescalings; a rescaling of $f$ amounts to a redefinition of the integration constant $k_0$. Similarly, starting from the $n$-soliton solution on flat space for even $n$, we first substitute $k_0\rightarrow C_n^{-2}k_0$ and set the BZ parameter $C_n$ as
\begin{align}
C_{n}\rightarrow +\infty\,.
\label{even-to-odd_BZ}
\end{align}
If we further rescale the metric by making the substitutions
\begin{align}
\tau\rightarrow ({\mu_{n}})^{\frac{1}{2}}\tau\,,\quad \phi\rightarrow({\mu_{n}})^{-\frac{1}{2}}\phi\,,\quad 
f\rightarrow(\mu_{n})^{1-n}f\,,
\label{even-to-odd}
\end{align}
and take the limit $\mu_{n}\rightarrow +\infty$, the $(n-1)$-soliton solution on flat space is recovered exactly. 

The transition limit can equally be obtained by setting $z_n\rightarrow -\infty$. The contradiction in this case with the ordering (\ref{ordering_solitons}) is not an issue since we can relabel all solitons if necessary. Notice that $\mu_n\simeq -\frac{\rho^2}{2z_n}\rightarrow 0^{+}$ as $z_n\rightarrow -\infty$ and contrast this with the case $\mu_n\simeq 2z_n\rightarrow +\infty$ as $z_n\rightarrow +\infty$. We see that the $\mu_n$'s in these two cases formally become the anti-soliton of each other. Based on the discussion in the proceeding paragraph for $z_n\rightarrow +\infty$ and the inversion symmetry (\ref{symmetry_inversion}), it is not difficult to work out how, in the case $z_n\rightarrow -\infty$, we should fix the BZ parameter $C_n$ and rescale the metric to obtain the $(n-1)$-soliton solution from the $n$-soliton solution.

\subsection{Lorentzian section of the $n$-soliton solution on flat space}

The $n$-soliton solution on flat space we constructed has the Euclidean signature satisfying $\det g=\rho^2$. It has a natural Wick rotation, given by setting
\begin{align}
\label{wick_rotation}
\tau\rightarrow -it\,,\quad C_k\rightarrow i\mathcal{C}_k\,.
\end{align}
From our ISM construction, it is straightforward to check that the Wick rotation leaves all metric components real and changes $\det g$ from $\rho^2$ to $-\rho^2$. Hence, the Wick rotation changes the signature of the solution from being Euclidean to being Lorentzian. Performing the inverse Wick rotation, one can then change the Lorentzian $n$-soliton solution on flat space to a Euclidean one. So, the $n$-soliton solution on flat space we constructed admits both Lorentzian and Euclidean sections.

The $n$-soliton solution on flat space with the Lorentzian signature can be directly constructed using the seeds (\ref{metric_seed_even_Lorentzian}) and (\ref{metric_seed_odd_Lorentzian}), with a (real) BZ parameter $\mathcal{C}_k$ for each soliton $\mu_k$, along similar lines to the construction we did in Secs.~\ref{sec_even} and \ref{sec_odd}. In fact, the above two approaches yield identical final solutions. Hence, we will not distinguish the Lorentzian $n$-soliton solution based on the seeds (\ref{metric_seed_even_Lorentzian}) and (\ref{metric_seed_odd_Lorentzian}), and the Lorentzian section of the Euclidean $n$-soliton solution based on the seeds (\ref{metric_seed_even_Euclidean}) and (\ref{metric_seed_odd_Euclidean}). This is the reason why we can use the seeds (\ref{metric_seed_even_Euclidean}) and (\ref{metric_seed_odd_Euclidean}), while completely ignoring (\ref{metric_seed_even_Lorentzian}) and (\ref{metric_seed_odd_Lorentzian}), in our ISM construction. As long as Wick rotations are included, the constructed $n$-soliton solution on flat space can have a demanded Lorentzian or Euclidean signature.

We remark that the sign of the constant $k_0$ is fixed by the condition that the metric component $f$ should remain positive, in both its Euclidean and Lorentzian sections. In the ``static limit'' of the Euclidean section, to be discussed in the following subsection, $k_0$ should take the positive sign. The $n$-soliton solution on flat space also admits sections with signatures $(-,-,+,+)$, $(-,-,-,+)$ or $(-,-,-,-)$. These cases will not be considered in this paper.

\subsection{Static limit and asymptotic structure of the $n$-soliton solution}

\label{sec_asymptotic_structure}

We now focus on the $n$-soliton solution on flat space constructed in Secs.~\ref{sec_even} and \ref{sec_odd}, with the Euclidean signature. We sometimes refer to it simply as the $n$-soliton solution in the rest of the paper, without mentioning the seeds we use. The reader is reminded that the flat-space seeds (\ref{metric_seed_even_Euclidean}) and (\ref{metric_seed_odd_Euclidean}) are assumed for even and odd $n$, respectively.

The static limit for the 2-soliton solution was known to be taken by first setting
$C_1\rightarrow 0$ and then setting $C_2\rightarrow \infty$, in which the Schwarzschild solution is recovered. It turns out that the static limit of the $n$-soliton solution (with the locations of solitons ordered as (\ref{ordering_solitons})) can be taken by first setting $C_{1,3,...,2\lfloor\frac{n+1}{2}\rfloor-1}\rightarrow 0$ and then setting $C_{2,4,...,2\lfloor\frac{n}{2}\rfloor}\rightarrow \infty$; we simply write this limit as 
\begin{align}
\label{limit_static} C_{1,3,...,2\lfloor\frac{n+1}{2}\rfloor-1}\rightarrow 0\,,\quad C_{2,4,...,2\lfloor\frac{n}{2}\rfloor}\rightarrow \infty\,.
\end{align}
To have a well-defined metric component $f$, one needs to first define $k_0=(C_2C_4...C_{2\lfloor\frac{n}{2}\rfloor})^{-2}$ and then take the limit (\ref{limit_static}). The static $n$-soliton solution can be written explicitly as follows:
\begin{align}
g_{\text{{static}}}&=\left[ \begin {array}{cc} \frac{\mu_1\mu_3...\mu_{2\lfloor\frac{n+1}{2}\rfloor-1}}{\mu_2\mu_4...\mu_{2\lfloor\frac{n}{2}\rfloor}}&0\\0&{\frac {{\rho}^{2}\mu_2\mu_4...\mu_{2\lfloor\frac{n}{2}\rfloor}}{{\mu_1\mu_3...\mu_{2\lfloor\frac{n+1}{2}\rfloor-1}}}}\end {array} \right],\nonumber\\
f_{\text{{static}}}&=\frac{\prod\limits_{q-p=0\text{ mod }2}^{1\le p<q\le n}\mu_{pq}^2\prod\limits_{s-r=1\text{ mod }2}^{1\le r<s\le n}R_{rs}^2}{\prod\limits_{1\leq i\leq n}(\mu_i^{n-2}R_{ii})}\times \Bigg\{\begin{array}{ll}
1& \text{for odd }n\\ \frac {\mu_2\mu_4...\mu_{2\lfloor\frac{n}{2}\rfloor}}{{\mu_1\mu_3...\mu_{2\lfloor\frac{n+1}{2}\rfloor-1}}}&\text{for even }n,
\label{metric_limit_static}
\end{array}
\end{align}
where we have defined $\mu_{pq}\equiv\mu_p-\mu_q$.

\begin{figure}[!t]
\begin{center}
\includegraphics[]{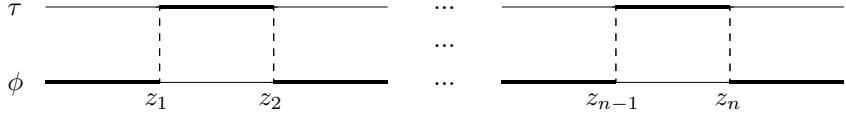}
\caption{Rod sources of the static $n$-soliton solution on flat space for even $n$, all with mass one half per unit length.}
\label{static_limit_even}
\end{center}
\end{figure}

The static $n$-soliton solution (\ref{metric_limit_static}) on flat space belongs to the well-known Weyl's class of solutions. Such solutions are characterised by linear rod sources on an auxiliary three-dimensional flat space with certain densities (masses per unit length), for each coordinate $\tau$ and $\phi$ (see, e.g., \cite{Emparan:2001wk}). The rod sources of the static $n$-soliton solution (\ref{metric_limit_static}) for even $n$ are shown in Fig.~\ref{static_limit_even}. In the Lorentzian section, each rod for the coordinate $\tau$ represents a Killing horizon and each rod for the coordinate $\phi$ represents an axis \cite{Emparan:2001wk}. We thus see that, when $n$ is even, the static $n$-soliton solution describes a number $\frac{n}{2}$ of Schwarzschild black holes in superposition.  

It is easy to see that in the static $n$-soliton solution for even $n$, $g_{\tau\tau}\rightarrow 1$ at infinity $(\rho,z)\rightarrow (\infty,\infty)$. This continues to hold in the general non-diagonal case when all BZ parameters are finite and non-zero. In this case, the asymptotic structure of the $n$-soliton solution can be found by first deriving the asymptotic forms of the $m$- and $n$-vectors and then substituting them into the metric components. It was shown in \cite{Belinski:2001ph} that, if we perform linear coordinate transformations
\begin{align}
\label{linear_transformation_even}
\tau= \tilde{\tau}+b\tilde{\phi}\,,\quad \phi=\pm \tilde{\phi}\,,
\end{align}
define coordinates $(r,\theta)$ as
\begin{align}
\rho=r\sin\theta\,,\quad z=r\cos\theta\,,
\label{radial_coordinate_even}
\end{align}
and appropriately choose the constants $b$ and $k_0$, the $n$-soliton solution for even $n$ approaches
\begin{align}
\label{ALF_Euclidean}
\dif s^{2\text{(even)}}\rightarrow \left(\dif \tilde{\tau}+2n_{\text{nut}}\cos\theta\,\dif \tilde{\phi}\right)^2+\dif r^2+r^2\left(\dif\theta^2+\sin^2\theta\,\dif\tilde{\phi}^2\right),
\end{align}
at infinity $r\rightarrow \infty$. The radial coordinate $r$ in (\ref{radial_coordinate_even}) can be found by equating $\det g$ of (\ref{ALF_Euclidean}) and that of the $n$-soliton solution, which is $\rho^2$. Notice that the linear coordinate transformations (\ref{linear_transformation_even}) do not change $g_{11}$, so $g_{\tilde{\tau}\tilde{\tau}}=g_{\tau\tau}$. The so-called Newman--Unti--Tamburino (NUT) charge $n_{\text{nut}}$ is a function of all of the parameters $z_k$ and $C_k$. In view of (\ref{ALF_Euclidean}), the $n$-soliton solution for even $n$ is said to be asymptotically locally flat (ALF) according to the classification of \cite{Gibbons:1979gd} when global topology is not of concern. The special case with $n_{\text{nut}}=0$ is said to be asymptotically flat (AF). In the Lorentzian section, the $n$-soliton solution for even $n$ approaches
\begin{align}
\label{ALF_Lorentzian}
\dif s^{2\text{(even)}}\rightarrow -\left(\dif \tilde{t}+2n_{\text{nut}}\cos\theta\,\dif \tilde{\phi}\right)^2+\dif r^2+r^2\left(\dif\theta^2+\sin^2\theta\,\dif\tilde{\phi}^2\right),
\end{align}
at $r\rightarrow \infty$. The NUT charge is now a function of the parameters $z_k$ and $\mathcal{C}_k$.

\begin{figure}[!t]
\begin{center}
\includegraphics[]{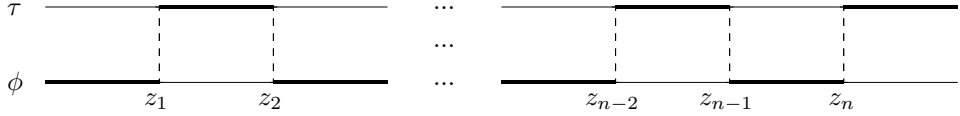}
\caption{Rod sources of the static $n$-soliton solution on flat space for odd $n$, all with mass one half per unit length.}
\label{static_limit_odd}
\end{center}
\end{figure}

When $n$ is odd, the rod sources of the static $n$-soliton solution (\ref{metric_limit_static}) are shown in Fig.~\ref{static_limit_odd}. Compared to Fig.~\ref{static_limit_even}, the last semi-infinite rod is now on the coordinate $\tau$, which is known to describe an acceleration horizon in the Lorentzian section \cite{Emparan:2001wk}. It will be clear later that the static $1$-soliton solution is the Minkowski space in an accelerating frame, or the ``accelerating nothing''. In this frame, the space-time possesses an acceleration horizon. The static $3$-soliton solution is the C-metric, which describes an accelerating Schwarzschild black hole \cite{Kinnersley:1970zw}. It is thus natural to interpret the static $n$-soliton solution for odd $n$ as describing an accelerating system of a number $\lfloor\frac{n}{2}\rfloor$ of Schwarzschild black holes, in the presence of an acceleration horizon.

Similar lines as in the case of even $n$ can be followed to study the asymptotic structure of the $n$-soliton solution in the case of odd $n$. When global topology is not of concern, it is asymptotically locally Euclidean (ALE), according to the classification of \cite{Gibbons:1979gd}. This means that if we perform appropriate linear coordinate transformations from $(\tau,\phi)$ to $(\tilde{\tau},\tilde{\phi})$, define coordinates $(r,\theta)$ as
\begin{align}
\rho= \frac{1}{2}r^2\sin 2\theta\,,\qquad z=\frac{1}{2}r^2\cos 2\theta\,,
\label{radial_coordinate_odd}
\end{align}
and appropriately choose the constant $k_0$, the $n$-soliton solution for odd $n$ approaches
\begin{align}
\label{ALE_Euclidean}
\dif s^{2\text{(odd)}}\rightarrow \dif r^2+r^2\left(\dif\theta^2+\sin^2\theta\,\dif \tilde{\tau}^2+\cos^2\theta\,\dif\tilde{\phi}^2\right),
\end{align}
at infinity $r\rightarrow \infty$. The static $n$-soliton solution for odd $n$ obviously has this asymptotic structure. The radial coordinate $r$ in (\ref{radial_coordinate_odd}) can be found by equating $\det g$ of (\ref{ALE_Euclidean}) and that of the $n$-soliton solution. It will be clear later that the linear coordinate transformations we need to perform have the form
\begin{align} \label{linear_transformation_odd}
	\tau= b_1\tilde{\tau}+a_{1}\tilde{\phi}\,,\quad \phi= a_1\tilde{\tau}+b_{1}\tilde{\phi}\,,
\end{align}
for some constants $a_1$ and $b_1$. In the Lorentzian section, the $n$-soliton solution for odd $n$ approaches
\begin{align}
\label{ALE_Lorentzian}
\dif s^{2\text{(odd)}}\rightarrow-r^2\sin^2\theta\,\dif \tilde{t}^2+\dif r^2+r^2\left(\dif\theta^2+\cos^2\theta\,\dif\tilde{\phi}^2\right),
\end{align}
at infinity $r\rightarrow \infty$. This asymptotic space-time is just flat space in an accelerating frame, with $\tilde{t}$ being the Rindler time \cite{Rindler:1966zz}.

The coefficients in (\ref{linear_transformation_odd}) are determined by the rod structure of the $n$-soliton solution to be studied in the next section. In fact, $(a_1,b_1)$ is the direction of the first rod. The constant $b$ in (\ref{linear_transformation_even}) and $n_{\text{nut}}$ in (\ref{ALF_Euclidean}) are also fixed by the rod structure. The derivation of these constants and $k_0$ can be found in Sec.~\ref{subsec_rod_structure}. We say that the linear coordinate transformations (\ref{linear_transformation_even}) and (\ref{linear_transformation_odd}) bring the $n$-soliton solution to the natural orientation,\footnote{For ALE solutions, the natural orientation defined here coincides with the standard orientation defined in \cite{Chen:2010zu}, while for ALF solutions, these two orientations are different and it turns out that the natural orientation is slightly simpler.} in the coordinates $(\tilde{\tau},\tilde{\phi},\rho,z)$. These coordinates are invariantly defined for ALF and ALE solutions and are useful when we compare two solutions with the same asymptotic structure. They will be frequently used implicitly when we study several concrete examples in Sec.~\ref{sec_examples}. The tildes in these coordinates will be omitted when no confusion is caused.

In the general non-diagonal case, the Lorentzian $n$-soliton solution for even $n$ can be interpreted as a number $\frac{n}{2}$ of Kerr-NUT black holes in superposition \cite{Belinsky:1971nt}. Each black hole now carries a certain amount of rotation and NUT charge. In view of the above results, it is natural to interpret the Lorentzian $n$-soliton solution for odd $n$ as an accelerating system of a number $\lfloor\frac{n}{2}\rfloor$ of Kerr-NUT black holes, in the presence of an acceleration horizon. So the $n$-soliton solution on flat space is one explicit representation of the multi-Kerr-NUT solution and its accelerating generalisation, with a demanded Lorentzian or Euclidean signature.

\section{$[n-m]$-soliton solutions on flat space}

\label{sec_n-m-soliton}

Having understood the asymptotic structure of the $n$-soliton solution, we now proceed to examine in detail the local geometry around its interior points, i.e., those points with finite coordinates $\rho$ and $z$. This is done with the help of the rod-structure formalism. We find from the rod structure of the $n$-soliton solution how one or more of its solitons can be eliminated non-trivially. This allows us to define and study $[n-m]$-soliton solutions on flat space.

\subsection{The rod-structure formalism}

The $n$-soliton solution possesses two commuting and non-orthogonal Killing vectors. In the Lorentzian section, they generate the stationary time flow and an axial symmetry, while in the Euclidean section, they generate two axial symmetries. In the static limit when the two Killing vectors become orthogonal, the $n$-soliton solution belongs to the Weyl's class, and we have shown in Sec.~\ref{sec_asymptotic_structure} that the rod-source characterisation can be used to study it. When the two Killing vectors are non-orthogonal, a generalisation of the rod-source characterisation called the rod-structure formalism \cite{Harmark:2004rm,Hollands:2007aj,Chen:2010zu} will prove to be useful. The rod structure contains certain invariants of a solution by studying the behaviour of the metric near $\rho\rightarrow 0$ and encodes useful geometrical and topological information of that solution. We now study the rod structure of the (Euclidean) $n$-soliton solution.

\begin{figure}[!t]
	\begin{center}
		\includegraphics[]{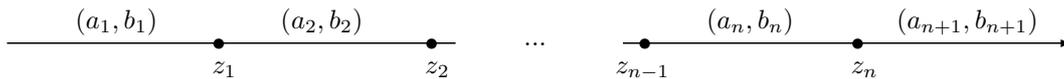}
		\caption{The rod structure of the $n$-soliton solution: the $n$ turning points $z_{1,...,n}$, corresponding to the locations of the $n$ solitons, divide the $z$-axis into $n+1$ rods, each assigned with a direction placed above it. The directions $(a_k,b_k)$ for $1\le k\le n+1$ are explicitly given by (\ref{first_direction}) and (\ref{recursive_relation}).}
		\label{fig_rod_structure}
	\end{center}
\end{figure}

For the $n$-soliton solution, two apparent Killing vectors are $\partial_\tau$ and $\partial_\phi$. The metric is non-degenerate if $\rho>0$ and becomes degenerate at $\rho=0$, i.e., on the $z$-axis, due to the condition (\ref{rho}). One then looks for locations on the $z$-axis where the $g$-matrix vanishes with a two-dimensional kernel. It turns out that these locations are exactly where the solitons are, i.e., at $z=z_k$ (and $\rho=0$). There are $n$ of them of course, and they are called turning points of the solution. They divide the $z$-axis into $n+1$ rods. These rods are labelled from left to right as Rod $1$, Rod 2, ..., Rod $n+1$. So Rod $k$ refers to the part of the $z$-axis with coordinates $(\rho=0,z_{k-1}\leq z\leq z_k)$, where we define $z_0=-\infty, z_{n+1}=\infty$ for convenience. Each rod say Rod $k$, is assigned a direction say $l_k$, i.e., a Killing vector in the form
\begin{align} l_k=a_k\partial_{\tau}+b_k\partial_\phi\,,
\end{align}
abbreviated as $l_k= (a_k,b_k)$, where $a_k$ and $b_k$ are two constants to be determined. The specification of the locations and directions of all rods of a solution under study is known as the rod structure of that solution. The rod structure of the $n$-soliton solution is illustrated in Fig.~\ref{fig_rod_structure}.

The direction of a rod say Rod $k$ is determined as follows. Within Rod $k$ $(\rho=0,z_{k-1}<z<z_k)$, the $g$-matrix has \emph{exactly} one-dimensional kernel, to which its direction belongs; in other words, within this rod we have the matrix equation $gl_k=0$. Moreover, the direction of a rod is normalised (up to an overall minus sign) as follows: if one defines a coordinate $\eta_k$ by defining
\begin{align}
\partial_{\eta_k}= a_k\partial_\tau+b_k\partial_\phi(=l_k)\,,
\end{align}
the metric near Rod $k$, i.e., for $\rho\rightarrow 0$ and $z_{k-1}<z<z_k$, has a behaviour as
\begin{align}
\dif s^2\rightarrow \dif\rho^2+\rho^2\dif\eta_k^2\,,
\label{bolt_geometry}
\end{align}
up to a conformal factor independent on $\rho$. Metric components involving two other coordinates are also independent on $\rho$ and are omitted here. We remark that this metric behaviour implies, in particular, that all metric components are finite when $\rho\rightarrow 0$ is approached (with $z$ fixed).  Obviously, the Killing vector $\partial_{\eta_k}$ vanishes on Rod $k$, and the coordinate $\eta_k$ should be $2\pi$-periodic to avoid a conical singularity along this rod. In terms of the original coordinates $(\tau,\phi)$, the following identification should be made:
\begin{align}
(\tau,\phi)\sim (\tau+2\pi a_k,\phi+2\pi b_k)\,,
\label{rod_identifications}
\end{align}
to avoid the conical singularity. So a rod is in fact an axis, with the axial symmetry generated by its direction. Rods of this type are called space-like.

It is clear from our definition that a turning point is the meeting point of the two axes represented by its two neighbouring rods. So the turning point $z_k$ is the meeting point of Rod $k$ with direction $l_k$ and Rod $k+1$ with direction $l_{k+1}$. If we define coordinates $\eta_k$ and $\eta_{k+1}$ such that $l_k=\partial_{\eta_k}$ and $l_{k+1}=\partial_{\eta_{k+1}}$, the metric near this turning point approaches
\begin{align}
\dif s^2\rightarrow \dif r_k^2+r_k^2\dif \eta_k^2+\dif r_{k+1}^2+r_{k+1}^2\dif \eta_{k+1}^2\,,
\label{nut_geometry}
\end{align}
for appropriately defined coordinates $r_{k}$ and $r_{k+1}$. We see that if both $\eta_k$ and $\eta_{k+1}$ are $2\pi$-periodic, i.e., if the identifications
\begin{align}
(\tau,\phi)\sim (\tau+2\pi a_k,\phi+2\pi b_k)\sim (\tau+2\pi a_{k+1},\phi+2\pi b_{k+1})\,,
\label{turning_point_identifications}
\end{align}
are made, $\partial_{\eta_k}$ and $\partial_{\eta_{k+1}}$ generate the two axial symmetries of $\mathbb{R}^4$ near its origin.

The metric behaviours (\ref{bolt_geometry}) and (\ref{nut_geometry}) near the rods and turning points incorporate our demand that all axes together with their meeting points of the solution under study are locally regular. If the identification (\ref{rod_identifications}) is made, the solution near Rod $k$ is regular; if the identifications (\ref{turning_point_identifications}) are made, the solution near the turning point $z_k$ is regular. In these cases, Rod $k$ and the turning point $z_k$ are made part of the space described by the solution. The $n$-soliton solution on flat space is locally regular in this sense and is picked out among soliton solutions on the general Levi-Civita seed (\ref{LeviC}). For other choices of the parameter $d$ and the number $n$ as far as the author had examined, soliton solutions on (\ref{LeviC}) do not have the above demanded metric behaviours and always possess a curvature singularity somewhere along the $z$-axis.

Now that the locations of all the turning points and rods are known for the $n$-soliton solution, the task left is to calculate all rod directions from its metric components. The calculation is standard \cite{Chen:2010zu}. To calculate $l_k$, one first chooses a vector $v_k$ proportional to it in the kernel of the $g$-matrix within Rod $k$
\begin{align}
l_k\propto v_k=(\alpha_k,\beta_k)\in \ker(g|_{\text{Rod $k$}})\,.
\end{align}
The inverse of the proportional constant is just the Euclidean surface gravity, which can be computed as
\begin{align} \kappa_k=\sqrt{\frac{G_{ij}v^iv^j}{\rho^2f}{}\Big|_{\text{Rod $k$}}}\,\,.\label{surface_gravity_general}
\end{align}
The normalised direction is then given by
\begin{align} l_k=\left(\frac{\alpha_k}{\kappa_k},\frac{\beta_k}{\kappa_k}\right).
\label{general_direction}
\end{align}
An overall minus sign can be added to (\ref{general_direction}), which will be implemented whenever it is convenient. It is worthwhile to remark that whenever $\rho=0$ is set in an expression in this paper, it should really be understood as taking the limit $\rho\rightarrow 0$. So one needs to take the limit $\rho\rightarrow 0$ for fixed $z$ within the range $z_{k-1}<z<z_k$ to calculate $\kappa_k$ in (\ref{surface_gravity_general}).

With all metric components of the $n$-soliton solution known from our ISM construction, it is, in principle, a straightforward task to calculate all the rod directions. In practice, however, calculations using the above formulas can be sufficiently complicated. Here, we prove some simplified formulas to achieve the goal.

Notice that the solutions under study can be written in the form
\begin{align}
\label{diagnised}\dif s^2=g_{\phi\phi}\left(\dif \phi+\frac{g_{\tau\phi}}{g_{\phi\phi}}\dif\tau\right)^2+\frac{\rho^2}{g_{\phi\phi}}\dif\tau^2+f\left(\dif\rho^2+\dif z^2\right).
\end{align}
We see that the first term  on the right-hand side is finite on Rod $k$, while the second term vanishes. This means that the metric is diagonalised with a zero eigenvalue on Rod $k$, by transforming coordinates from $(\tau,\phi)$ to $(\tau, \phi+\tau\frac{g_{\tau\phi}}{g_{\phi\phi}}|_{\text{Rod $k$}})$. Thus a possible choice of $v_k$ is 
\begin{align}
\label{direction} v_k=\left(1,-\frac{g_{\tau\phi}}{g_{\phi\phi}}\Big|_{\text{Rod $k$}}\right),
\end{align}
for which $\dfrac{g_{ij}v_k^iv_k^j}{\rho^2}\Big|_{\text{Rod }k}=\frac{1}{g_{\phi\phi}}\Big|_{\text{Rod }k}$. The Euclidean surface gravity for $v_k$ is thus
\begin{align}
\label{surface_gravity1} \kappa_k=\frac{1}{\sqrt{(fg_{\phi\phi})|_{\text{Rod $k$}}}}\,.
\end{align}
The normalised direction $l_k$ can be obtained by substituting (\ref{direction}) and (\ref{surface_gravity1}) into (\ref{general_direction}). Similarly, writing the metric in the form
\begin{align}
\dif s^2=g_{\tau\tau}\left(\dif \tau+\frac{g_{\tau\phi}}{g_{\tau\tau}}\dif\phi\right)^2+\frac{\rho^2}{g_{\tau\tau}}\dif\phi^2+f\left(\dif\rho^2+\dif z^2\right),
\end{align}
one can choose a different $v_k$ and identify its surface gravity as
\begin{align}
\label{direction2} v_k'=\left(-\frac{g_{\tau\phi}}{g_{\tau\tau}}\Big|_{\text{Rod $k$}},1\right),\quad \kappa_k'=\frac{1}{\sqrt{(fg_{\tau\tau})|_{\text{Rod $k$}}}}\,.
\end{align}
When substituted into (\ref{general_direction}), Eq.~(\ref{direction2}) yields the same $l_k$ as (\ref{direction}) and (\ref{surface_gravity1}) do.

Now, observe that the condition (\ref{rho}) reduces to the equality $g_{\tau\tau}g_{\phi\phi}=g_{\tau\phi}^2$ on Rod $k$. Using either (\ref{direction}) and (\ref{surface_gravity1}) or (\ref{direction2}),  one finds that the direction of Rod $k$ has a simple expression in terms of metric components
\begin{align}
\label{direction3} l_k=\left(\sqrt{fg_{\phi\phi}},\pm \sqrt{fg_{\tau\tau}}\right)\Big|_{\text{Rod $k$}}\,,
\end{align}
where the sign of the second component has to be chosen in such a way that $l_k\propto v_k\text{ or }v_k'$. This is the expression that we will use to compute the rod directions of the $n$-soliton solution.

Let us briefly comment on the above formulas. We first observe that the metric components of the $n$-soliton solution fail to be analytic if $f=0$, since $f$ essentially appears in the denominators of the $g$-matrix. For physically sensible solutions, we thus demand that $f$ does not change sign on the whole plane $(\rho>0,-\infty<z<\infty)$. Since we are interested in solutions with either the Euclidean or Lorentzian signature, we in fact demand $f>0$. This condition is obviously satisfied in the static $n$-soliton solution and is usually used to constrain the ranges of BZ parameters in the general non-diagonal case. Continuity consideration ensures that $f>0$ within each rod.

The $g$-matrix of the $n$-soliton solution is thus analytic, so it admits Taylor expansions in terms of $\rho$ at $\rho\rightarrow 0$ for fixed $z$. The quantities appearing in (\ref{direction2}) are well defined except in the case $g_{\tau\tau}\rightarrow 0$ near Rod $k$. Since the Taylor expansions of all metric components contain only even powers of $\rho$, in this exceptional case we can deduce that $g_{\tau\tau}\rightarrow O(\rho^2)$. In view of the condition (\ref{rho}), we thus have $g_{\tau\phi}\rightarrow O(\rho^2)$ and $g_{\phi\phi}\rightarrow O(1)$, from which we conclude that the direction in this case is proportional to $\partial_\tau$. So (\ref{direction2}) is not applicable; one should instead use (\ref{direction}) and (\ref{surface_gravity1}). Similar analysis can be done in the case $g_{\phi\phi}\rightarrow 0$ near Rod $k$, and it can be shown that the direction is proportional to $\partial_\phi$. In this case (\ref{direction}) and (\ref{surface_gravity1}) are not well defined and (\ref{direction2}) should be used instead. These two situations exactly occur when one component of the direction becomes zero, as in the static $n$-soliton solution. The formula (\ref{direction3}) is, however, always applicable.

\subsection{The rod structure of the $n$-soliton solution}

\label{subsec_rod_structure}

To use the formula (\ref{direction3}) to calculate the direction of any rod of the $n$-soliton solution, one needs to find the leading terms of the metric components near that rod for small $\rho$. Notice that all metric components of the $n$-soliton solution are algebraic expressions in terms of the coordinate $\rho$ and the $n$ solitons $\mu_k$ for $k=1,...,n$. For any $k$, the leading order behaviour of the soliton $\mu_k$ at small $\rho$ is
\begin{align}
\mu_k\sim{\Big\{}\begin{array}{ll}
2(z_k-z), &\text{for fixed }z<z_k\\
-\frac{\rho^2}{2(z_k-z)},&\text{for fixed }z>z_k.
\end{array}
\label{soliton_behavior}
\end{align}
Upon substituting (\ref{soliton_behavior}), all metric components are now algebraic expressions in terms of the coordinates $\rho$ and $z$. The calculation of all rod directions using (\ref{direction3}) becomes an elementary exercise.

The direction $l_1$ for Rod 1 is the easiest to calculate, since near this rod, we can replace $\mu_k$ by $2(z_k-z$) for all $k$. So all solitons can be taken as constants in the limit when we approach Rod 1. After some algebra, the result is given by $l_1=(a_1,b_1)$ with
\begin{align}
a_1&=\epsilon_{a}\sqrt{k_0}\sum_{i_1<...<i_{\lfloor\frac{n}{2}\rfloor+1}}\left( (-1)^{\sum\limits_{k=1}^{\lfloor\frac{n}{2}\rfloor+1} i_k}C_{i_1}...C_{i_{\lfloor\frac{n}{2}\rfloor+1}}\prod_{i_k<i_l} 2z_{i_{k}i_{l}}\prod_{ j_r<j_s} 2z_{j_{r}j_{s}}\right),\nonumber\\
b_1&=\epsilon_{b}\sqrt{k_0}\sum_{i_1<...<i_{\lfloor\frac{n}{2}\rfloor}}\left((-1)^{\sum\limits_{k=1}^{\lfloor\frac{n}{2}\rfloor} i_k} C_{i_1}...C_{i_{\lfloor\frac{n}{2}\rfloor}}\prod_{i_k<i_l} 2z_{i_{k}i_{l}}\prod_{ j_r<j_s, } 2z_{j_{r}j_{s}}\right),
\label{first_direction}
\end{align}
where $\{j_r,j_s\}\bigcap \{i_1,...,i_{\lfloor\frac{n}{2}\rfloor+1}\}=\emptyset$ in the first line and $\{j_r,j_s\}\bigcap \{i_1,...,i_{\lfloor\frac{n}{2}\rfloor}\}=\emptyset$ in the second line. Here, $z_{i_ki_l}\equiv z_{i_k}-z_{i_l}$ and likewise for $z_{j_rj_s}$; when the index set contains less than two elements, $z_{i_ki_l}\equiv 1$ and likewise for $z_{j_rj_s}$. The signs $\epsilon_{a}$ and $\epsilon_{b}$ are given by
\begin{align}
(\epsilon_{a},\,\,\epsilon_{b})=\Big\{\begin{array}{ll}
            \left(i^{n-1},\,\,i^{n-1} \right), & \text{for odd $n$} \\
            \left(-1,\,\,1\right), & \text{for even $n$}.
          \end{array}
\end{align}
Up to an overall minus sign that is independent of $n$, these $\epsilon$'s are entirely determined by the transition limit and their values for $n=1$. We demand that $l_1$ of the $n$-soliton solution reduces to $l_1$ of the $(n-1)$-soliton solution in the transition limit.\footnote{We refer specifically to the transition limit in which $z_n\rightarrow +\infty$ (rather than $z_n\rightarrow -\infty$) is taken.} For the 1-soliton solution, we choose $\epsilon_b=1$, and direct calculation then yields $\epsilon_a=1$. These $\epsilon$'s play little role in our analysis below.

The directions of the rest rods can be found by an important recurrence relation. Note that when one goes from Rod $k$ to Rod $k+1$, the near-rod behaviour of the metric is caused by the change of the behaviour of $\mu_k$ from $2(z_k-z)$ to $-\frac{\rho^2}{2(z_k-z)}$, seen from (\ref{soliton_behavior}). Observe that it would be a symmetry if one additionally performs the substitutions $(C_k\rightarrow C_k^{-1}, k_0\rightarrow C_k^2k_0$) simultaneously, according to the inversion symmetry (\ref{symmetry_inversion}). It is then clear from the formula (\ref{direction3}) that, if $l_k$ is written in terms of the BZ parameters as
\begin{align}
l_k=l_k(C_1,C_2,...,C_n)\,,
\end{align}
then $l_{k+1}$ can be written as
\begin{align} l_{k+1}(C_1,C_2,...,C_n)=\left(-1\right)^{k}C_{k}l_k(C_1,...,C_{k}^{-1},...,C_n)\,.
\label{recursive_relation}
\end{align}
The overall sign choice $(-1)^k$ is made for later convenience. This recurrence relation, together with $l_1$ given in (\ref{first_direction}), determines the directions of all the $n+1$ rods unambiguously. This completes our calculation of the rod structure of the $n$-soliton solution.

A few comments are in order. We first note a remarkable fact that the quantities $a_1$ and $b_1$ in (\ref{first_direction}) are homogeneous polynomials in terms of the BZ parameters, and they are linear with respect to each individual BZ parameter. The recurrence relation (\ref{recursive_relation}) then guarantees that all rod-direction components $a_{1,...,n+1}$ and $b_{1,...,n+1}$ are polynomials in terms of the BZ parameters.

We see from (\ref{first_direction}) that $l_1$ becomes vanishing, namely, $l_1=(0,0)$, in the limit when all the BZ parameters are turned off (for $n\ge 2$), indicating that the $n$-soliton solution in this case is not locally regular. If, however, one takes the limit (\ref{limit_static}) (with $k_0$ set as $k_0=(C_2C_4...C_{2\lfloor\frac{n}{2}\rfloor})^{-2}$), all rod directions $l_k$ are well behaved, proportional to $\partial_\tau$ and $\partial_\phi$ alternatively as $k$ varies. This is the reason why the latter rather than the former corresponds to the static limit. The rod structure also gives an intuitive interpretation of the parameters of the $n$-soliton solution: the locations of the solitons determine the locations of various axes, while the BZ parameters characterise orientations of these axes.

One straightforward application of the rod structure of the $n$-soliton solution just obtained is the determination of the constants in the linear coordinate transformations (\ref{linear_transformation_even}) and (\ref{linear_transformation_odd}) and also the constant $k_0$ to bring the solution to the standard asymptotic structure (\ref{ALF_Euclidean}) and (\ref{ALE_Euclidean}). We begin by remarking that the rod directions are Killing vectors of the solution, so their components transform contravariantly when the Killing coordinates are transformed \cite{Chen:2010zu}. What we need to do is simply identify the directions of the first and last rods of the even- and odd-soliton solutions with those of the two asymptotic axes of (\ref{ALF_Euclidean}) and (\ref{ALE_Euclidean}), respectively.

For (\ref{ALF_Euclidean}), the two asymptotic axes are located at $\theta=\pi$ and $\theta=0$, with directions $\pm 2n_{\text{nut}}\partial_{\tilde{\tau}}+\partial_{\tilde{\phi}}$, respectively \cite{Chen:2010zu}. So we identify the following:
\begin{align}
&\text{First-rod directions:} &2n_{\text{nut}}\partial_{\tilde{\tau}}+\partial_{\tilde{\phi}} &=a_1\partial_\tau+b_1\partial_\phi\,,\nonumber\\
&\text{Last-rod directions:} & -2n_{\text{nut}}\partial_{\tilde{\tau}}+\partial_{\tilde{\phi}}&=a_{n+1}\partial_\tau+b_{n+1}\partial_\phi\,.
\label{determine_nut}
\end{align}
For even $n$, one can deduce from (\ref{first_direction}) and (\ref{recursive_relation}) the relation
\begin{align}
b_{1}=b_{n+1}\,.
\label{even_equalb}
\end{align}
The equations (\ref{determine_nut}) can then be written as
\begin{align}
\partial_{\tilde{\tau}}&=\frac{a_1-a_{n+1}}{4n_{\text{nut}}}\partial_\tau\,,\nonumber\\
\partial_{\tilde{\phi}}&=\frac{a_1+a_{n+1}}{2}\partial_\tau+\frac{b_1+b_{n+1}}{2}\partial_\phi\,.
\end{align}
The constants $b$, $n_{\text{nut}}$ and $k_0$ are obtained by comparing these relations with (\ref{linear_transformation_even}). It is easy to see that $b$ and $n_{\text{nut}}$ are given by
\begin{align}
b=\frac{a_1+a_{n+1}}{2}\,,\quad n_{\text{nut}}=\frac{a_1-a_{n+1}}{4}\,,
\end{align}
and $k_0$ is determined by the equation
\begin{align}
|b_1|=1\,.
\end{align}
This last equation fixes $k_0$ in terms of the soliton locations and BZ parameters.

The two asymptotic axes of (\ref{ALE_Euclidean}) are located at $\theta=\frac{\pi}{2}$ and $\theta=0$ and have directions $\partial_{\tilde{\phi}}$ and $\partial_{\tilde{\tau}}$, respectively \cite{Chen:2010zu}. So we identify the following:
\begin{align}
&\text{First-rod directions:} &\partial_{\tilde{\phi}}&=a_1\partial_\tau+b_1\partial_\phi\,,\nonumber\\
&\text{Last-rod directions:} & \partial_{\tilde{\tau}} &=a_{n+1}\partial_\tau+b_{n+1}\partial_\phi\,.
\label{rod_identification_even}
\end{align}
For odd $n$, from the expressions (\ref{first_direction}) and (\ref{recursive_relation}) one can deduce the relations
\begin{align}
a_{n+1}=b_1\,,\quad b_{n+1}=a_1\,.
\label{first_last_relation_even}
\end{align}
From (\ref{rod_identification_even}) and (\ref{first_last_relation_even}) one immediately gets (\ref{linear_transformation_odd}). The constant $k_0$ is determined by the requirement that this transformation has a determinant with unit modulus, i.e.,
\begin{align}
|b_1^2-a_1^2|=1\,.
\end{align}

\subsection{The rod structure of the Lorentzian section of the $n$-soliton solution}

\label{subsec_rod_structure_Lorentzian}

The rod-structure formalism can be applied to both Euclidean and Lorentzian solutions. For a Lorentzian solution, a rod, say Rod $k$, may become time-like, in the sense that the quantity in the square root of (\ref{surface_gravity1}) is negative \cite{Harmark:2004rm}. Hence, both components of the direction of this rod $\ell_k$ are purely imaginary. Such a rod represents a Killing horizon. For an asymptotically flat solution with an axial symmetry generated by $\partial_\phi$, one can write the direction $\ell_k$ as
\begin{align}
\ell_k=\frac{1}{i|\kappa_k|}(1,\Omega_k)\,.
\end{align}
Then, $\Omega_k$ has the interpretation of the angular velocity and $|\kappa_k|$ the surface gravity on the Killing horizon. If one defines $i\ell_k\equiv\partial_{\chi_k}$, the metric near this rod approaches
\begin{align}
\dif s^2\rightarrow \dif\rho^2-\rho^2\dif\chi_k^2,
\end{align}
where a finite conformal factor is omitted and metric components involving two other coordinates are not shown.

Now, we go to the Lorentzian section of the $n$-soliton solution and calculate its rod structure. First, recall that the Lorentzian section is obtained from the Euclidean solution via the Wick rotation (\ref{wick_rotation}). This transformation can formally be viewed as a linear coordinate transformation on the Killing coordinates, accompanied by a redefinition of the parameters $C_k$ in terms of $\mathcal{C}_k$. Under such a linear coordinate transformation, locations of all rods are unchanged, but the rod directions change (from $l_k=(a_k,b_k)$) contravariantly to \cite{Chen:2010zu}
\begin{align}
\label{directions_Lorentzian}
{\ell}_k=(ia_k,b_k)=(ia_k(C= i\mathcal{C}),b_k(C= i\mathcal{  C}))\,,
\end{align}
where we have denoted $\{C_1,...C_n\}$ and $\{\mathcal{C}_1,...,\mathcal{C}_n\}$ as $C$ and $\mathcal{C}$, respectively. This completes the calculation of the rod structure of the Lorentzian $n$-soliton solution. Several interesting results then follow. Firstly, we note that all rod directions are either real or purely imaginary (in terms of the BZ parameters $\mathcal{C}_k$) in the Lorentzian section. Secondly, from (\ref{recursive_relation}) and (\ref{directions_Lorentzian}) one can deduce the recurrence relation between adjacent directions in the Lorentzian section
\begin{align} \label{direction_recurrence_Lorentzian}
\ell_{k+1}(\mathcal{C}_1,\mathcal{C}_2,...,\mathcal{C}_n)=(-1)^ki\mathcal{C}_{k}\ell_k(\mathcal{C}_1,...,-\mathcal{C}_{k}^{-1},...,\mathcal{C}_n)\,.
\end{align}
This recurrence relation implies the interesting result that in the Lorentzian section, space-like and time-like rods, representing axes and horizons, respectively, appear alternatively. This result, though trivial in the static limit, is far from obvious in the general non-diagonal case, and it certainly does not hold in higher dimensions. For a time-like Rod $k$ and a neighbouring space-like Rod $k+1$, if we define coordinates $\chi_k$ and $\eta_{k+1}$ such that $i\ell_k=\partial_{\chi_k}$ and $\ell_{k+1}=\partial_{\eta_{k+1}}$, the metric near the turning point $z_k$ then approaches
\begin{align}
\dif s^2\rightarrow \dif r_k^2-r_k^2\dif \chi_k^2+\dif r_{k+1}^2+r_{k+1}^2\dif \eta_{k+1}^2\,,
\end{align}
for appropriately defined coordinates $r_{k}$ and $r_{k+1}$.

We remark that for the Lorentzian $n$-soliton solution, the special situation of a degenerate horizon can occur. In this case, the rod representing this horizon shrinks to zero length; the two components of its direction are divergent, but their ratio remains finite. The horizon in this case is extremal, having a finite angular velocity and zero temperature.

\subsection{$[n-m]$-soliton solutions on flat space}

Now, we return to the Euclidean $n$-soliton solution and explore in depth one interesting consequence of the inversion symmetry (\ref{symmetry_inversion}). We have shown that this symmetry guarantees the  recurrence relation (\ref{recursive_relation}) of adjacent rod directions. Now, observe that if the BZ parameter of the soliton $\mu_k$ is set to
\begin{align}
\label{condition_joinup_single}
C_k=\pm 1\,,
\end{align}
one finds that
\begin{align}
l_k=l_{k+1}\,,
\end{align}
up to an overall sign difference, which we recall is irrelevant in our formalism. The physical implication of this condition is  interesting: Now, Rod $k$ and Rod $k+1$ have identical directions, so the axes they represent share the same generator and merge into a single and larger axis. We say that Rod $k$ and Rod $k+1$ are joined up, and the turning point where these two rods originally meet is effectively eliminated.

Let us study this phenomenon in terms of the behaviour of the $g$-matrix near the point $(\rho= 0,z= z_k)$. We know that for a general BZ parameter $C_k$, both $l_k$ and $l_{k+1}$ are in the kernel of $g$ at $(\rho= 0,z= z_k)$, implying that $g$ should vanish at this point. One can transform to an appropriate coordinate system and show that the metric can be brought to (\ref{nut_geometry}) near this point. Now, by setting (\ref{condition_joinup_single}), the two rod directions $l_k$ and $l_{k+1}$ become identical, making the kernel of the $g$-matrix only one-dimensional at $(\rho= 0,z= z_k)$. This means that in this situation the $g$-matrix itself necessarily becomes non-vanishing. The point $(\rho= 0,z= z_k)$ is now a normal point on the larger rod $(\rho=0,z_{k-1}<z<z_{k+1})$, whose near-rod geometry approaches (\ref{bolt_geometry}) in appropriate coordinates.

To better understand this phenomenon, let us study a concrete example. Consider the Taub-NUT metric in the following form:
\begin{align}
\label{example_Taub-NUT}
\dif s^{2}=\frac{r}{r+2n} { \left( {\dif\tau}+2n\cos  \theta\,
{\dif\phi} \right) ^{2}}+(r^2+2nr) \sin ^2 \theta \,{{\dif\phi}}^{2}+\frac{r+2n}{r} ( {\dif r}^{2}+{r}^{2}{{\dif\theta}}^{2} )\,,
\end{align}
where $n$ is the NUT-charge parameter. The rod structure of this metric has been studied in \cite{Chen:2010zu}, where the reader is referred to for details. It consists of two rods, which meet at the only turning point at $r=0$. The two rods correspond to the two axes of the space and has directions
\begin{align}
l_1=2n\partial_\tau+\partial_\phi\,,\quad l_2=-2n\partial_\tau+\partial_\phi\,.
\end{align}
The turning point for the Taub-NUT metric is called a nut in the language of \cite{Gibbons:1979xm} (with respect to the Killing vector $\partial_\tau$). Around the nut, the geometry locally resembles (\ref{nut_geometry}) for any $n\neq 0$, if one identifies $l_1=\partial_{\eta_1}$ and $l_2=\partial_{\eta_2}$ and appropriately redefines $r$ and $\theta$. In particular, it is direct to show that the $g$-matrix vanishes at the turning point $r=0$. Now, consider the special limit $n\rightarrow 0$ of the metric (\ref{example_Taub-NUT}). One then has $l_1=l_2$, and the two rods are joined up. In this example, it is clear that upon taking the limit $n\rightarrow 0$, the nut (an \emph{isolated} fixed point of $\partial_\tau$) disappears, and the metric locally resembles (\ref{bolt_geometry}) if one identifies $\partial_\phi=\partial_\eta$ and $\rho=r\sin\theta$. In particular, the $g$-matrix is no longer vanishing at $r=0$. We see that, in this example, the joining up of two adjacent rods corresponds to setting the NUT charge to zero. In this process, the local geometry near the nut is changed from (\ref{nut_geometry}) to (\ref{bolt_geometry}). This last feature is generic in our formalism when two adjacent rods are joined up.  

In terms of the solitons, by setting (\ref{condition_joinup_single}), the soliton $\mu_k$ is effectively eliminated from the solution. Although $\mu_k$ still appears in the metric explicitly, there is actually nothing special about the point $(\rho=0,z=z_k)$. The metric can in fact be completely written in terms of the rest solitons, which will be shown in a few examples in the next section. Saying this, we do not mean that the effect of the soliton $\mu_k$ is completely removed. As a matter of fact, the parameter $z_k$ continues to enter the solution non-trivially. Contrast this with the transition limit of the $n$-soliton solution. There, a soliton ($\mu_n$) is eliminated by sending its location ($z_n$) to infinity while fixing its BZ parameter ($C_n$) appropriately as either zero or infinity. The effect of this soliton then completely disappears: the $n$-soliton solution is reduced to the $(n-1)$-soliton solution.

For convenience, for the $n$-soliton solution, the soliton $\mu_k$ will be called a positive phantom soliton if its BZ parameter is fixed as $C_k=1$, with the resulting solution denoted by $n\ominus 1^+$. It will be called a negative phantom soliton if $C_k=-1$ is fixed instead with the resulting solution denoted by $n\ominus 1^-$. A soliton whose BZ parameter is a free parameter is called a free soliton. 

It becomes more interesting when multiple solitons, say with a number of $m$, are eliminated simultaneously. This can be done by setting
\begin{align}
\label{condition_joinup} C_{i_1}=\pm 1\,,\quad C_{i_2}=\pm 1\,,\quad...,\quad C_{i_m}=\pm 1\,,
\end{align}
where $\{i_1,i_2,...,i_m\}$ is an $m$-element subset of $\{1,2,...,n\}$. The corresponding solitons $\mu_{i_1,i_2,...,i_m}$ are then phantom solitons. Denote the number of positive phantom solitons as $m_1$ and that of negative phantom solitons as $m_{2}$. A solution is then unambiguously given by the three integers $(n,m_{1},m_{2})$, defined as the $n$-soliton solution with $m_{1}+m_2$ solitons eliminated to have $m_1$ positive phantom solitons and $m_{2}$ negative phantom solitons. Clearly the total number of phantom solitons is $m=m_{1}+m_{2}$, and the number of free solitons is $n-m$. This solution, solitonically represented as $n\ominus m_{1}^+\ominus m_{2}^-$, is called the $[n-m_{1}-m_{2}]$-soliton solution, i.e.,
\begin{align}
[n-m_{1}-m_{2}]\equiv n\ominus m_{1}^+\ominus m_{2}^-\,.
\end{align}
Due to the symmetries of the $n$-soliton solution, it does not matter which $m_1$ solitons become positive phantom solitons and which $m_2$ solitons become negative phantom solitons.

Note that the reflection symmetry (\ref{symmetry_reflection}) sends a positive phantom soliton to a negative one, and vice versa. This allows us to assume $m_{1}\geq m_{2}$ without loss of generality. We further note that there is a restriction on $m_{1,2}$ that originates from the ISM integration scheme: $\max\{m_{1},m_{2}\}\leq \lfloor\frac{n}{2}\rfloor$; if this constraint is violated, the $\Gamma$-matrix will be singular and our ISM construction breaks down. Hence, the $[n-m_{1}-m_{2}]$-soliton solution satisfies
\begin{align}
\big\lfloor\frac{n}{2}\big\rfloor\geq m_{1}\geq m_{2}\,.
\label{nmmplus}
\end{align}
The $[n-m_{1}-m_{2}]$-soliton solutions with the same $n$ and $m=m_{1}+m_{2}$ but different $m_1$ are collectively denoted as $[n-m]$-soliton solutions. They all have the same number of free solitons. When $m\geq 2$, they are really a collection of solutions. When $m=1$, we necessarily have $m_{1}=1, m_{2}=0$, and we denote the only solution $[n-1-0]$ simply as $[n-1]$. We define the $[n-0-0]$-soliton solution as the $n$-soliton solution itself. We remind the reader that for $m\geq 1$, $[n-m]$-soliton solutions are different from the $(n-m)$-soliton solution. The latter is defined here as the $s$-soliton solution, with $s=n-m$. So starting from the $n$-soliton solution, by eliminating a certain number of solitons simultaneously, we can obtain various subclasses of locally regular solutions with two axial symmetries. In the next section, we study a few examples of these multi-soliton solutions---the $n$- and $[n-m]$-soliton solutions---on flat space.

We end this subsection by remarking that, unlike the $n$-soliton solution, an $[n-m]$-soliton solution with $m\ge 1$ no longer has a natural Lorentzian section. The Wick rotation (\ref{wick_rotation}) that realises the Lorentzian section of the $n$-soliton solution is spoiled by the joining-up condition (\ref{condition_joinup}) in an $[n-m]$-soliton solution. However, we must emphasise that this does not mean that no $[n-m]$-soliton solution possesses a Lorentzian section. It may be possible that a Lorentzian section is obtained by more complicated analytical continuations than (\ref{wick_rotation}). We will demonstrate that the Kerr-NUT solution, which has a Lorentzian section, is also an $[n-m]$-soliton solution (it is the $[4-1-1]$-soliton solution, for example). A related issue is that the operation of eliminating a soliton by fixing its BZ parameter described above does not extend to the Lorentzian $n$-soliton solution. This is because in the Lorentzian section, time-like rods and space-like rods appear alternatively as we have shown. Joining up adjacent rods then becomes impossible.\footnote{In higher dimensions, space-like rods and time-like rods may not necessarily appear alternatively. So it is possible in certain situations to eliminate a turning point or a soliton and thus join up two adjacent rods of a higher-dimensional Lorentzian solution. An example can be found in \cite{Iguchi:2006rd,Tomizawa:2006vp}.} This can be seen from the relation (\ref{direction_recurrence_Lorentzian}), which implies that joining up two adjacent rods, say Rod $k$ and Rod $k+1$, requires that $\mathcal{C}_k=\pm i$. This, however, will surrender the metric in the Lorentzian section complex, in general.

\section{Examples}

\label{sec_examples}

After the above lengthy discussions of some general properties of the $n$- and $[n-m]$-soliton solutions on flat space, it is now time to study several examples in detail. We will be restricted to the simpler cases with $n\le 6$ and $n-m\leq 3$. Our focus is to show how various known solutions are recovered and to explicitly present the necessary coordinate transformations. Readers who are not interested in the calculation details can skip to the next section, where major results of this section are summarised. Again, we work in the Euclidean sections of these solutions. Since the $0$-soliton solution is trivial, we start from the $1$-soliton solution.

\subsection{The $1$-soliton solution}

\label{sec_1-soliton}

The $1$-soliton solution can be explicitly written in the following form:
\begin{align}
\label{metric_1-soliton}
\dif s^2=&\frac{C_1^2\rho^2+\mu_1^2}{\mu_1(1-C_1^2)}\,\dif \tau^2+ \frac{2C_1(\rho^2+\mu_1^2)}{\mu_1(1-C_1^2)}\,\dif\tau\dif\phi+ \frac{\rho^2+C_1^2\mu_1^2}{\mu_1(1-C_1^2)}\,\dif\phi^2\nonumber\\
&+\frac{k_0(1-C_1^2)\mu_1}{R_{11}}\left(\dif\rho^2+\dif z^2\right).
\end{align}
The reader is reminded that $R_{11}= \rho^2+\mu_1^2$ as defined in (\ref{Gamma_even}).

The first observation of this solution is that the BZ parameter $C_1$ is redundant. This can be proved by bringing the solution to the natural orientation (see Sec.~\ref{sec_asymptotic_structure}). So we calculate the rod structure of the solution. There is one turning point located at $(\rho=0,z=z_{1})$, which divides the $z$-axis into two rods whose directions are, respectively,
\begin{align} l_1=\left(-C_1\sqrt{k_0},\sqrt{k_0}\right),\quad l_2=\left(\sqrt{k_0},-C_1\sqrt{k_0}\right),
\end{align}
consistent with the general analysis done in Sec.~\ref{subsec_rod_structure}. The linear coordinate transformation (\ref{linear_transformation_odd}), which brings the solution to the natural orientation, is then given by the matrix
\begin{align}
A= \left[ \begin {array}{cc} \sqrt{k_0}&-{C_1}\sqrt{k_0}\\ -{C_1}\sqrt{k_0}& \sqrt{k_0}\end {array} \right].
\end{align}
Under this transformation, the $g$-matrix of the 1-soliton solution is brought to $g_{\text{acc}}=A^T g A$. We also fix $k_0$ as $k_0= 1/(1-C_1^2)$ so that $|\det A|=1$ and denote the function $f$ in this case as $f_{\text{acc}}$. We find that $g_{\text{acc}}$ is now diagonal and that, being the same as $f_{\text{acc}}$, it has no explicit $C_1$-dependence.

The solution in its natural orientation described above is also the ``static limit'' of (\ref{metric_1-soliton}), obtained by setting $k_0=1$ and taking
\begin{align}
 C_1\rightarrow 0\,,
\end{align}
with a resulting metric given by
\begin{align}
g_{\text{{acc}}}=\left[ \begin {array}{cc} {\mu_1}&0\\0&{\frac {{\rho}^{2}}{{\mu_1}}}\end {array} \right],\qquad
f_{\text{acc}}=\frac{\mu_1}{R_{11}}\,.
\label{1-soliton-static}
\end{align}
We recognise that this is flat space with two manifest axial symmetries. Indeed, if we perform the following coordinate transformations:
\begin{align} \rho=\frac{1}{2}r^2\sin 2\theta\,,\qquad z-z_1=\frac{1}{2}r^2\cos 2\theta\,,
\end{align}
the metric (\ref{1-soliton-static}) is cast in the familiar form of flat space
\begin{align}
\dif s_{\text{acc}}^2=\dif r^2+r^2\left(\dif\theta^2+\sin^2\theta\,\dif\tau^2+\cos^2\theta\,\dif\phi^2\right).
\label{metric_accelerating_nothing}
\end{align}
If $\tau$ is analytically continued to the time variable $t$, this metric describes the space-time of the accelerating nothing. In this case, the axis $\theta=0$ of flat space becomes an acceleration horizon. The entire space-time outside this horizon contains nothing and forms a part of the Minkowski space-time. The metric (\ref{metric_accelerating_nothing}) has been constructed using the ISM in one form or another in the literature (see, e.g., \cite{Belinski:2001ph,Letelier:1985mba}).

The $1$-soliton solution (\ref{metric_1-soliton}) is thus ALE. Since it is essentially flat space with one turning point, it must be equivalent to the 1-centred Gibbons--Hawking solution. To show this, we first appropriately shift the coordinate $z$ such that $z_1=0$, define coordinates $(r,\theta)$ as
\begin{align}
\label{weyltortheta} 
\rho=r\sin\theta\,,\qquad z=r\cos\theta\,,
\end{align}
and perform linear coordinate transformations by doing the simultaneous substitutions as follows:
\begin{align}
\label{linear_transformation_1-soliton}
\tau\rightarrow -\tau+\phi/2\,,\qquad \phi\rightarrow \tau+\phi/2\,.
\end{align}
If we choose $k_0=(1-C_1)^{-2}$, the 1-soliton solution (\ref{metric_1-soliton}) is then brought to
\begin{align}
\label{1-centred_GH}
\dif s^{2}=V^{-1} { \left( {\dif\tau}+2n\cos  \theta\,
  {\dif\phi} \right) ^{2}}+V \left( {\dif r}^{2}+{r}^{2}{
{\dif\theta}}^{2}+{r}^{2} \sin ^2 \theta \,{{\dif\phi}}^{2} \right),\quad V=\frac{2 n}{r}\,,
\end{align}
with the parameter $n$ identified as
\begin{align}
n=\frac{1+C_1}{4(1-C_1)}\,.
\end{align}
This is the familiar form of the $1$-centred Gibbons--Hawking solution. Since $n$ is known to be redundant and can be set to unity in (\ref{1-centred_GH}), we see again that the BZ parameter $C_1$ in (\ref{metric_1-soliton}) is redundant.

We remark that the redundancy of the BZ parameter $C_1$ of the $1$-soliton solution has one immediate implication. Since the $n$-soliton solution on flat space for odd $n$ is in fact an even-soliton solution on the seed of the $1$-soliton solution, the above one-parameter ($C_1$) redundancy pervades the former. Hence, for odd $n$, the number of independent parameters of the $n$-soliton solution is $2n-2$. The accelerating multi-Kerr-NUT solution is thus a one-parameter generalisation of the multi-Kerr-NUT solution. This is also noticed in \cite{Jantzen:1983mba}.

To end this subsection, we present the explicit form of the multi-centred Gibbons--Hawking solution:
\begin{align}
\label{multi_GH}
{\dif s}^{2}&= V^{-1} \left( {\dif\tau}+A \right) ^{2}+V
\left( {\dif r}^{2}+{r}^{2} {{\dif\theta}}^{2}+ r^2 \sin ^{2}
\theta\, {{\dif\phi}}^{2}  \right),\quad
V=\sum_{i=1}^s \frac{2n_i}{r_i}\,,\nonumber\\ A&= \sum_{i=1}^s \frac{2 n_i (r \cos \theta-a_i)}{r_i}\,\dif\phi\,,\quad r_i=\sqrt {{r}^{2}+{{a_i}}^{2}-2 {a_i} r \cos \theta}\,.
\end{align}
This is the multi-centred generalisation of (\ref{1-centred_GH}) and is ALE. The centres are located at $r_i=0$, or in the Weyl--Papapetrou coordinates at $(\rho=0,z=a_i)$ \cite{Chen:2010zu}. We explicitly show how the $2$- and $3$-centred Gibbons--Hawking solutions are obtained from multi-soliton solutions on flat space in subsequent subsections.

\subsection{The $2$-soliton solution}

\label{sec_2-soliton}

The $2$-soliton solution can be explicitly written in the following form:
\begin{align}
\dif s^2&=\frac{\mu_1\mu_2F}{H}\left(\dif \tau+\omega\dif \phi\right)^2+\frac{\rho^2 H}{\mu_1\mu_2F}\dif\phi^2+\frac{k_0H}{\mu_1\mu_2R_{11}R_{22}}\left(\dif \rho^2+\dif z^2\right),\nonumber\\
H&=(C_1^2\mu_1^2+C_2^2\mu_2^2)R_{12}^2-C_1^2C_2^2\mu_1^2\mu_2^2\mu_{12}^2-2C_1C_2\mu_1\mu_2R_{11}R_{22}-\rho^4\mu_{12}^2\,,\nonumber\\
F&=(1+C_1^2C_2^2)\rho^2\mu_{12}^2-2C_1C_2R_{11}R_{22}+(C_1^2+C_2^2)R_{12}^2\,,\nonumber\\
\omega&=\frac{\mu_{12}R_{12}}{\mu_1\mu_2F}\left[C_1^2C_2\mu_1^2R_{22}-C_1C_2^2\mu_2^2R_{11}+\rho^2(C_2R_{22}-C_1R_{11})\right].
\label{metric_2-soliton}
\end{align}
The reader is reminded that $\mu_{ij}= \mu_i-\mu_j$ as defined in (\ref{metric_limit_static}).

The static limit of the 2-soliton solution is obtained by setting $k_0=C_2^{-2}$ and taking
\begin{align}
C_1\rightarrow 0\,,\quad C_2\rightarrow \infty\,,
\end{align}
with a resulting metric given by
\begin{align}
g_{\text{Sch}}=\left[ \begin {array}{cc} {\frac{\mu_1}{\mu_2}}&0\\0&{\frac {{\rho}^{2}\mu_2}{{\mu_1}} }\end {array} \right],
\qquad f_{\text{Sch}}=\frac{\mu_2 R_{12}^2}{\mu_1 R_{11} R_{22}}\,.
\label{metric_Sch}
\end{align}
This is the Schwarzschild solution.

To cast the $2$-soliton solution (\ref{metric_2-soliton}) in more familiar forms, we prefer to use prolate spheroidal coordinates $(p,q)$ defined in Appendix~\ref{sec_prolate_sph}. Expressed in these coordinates, the solitons $\mu_{1,2}$ are algebraic expressions and contain no cumbersome square roots. So we define
\begin{align}
z_1=z_{\text{I}}\equiv -\iota\,,\quad z_2=z_{\text{II}}\equiv \iota\,,
\end{align}
for some positive constant $\iota$. This identifies $\mu_{1,2}$ as
\begin{align}
\mu_1=\mu_{\text{I}}\,,\quad \mu_2=\mu_{\text{II}}\,,
\end{align}
respectively, whose explicit expressions are given in (\ref{mu_III}). The 2-soliton solution is then cast in the coordinate system $(\tau,\phi,p,q)$ and parameterised by $\iota$ and $C_{1,2}$. The metric components are purely algebraic, and their explicit form can be obtained straightforwardly and will not be presented.

The $2$-soliton solution is the Kerr-NUT solution, first demonstrated by Belinski and Zakharov \cite{Belinsky:1971nt}. To show this, we perform a linear coordinate transformation by doing the substitution as follows:
\begin{align}
\tau\rightarrow \tau+\frac{2\iota(1-C_1C_2)}{C_2-C_1}\phi\,,
\end{align}
and choose $k_0=(C_1-C_2)^{-2}$. The $2$-soliton solution is then brought to the Kerr-NUT solution in the familiar form
\begin{align}
\label{metric_Kerr-NUT}
{\dif s}^{2}=&{\frac {{\Delta}}{{\Sigma}}} [{\dif\tau}+ \left( 2n
\cos \theta +a \sin  ^{2} \theta
  \right) {\dif\phi}] ^{2}+{\frac {
  \sin ^{2} \theta  }{{\Sigma}}}[a{\dif\tau}-
 \left( {r}^{2}-{n}^{2}-{a}^{2} \right) {\dif\phi}] ^{2}\nonumber\\
 &+{\Sigma} \left( {\frac {{\dif r}^{2}}{{\Delta}}}
+{{\dif\theta}}^{2} \right),\nonumber\\
{\Sigma}=&{r}^{2}- \left( n-a\cos \theta  \right) ^{
2},\qquad
{\Delta}={r}^{2}-2mr-{a}^{2}+{n}^{2}\,,
\end{align}
with the coordinates $(r,\theta)$ identified as
\begin{align}
\label{rtheta1topq}
r=\iota q+m\,,\quad \cos\theta=p\,,
\end{align}
and the parameters $m$, $n$, and $a$ identified as
\begin{align}  m=\frac{\iota(C_1+C_2)}{C_2-C_1}\,,\quad n=\frac{\iota(1+C_1C_2)}{C_1-C_2}\,,\quad a=\frac{\iota(1-C_1C_2)}{C_2-C_1}\,.
\end{align}

\subsubsection{The $[2-1]$-soliton solution}

\label{sec_2m1}

For completeness, we calculate the rod structure of the 2-soliton solution. There are two turning points located at $(\rho=0,z=z_{1})$ and $(\rho=0,z=z_2)$, respectively, which divide the $z$-axis into three rods. The directions of these three rods, given by (\ref{first_direction}) and (\ref{recursive_relation}) with $n=2$, are, respectively,
\begin{align}
l_1&=\left(2\sqrt{k_0}z_{12}C_1C_2,\sqrt{k_0}(-C_1+C_2)\right),\quad 
l_2=\left(-2\sqrt{k_0}z_{12}C_2,-\sqrt{k_0}(-1+C_1C_2)\right),\nonumber\\
l_3&=\left(-2\sqrt{k_0}z_{12},\sqrt{k_0}(-C_1+C_2)\right).
\end{align}

There is only one $[2-1]$-soliton solution, namely, the $[2-1-0]$-soliton solution. Starting from the 2-soliton solution in the coordinates $(p,q)$, we choose to eliminate $\mu_2$ by setting
\begin{align}
C_2=1\,,
\end{align}
in which case Rod 2 and Rod 3 are joined up: $l_2=l_3$. The resulting solution is ALF with one turning point. The only known solution with these properties is the 1-centred Taub-NUT solution. They are indeed equivalent. To show this, we perform linear coordinate transformations on the $[2-1]$-soliton solution by doing the simultaneous substitutions as follows:
\begin{align}
\label{linear_transformation_2m1}
\tau\rightarrow \tau/\sqrt{\epsilon}+\sqrt{\epsilon}z_{12}\phi\,,\quad \phi\rightarrow -\sqrt{\epsilon}\phi\,,
\end{align}
and choose $k_0=\epsilon(1-C_1)^{-2}$, where $z_{ij}\equiv z_i-z_j$ and $\epsilon$ is an arbitrary constant. The $[2-1]$-soliton solution is then brought to the 1-centred Taub-NUT solution in the familiar form
\begin{align}
\label{self-dual Taub-NUT}
\dif s^{2}=V^{-1} { \left( {\dif\tau}+2n\cos  \theta\,
  {\dif\phi} \right) ^{2}}+V \left( {\dif r}^{2}+{r}^{2}{
{\dif\theta}}^{2}+{r}^{2} \sin ^2 \theta \,{{\dif\phi}}^{2} \right),\quad V=\epsilon+\frac{2 n}{r}\,,
\end{align}
with the coordinates $(r,\theta)$ identified as
\begin{align}
\label{rtheta} r\sin\theta=\iota\sqrt{(1-p^2)(q^2-1)}\,,\quad r\cos\theta=\iota pq+\iota\,, 
\end{align}
which are equivalent to (\ref{weyltortheta}) up to a shift of $z$, and the parameter $n$ identified as
\begin{align}
n=-\frac{\epsilon z_{12}(1+C_1)}{2(1-C_1)}\,.
\label{nut_charge_1TN}
\end{align}

We see from the above expression of $n$ that, among the parameters $\iota$ and $C_1$, one is redundant. In hindsight, $C_1$ can be set to an arbitrary value (except the degenerate cases $C_1=\pm 1$), say zero or infinity, to simplify calculations. We will see that it is a general feature of the $n$-soliton solution that whenever a soliton is eliminated and becomes a phantom soliton, the resulting solution possesses a one-parameter redundancy.

We now explicitly demonstrate that the soliton $\mu_2=\mu_{\text{II}}$ is indeed eliminated and the $[2-1]$-soliton solution can be written completely in terms of the soliton $\mu_1=\mu_{\text{I}}$. We start from the $[2-1]$-soliton solution in the form of the Taub-NUT metric (\ref{self-dual Taub-NUT}). The expressions $(r,\cos\theta)$ can be written in terms of the Weyl--Papapetrou coordinates and the soliton $\mu_{\text{I}}$ defined in Appendix~\ref{sec_prolate_sph} as follows:
\begin{align}
r=\frac{\rho^2+\mu_{\text{I}}^2}{2\mu_{\text{I}}}\,,\quad \cos\theta=\frac{\rho^2-\mu_{\text{I}}^2}{\rho^2+\mu_{\text{I}}^2}\,,
\label{rtheta-mu1}
\end{align}
which can be directly verified using (\ref{rtheta}) and (\ref{mu_III}). Substituting (\ref{rtheta-mu1}) into (\ref{self-dual Taub-NUT}), we obtain the $[2-1]$-soliton solution (or the $1$-centred Taub-NUT solution) in the form
\begin{align}
\dif s^2={V}^{-1}\left[\dif \tau+\frac{2n(\rho^2-\mu_{\text{I}}^2)}{\rho^2+\mu_{\text{I}}^2}\dif \phi\right]^2+{V}\left(\dif \rho^2+\dif z^2+\rho^2\dif\phi^2\right),\quad {V}=\epsilon+\frac{4n\mu_{\text{I}}}{\rho^2+\mu_{\text{I}}^2}\,,
\label{TN-1-soliton}
\end{align}
containing only the soliton $\mu_{\text{I}}$.

The 1-centred Taub-NUT solution (\ref{self-dual Taub-NUT}) has a multi-centred generalisation, which is also ALF and can be written as
\begin{align}
\label{multi_TN}
{\dif s}^{2}&= V^{-1} \left( {\dif\tau}+A \right) ^{2}+V
\left( {\dif r}^{2}+{r}^{2} {{\dif\theta}}^{2}+ r^2 \sin ^{2}
\theta\, {{\dif\phi}}^{2}  \right),\quad
 V=\epsilon+\sum_{i=1}^s \frac{2n_i}{r_i}\,,\nonumber\\
 A&= \sum_{i=1}^s \frac{2 n_i (r \cos \theta-a_i)}{r_i}\,\dif\phi\,,\quad r_i=\sqrt {{r}^{2}+{{a_i}}^{2}-2 {a_i} r \cos \theta}\,.
\end{align}
The centres are located at $r_i=0$, each carrying a NUT charge $n_i$. The $s$-centred Gibbons--Hawking solution (\ref{multi_GH}) can be recovered from the $s$-centred Taub-NUT solution (\ref{multi_TN}) in the ALE limit $\epsilon=0$. Reversely, the $s$-centred Taub-NUT solution (\ref{multi_TN}) can be recovered from the Gibbons--Hawking solution (\ref{multi_GH}) with $s+1$ centres, by taking the limit $a_{s+1}\rightarrow \infty$ while keeping the ratio $\frac{n_{s+1}}{a_{s+1}}$ fixed. In this limit, the constant $\epsilon$ and the one-form $A$ should also be appropriately redefined.  In subsequent subsections, we explicitly show how the $2$- and $3$-centred Taub-NUT solutions are obtained from multi-soliton solutions on flat space.

The only $[2-2]$-soliton solution is the $[2-1-1]$-soliton solution. It can be obtained from the $[2-1]$-soliton solution by further setting $C_1=-1$. So it is equivalent to (\ref{self-dual Taub-NUT}) with $n=0$ (see (\ref{nut_charge_1TN})), which is flat space (\ref{metric_seed_even_Euclidean}). 

\subsection{The $3$-soliton solution}

\label{sec_3-soliton}

The 3-soliton solution with the BZ parameter $C_3=0$ can be explicitly written in the following form:
\begin{align}
\label{solution-3-soliton}
\dif s^2&=\frac{\mu_3F}{\mu_1\mu_2H}(\dif \tau+\omega\dif \phi)^2+\frac{\rho^2\mu_1\mu_2 H}{\mu_3F}\dif\phi^2+\frac{k_0H}{\mu_1\mu_2\mu_3R_{11}R_{22}R_{33}}\left(\dif \rho^2+\dif z^2\right),\nonumber\\
H&=-C_1^2C_2^2\mu_{12}^2R_{13}^2R_{23}^2+C_1^2\mu_{23}^2R_{12}^2R_{13}^2+C_2^2\mu_{13}^2R_{12}^2R_{23}^2-2C_1C_2\mu_{13}\mu_{23}R_{11}R_{22}R_{13}R_{23}\nonumber\\
&-\rho^4\mu_{12}^2\mu_{13}^2\mu_{23}^2\,,\nonumber\\
F&=C_1^2C_2^2\rho^2\mu_{12}^2R_{13}^2R_{23}^2+C_1^2\mu_2^2\mu_{23}^2R_{12}^2R_{13}^2+C_2^2\mu_1^2\mu_{13}^2R_{12}^2R_{23}^2+\rho^2\mu_1^2\mu_2^2\mu_{12}^2\mu_{13}^2\mu_{23}^2\nonumber\\
&-2C_1C_2\mu_1\mu_2\mu_{13}\mu_{23}R_{11}R_{22}R_{13}R_{23}\,,\nonumber\\
\omega&=\frac{1}{\mu_3F}\Big[C_1C_2\mu_{12}R_{12}R_{13}R_{23}(C_2\mu_2\mu_{13}R_{11}R_{23}-C_1\mu_1\mu_{23}R_{22}R_{13})\nonumber\\
&+\rho^2\mu_{12}\mu_{13}\mu_{23}R_{12}(C_1\mu_2\mu_{23}R_{11}R_{13}-C_2\mu_1\mu_{13}R_{22}R_{23}\Big].
\end{align}
We will show in a moment that $C_3$ is in fact a redundant parameter, so it can be set to zero without loss of generality.

The static limit of the 3-soliton solution is obtained by setting $k_0=C_2^{-2}$ and taking
\begin{align}
C_{1,3}\rightarrow 0\,,\quad C_2\rightarrow \infty\,. 
\end{align}
Equivalently, it can be obtained by setting $k_0=C_2^{-2}$ and taking the limit $C_1\rightarrow 0$ followed by $C_2\rightarrow \infty$ from the metric (\ref{solution-3-soliton}), with a resulting metric given by
\begin{align}
g_{\text{C}}=\left[ \begin {array}{cc} {\frac{\mu_1\mu_3}{\mu_2} }&0\\0&{\frac{\rho^2\mu_2}{\mu_1\mu_3}}\end {array} \right],\qquad
f_{\text{C}}=\frac{\mu_{13}^2 R_{12}^2 R_{23}^2}{\mu_1\mu_2\mu_3  R_{11} R_{22} R_{33}}\,.
\end{align}
We recognise this solution as the C-metric. The C-metric was also constructed using the ISM in \cite{Jantzen:1983mba,Letelier:1985mba}.

The 3-soliton solution is ALE and has three turning points in its rod structure. One can also check that its Riemann tensor is neither self-dual nor anti-self-dual. The only known solution with these properties is the Pleba\'nski--Demia\'nski solution \cite{Plebanski:1976gy},\footnote{In this paper, the Pleba\'nski--Demia\'nski solution exclusively refers to the Ricci-flat subclass of the solution presented in \cite{Plebanski:1976gy}, obtained by setting the cosmological constant, electric and magnetic charges to zero.} well known as the generalisation of the C-metric with the inclusion of rotation and NUT charge. We now show that they are indeed equivalent. To cast the $3$-soliton solution in more familiar forms, we prefer to use the C-metric-like coordinates $(u,v)$ defined in Appendix~\ref{sec_C-metric-coordinates}. The major advantage of using these coordinates is that all the three solitons can be written as algebraic expressions, avoiding cumbersome square roots. We refer the reader to Appendix~\ref{sec_C-metric-coordinates} for more details of many properties of these coordinates. So we define
\begin{align}
z_1=z_{\text{A}}\equiv -c\varkappa^2\,,\qquad z_2=z_{\text{B}}\equiv c\varkappa^2\,,\qquad z_3=z_{\text{C}}\equiv \varkappa^2\,,
\end{align}
for some positive constants $c$ and $\varkappa$ with $0<c<1$. This identifies the three solitons $\mu_{1,2,3}$ as
\begin{align}
\mu_1=\mu_{\text{A}}\,,\qquad \mu_2=\mu_{\text{B}}\,,\qquad \mu_3=\mu_{\text{C}}\,,
\end{align}
respectively, whose explicit expressions are given in (\ref{mu_ABC}). The 3-soliton solution is then cast in the coordinates $(u,v)$ straightforwardly, with all metric components being algebraic.

As already pointed out, all odd-soliton solutions possess a one-parameter redundancy. We now prove this explicitly for the 3-soliton solution, in the coordinates $(u,v)$. We proceed as follows: we first set $C_3=0$ in the solution and then show that, by appropriate coordinate transformations and redefinitions of parameters, the original solution with $C_3$ arbitrary is recovered. So we set $C_3=0$ first and reparameterise $C_{1,2}$ in terms of new parameters $C_{1,2,3}$ by doing the simultaneous substitutions
\begin{align}
C_1\rightarrow \frac{C_1-C_3}{1-C_1C_3}\,,\quad C_2\rightarrow \frac{C_2-C_3}{1-C_2C_3}\,.
\end{align}
Now, if we perform linear coordinate transformations by doing the simultaneous substitutions as follows:
\begin{align}
\tau\rightarrow \frac{(\tau+C_3\phi)}{\sqrt{1-C_3^2}}\,,\qquad \phi\rightarrow \frac{(C_3\tau+\phi)}{\sqrt{1-C_3^2}}\,,
\end{align}
and choose $k_0$ as
\begin{align}
k_0\rightarrow \frac{(1-C_2C_3)^2(1-C_1C_3)^2k_0}{(1-C_3^2)}\,,
\end{align}
the metric of the 3-soliton solution (with $C_3=0$) is brought to its original form with $C_3$ an arbitrary parameter. This proves that we can set $C_3=0$ without loss of generality. The metric (\ref{solution-3-soliton}) is then the most general 3-soliton solution.

At this point, the 3-soliton solution (\ref{solution-3-soliton}) is cast in the coordinates $(u,v)$ with a cubic structure function $G(u)$ (see Appendix~\ref{sec_C-metric-coordinates}) and is parameterised by $c$, $\varkappa$ and $C_{1,2}$. Now, we want to write it in the form (\ref{C-metric_metrics}) with a quartic structure function and attempt to use the coefficients of this structure function as the fundamental parameters. We bear in mind the remarkably compact form of the Pleba\'nski--Demia\'nski solution (\ref{metric_PD}) in doing so. A quartic structure function can be obtained by performing a M\"obius transformation on the coordinates $(u,v)$ as follows:
\begin{align} u=\frac{\delta_2x+\delta_3}{x+\delta_1}\,,\quad v=\frac{\delta_2y+\delta_3}{y+\delta_1}\,,
\end{align}
where $\delta_{1,2,3}$ are free parameters which can be chosen arbitrarily, and $(x,y)$ are newly introduced coordinates. Now, we appropriately fix $\delta_{1,2,3}$ such that the numerator of the metric component $g_{xx}$ contains a factor $1-x^2y^2$. This can be done by equating
\begin{align}
C_1&=\frac{\delta_1(\delta_1-\delta_3)(\delta_1^2-\delta_3^2)+(1-\delta_2)(\delta_2^2-1)}{(\delta_2-1)(\delta_1-\delta_3)(\delta_1\delta_2-\delta_3)}\,,\quad c=\frac{\delta_1^2(\delta_1^2-\delta_3^2)+\delta_2^2-1}{\delta_1\delta_3(\delta_3^2-\delta_1^2)+\delta_2(1-\delta_2^2)}\,,\nonumber\\
C_2&=\frac{(\delta_2+1)(\delta_1+\delta_3)(\delta_1\delta_2-\delta_3) }{\delta_1(\delta_1+\delta_3)(\delta_1^2-\delta_3^2)+(\delta_2+1)(\delta_2^2-1)}\,.
\label{parameters_redefinitions_3-soliton}
\end{align}
These equations, viewed as relations between the parameters $\{c, C_{1,2}\}$ and the parameters $\{\delta_{1,2,3}\}$, are invertible. So by imposing (\ref{parameters_redefinitions_3-soliton}), we choose to use $\delta_{1,2,3}$ and $\varkappa$, instead of $c$, $C_{1,2}$ and $\varkappa$, as fundamental parameters. The 3-soliton solution in this parameterisation is much simpler.

Now, the denominator of $g_{xx}$ contains a quartic polynomial of $x$ with roots
\begin{align}
x_1=-\delta_1\,,\quad x_2=-\frac{\delta_1+\delta_3}{1+\delta_2}\,,\quad x_3=\frac{\delta_1-\delta_3}{\delta_2-1}\,,\quad x_4=\frac{\delta_2^2-1}{\delta_1(\delta_1^2-\delta_3^2)}\,,
\label{roots_PD}
\end{align}
which satisfy
\begin{align}
\label{x1234e1}
x_1x_2x_3x_4=1\,.
\end{align}
The reader is reminded that this polynomial is the structure function of the solution in the coordinates $(x,y)$. We can now solve $\delta_{1,2,3}$ from (\ref{roots_PD}) in terms of $x_{1,...,4}$ (subject to (\ref{x1234e1})) and substitute them into the metric components. We identify the four roots $x_{1,...,4}$ with those of a quartic polynomial
\begin{align}
P=\gamma+2nx-\epsilon x^2+2mx^3+\gamma x^4\,,
\end{align}
by equating
\begin{align}
2n/\gamma&=-(x_1x_2x_3+x_1x_2x_4+x_1x_3x_4+x_2x_3x_4)\,,\quad 2m/\gamma=-(x_1+x_2+x_3+x_4)\,,\nonumber\\
\epsilon/\gamma&=-(x_1x_2+x_1x_3+x_1x_4+x_2x_3+x_2x_4+x_3x_4)\,.
\end{align}
This redefines the parameters $x_{1,...,4}$ (subject to (\ref{x1234e1})) in terms of $n$, $m$ and $\epsilon$, but the explicit expressions cannot be written down.

The metric components of the 3-soliton solution are parameterised in terms of $x_{1,...,4}$ in a non-symmetric way and cannot be explicitly written down in terms of the parameters $n$, $m$ and $\epsilon$. The last key step to recover the Pleba\'nski--Demia\'nski solution is to redefine coordinates and parameters by doing the simultaneous substitutions as follows:
\begin{align}
\label{linear_transformation_3-soliton} \tau&\rightarrow \tau-x_2x_3\phi\,,\qquad \phi\rightarrow x_2x_3\tau-\phi\,,\nonumber\\
\varkappa^2&=\frac{\gamma(x_1x_2+x_1x_3+x_2x_4+x_3x_4-2x_1x_4-2x_2x_3)}{4(1-x_2^2x_3^2)}\,.
\end{align}
The parameters $x_{1,...,4}$ obviously do not appear symmetrically in these transformations, but they enter the transformed metric components $g_{ab}$ symmetrically. So now $g_{ab}$ can be explicitly written in terms of the parameters $\gamma$, $n$, $\epsilon$ and $m$. If we appropriately choose $k_0$ as
\begin{align}
k_0=\frac{4x_3^2(1-x_2^2x_3^2)^2}{\gamma^4(x_1-x_3)^2(x_1-x_4)^2(x_2-x_3)^2(x_3-x_4)^2}\,,
\end{align}
$g_{xx}$ and $g_{yy}$ can also be written in terms of these parameters. The 3-soliton solution is then brought to the Pleba\'nski--Demia\'nski solution \cite{Plebanski:1976gy} in the familiar form
\begin{align}
\label{metric_PD}
\dif s^2&=\frac{1}{(x-y)^2}\Big[\frac{1-x^2y^2}{P}\dif x^2+\frac{P}{1-x^2y^2}(\dif \phi-y^2\dif \tau)^2\nonumber\\
&\hspace{0.56in}-\frac{1-x^2y^2}{Q}\dif y^2-\frac{Q}{1-x^2y^2}(\dif \tau-x^2\dif \phi)^2\Big],\nonumber\\
P&=\gamma+2nx-\epsilon x^2+2mx^3+\gamma x^4\,,\quad Q=\gamma+2ny-\epsilon y^2+2my^3+\gamma y^4\,.
\end{align}
We note that all free parameters are now encoded in the structure function $P$, with $\gamma$ setting the scale of the solution.

\subsubsection{The $[3-1]$-soliton solution}

\label{sec_3m1m1}

For completeness, we calculate the rod structure of the 3-soliton solution in this subsection. To preserve the symmetries, we restore the parameter $C_3$ and keep it arbitrary. There are three turning points located at $(\rho=0,z=z_{1})$, $(\rho=0,z=z_2)$ and $(\rho=0,z=z_3)$, respectively, which divide the $z$-axis into four rods. The directions of these four rods, given by (\ref{first_direction}) and (\ref{recursive_relation}) with $n=3$, are, respectively,
\begin{align}
l_1&=\left(-2\sqrt{k_0}(-C_1C_2z_{12}+C_1C_3z_{13}-C_2C_3z_{23}),-2\sqrt{k_0}(-C_1z_{23}+C_2z_{13}-C_3z_{12})\right),\nonumber\\
l_2&=\left(2\sqrt{k_0}(-C_2z_{12}+C_3z_{13}-C_1C_2C_3z_{23}),2\sqrt{k_0}(-z_{23}+C_1C_2z_{13}-C_1C_3z_{12})\right),\nonumber\\
l_3&=\left(2\sqrt{k_0}(-z_{12}+C_2C_3z_{13}-C_1C_3z_{23}),2\sqrt{k_0}(-C_2z_{23}+C_1z_{13}-C_1C_2C_3z_{12})\right),\nonumber\\
l_4&=\left(-2\sqrt{k_0}(-C_3z_{12}+C_2z_{13}-C_1z_{23}),-2\sqrt{k_0}(-C_2C_3z_{23}+C_1C_3z_{13}-C_1C_2z_{12})\right).
\end{align}

There is only one $[3-1]$-soliton solution, namely, the $[3-1-0]$-soliton solution. We choose to eliminate $\mu_2$ by setting
\begin{align}
C_2=1\,,
\label{join_up_three-soliton}
\end{align}
in which case Rod 2 and Rod 3 are joined up: $l_2=l_3$. The resulting solution is ALE with two turning points. The only known solution with these properties is the 2-centred Gibbons--Hawking solution (with the Eguchi--Hanson instanton as a special case \cite{Eguchi:1978xp}). Indeed, they are equivalent as we will show. This is consistent with the known result that the Pleba\'nski--Demia\'nski solution (\ref{metric_PD}) reduces to the 2-centred Gibbons--Hawking solution if one of its turning points is eliminated. 

Since the $[3-1]$-soliton solution has two free solitons, we prefer to cast it in the prolate spheroidal coordinates $(p,q)$ (see Appendix~\ref{sec_prolate_sph}). So we define
\begin{align}
z_1=-\iota\,,\qquad z_2=d\iota\,,\qquad z_3=\iota\,,
\label{three-soliton_turning-points}
\end{align}
for some constants $\iota$ and $d$. This identifies the two free solitons $\mu_{1,3}$ as
\begin{align}
\mu_1=\mu_{\text{I}}\,,\quad \mu_3=\mu_{\text{II}}\,,
\end{align}
and the phantom soliton $\mu_2$ as
\begin{align}
\mu_{2}=\iota(W-pq+d)\,,
\end{align}
where $W$ is a function of $p$ and $q$ whose square is
\begin{align}
W^2=p^2+q^2+d^2-2dpq-1\,.
\label{W}
\end{align}
If $C_2$ is kept general, all metric components of the 3-soliton solution are fractions in terms of $W$ and can be expressed in the form $\frac{w_1+w_2W}{w_3+w_4W}$, where $w_{1,...,4}$ are polynomials of $p$ and $q$. When the joining-up condition (\ref{join_up_three-soliton}) is imposed, one finds that $w_1w_4=w_2w_3$. So $W$ cancels in all the metric components, indicating that $\mu_2$ is indeed eliminated (see comments around (\ref{pq_metrics})). The resulting $[3-1]$-soliton solution is then written purely algebraically in the coordinates $(p,q)$ and is parameterised by $\iota$, $d$ and $C_{1,3}$. The explicit form of the $[3-1]$-soliton solution in this parameterisation will not be presented.

In our derivation, we choose to eliminate $\mu_2$ to obtain the $[3-1]$-soliton solution. From (\ref{ordering_solitons}) and (\ref{three-soliton_turning-points}), we could deduce the range for the parameter $d$ as $-1<d<1$. Recall that we can choose to eliminate $\mu_1$ or $\mu_3$, instead of $\mu_2$, to obtain the $[3-1]$-soliton solution, in which case the range $d<-1$ or $d>1$ could be obtained. This means that, in order not to lose generality, the parameter $d$ should be treated as an arbitrary parameter in the $[3-1]$-soliton solution, even though in the present case a particular soliton $\mu_2$ is chosen to be eliminated. Similar comments apply to all $[n-m]$-soliton solutions when one or more solitons are eliminated $m\ge 1$.

Now, we perform linear coordinate transformations by doing the simultaneous substitutions (\ref{linear_transformation_1-soliton}) and choose $k_0=[2z_{13}(1-C_1)(1-C_3)]^{-2}$. The $[3-1]$-soliton solution is then brought to the 2-centred Gibbons--Hawking solution in the familiar form, namely, the metric (\ref{multi_GH}) with $s=2$, with the coordinates $(r,\theta)$ identified as (\ref{weyltortheta}) or equivalently 
\begin{align}
\label{rtheta2topq}
r\sin\theta=\iota\sqrt{(1-p^2)(q^2-1)}\,,\quad r\cos\theta=\iota pq\,,
\end{align}
and the parameters $a_{1,2}$ and $n_{1,2}$ identified as $a_{1,2}=z_{\text{I,II}}$ and
\begin{align}
n_1=\frac{z_{12}(1+C_1)}{4z_{13}(1-C_1)}\,,\quad n_2=\frac{z_{23}(1+C_3)}{4z_{13}(1-C_3)}\,.
\label{nut-charge-identification-3m1}
\end{align}
From the above relations between parameters, we see that $C_{1,3}$ are redundant and can be set to zero without loss of generality. This is consistent with previous observations: among the two parameters $C_{1,3}$, one is generally redundant in the 3-soliton solution, and the other is redundant due to the elimination of $\mu_2$.

There is only one $[3-2]$-soliton solution, namely, the $[3-1-1]$-soliton solution. It can be obtained from the above $[3-1]$-soliton solution by further eliminating a soliton, say $\mu_1$, by setting its BZ parameter $C_1$ to $-1$. Now, we have shown that the $[3-1]$-soliton solution is the 2-centred Gibbons--Hawking solution. At this point, we can calculate the rod structure of the 2-centred Gibbons--Hawking solution (see \cite{Chen:2010zu}). From this rod structure, we conclude that one turning point can be eliminated from the 2-centred Gibbons--Hawking solution, namely, (\ref{multi_GH}) with $s=2$, by setting $n_1$ or $n_2$ to zero. The solution thus obtained is obviously the $1$-centred Gibbons--Hawking solution. The elimination of one turning point from the 2-centred Gibbons--Hawking solution corresponds to the elimination of one free soliton from the $[3-1]$-soliton solution. This identifies the $[3-1-1]$-soliton solution as the 1-centred Gibbons--Hawking solution.

\subsection{The $4$-soliton solution}

The explicit form of the 4-soliton solution is rather complicated, so we do not present it here. The static limit is obtained by setting $k_0=(C_2C_4)^{-2}$ and taking
\begin{align}
C_{1,3}\rightarrow 0\,,\quad C_{2,4}\rightarrow \infty\,,
\end{align}
with a resulting metric given by
\begin{align}
g_{{\text{2-Sch}}}=\left[ \begin {array}{cc} {\frac{\mu_1\mu_3}{\mu_2\mu_4}}&0\\0&{\frac{\rho^2 \mu_2\mu_4}{\mu_1\mu_3}}\end {array} \right],\qquad f_{\text{2-Sch}}=\frac{\mu_{13}^2 \mu_{24}^2 R_{12}^2 R_{23}^2 R_{34}^2R_{14}^2 }{\mu_1^3\mu_2 \mu_3^3\mu_4  R_{11} R_{22} R_{33} R_{44}}\,.
\end{align}
We recognise this solution as the double-Schwarzschild solution.

The 4-soliton solution has previously been constructed using the ISM in equivalent forms in \cite{Letelier:1998ft,Herdeiro:2008kq}. It is known to describe a system consisting of two Kerr-NUT black holes \cite{Kramer:1980mba}. Our primary interest in this section is, however, multi-soliton solutions with up to three free solitons. These solutions can be cast in purely algebraic forms in the coordinates $(p,q)$ and $(u,v)$ in the case of two and three free solitons, respectively, whose analysis can be well handled using a computer. So now we study the $[4-1]$- and $[4-2]$-soliton solutions.

\subsubsection{The $[4-1]$-soliton solution}

\label{sec_4m1}

There is only one $[4-1]$-soliton solution, namely, the $[4-1-0]$-soliton solution. We choose to eliminate $\mu_4$ by setting
\begin{align}
\label{condition_joinup_double_Sch1}
C_4=1\,.
\end{align}
The resulting solution is ALF with three turning points. One can also check that its Riemann tensor is neither self-dual nor anti-self-dual. The only known solution with these properties was first discovered in \cite{Chen:2015vva}. They are indeed equivalent. We now show how the latter solution can be recovered from the $[4-1]$-soliton solution.

Since the $[4-1]$-soliton solution contains three free solitons, the first step is to cast it in the coordinates $(u,v)$ (see Appendix~\ref{sec_C-metric-coordinates}). So we define
\begin{align}
z_1=z_{\text{A}}\,,\qquad z_2=z_{\text{B}}\,,\qquad z_3=z_{\text{C}}\,,\qquad z_4=d\varkappa^2\,,
\end{align}
for some constants $c$, $\varkappa$ and $d$. This identifies the three free solitons $\mu_{1,2,3}$ as
\begin{align}
\mu_1=\mu_{\text{A}}\,,\quad \mu_2=\mu_{\text{B}}\,,\quad \mu_3=\mu_{\text{C}}\,,
\end{align}
and the phantom soliton $\mu_4$ as
\begin{align} \mu_4=\frac{\varkappa^2Q}{u-v}-\left[\frac{\varkappa^2(1-uv)(2+cu+cv)}{(u-v)^2}-d\varkappa^2\right],
\end{align}
where $Q$ is a function of $u$ and $v$ whose square is
\begin{align}
Q^2=(u-v)^2d^2-2(1-uv)(2+cu+cv)d+c^2(1+uv)^2+4(1+cu+cv)\,.
\label{Q}
\end{align}
If $C_4$ is kept general, all metric components of the 4-soliton solution are fractions in terms of $Q$ and can be expressed in the form $\frac{q_1+q_2Q}{q_3+q_4Q}$, where $q_{1,...,4}$ are polynomials of $p$ and $q$. When the joining-up condition (\ref{condition_joinup_double_Sch1}) is imposed, one finds that $q_1q_4=q_2q_3$. So $Q$ cancels in all the metric components, indicating that $\mu_4$ is indeed eliminated (see comments around (\ref{C-metric_metrics})). The $[4-1]$-soliton solution is then written purely algebraically in terms of $(u,v)$. The situation is similar to the $[3-1]$-soliton solution, which can be written purely algebraically in terms of $(p,q)$.

The conversion of the $[4-1]$-soliton solution to the coordinates $(u,v)$ can be computationally so complicated that it cannot be done using a computer within a typical amount of time and memory usage. We used a trick that dramatically simplifies the computation: instead of converting the components $g_{ab}$ directly, we first convert expressions $f{g_{ab}}$ and $f$. The latter expressions are quadratic polynomials in terms of the independent BZ parameters $C_{1,2,3}$. The point to note is that, after imposing the joining-up condition (\ref{condition_joinup_double_Sch1}), $Q$ drops out in each coefficient of these polynomials. So we can convert each coefficient individually and assemble them to get $f{g_{ab}}$ and $f$, from which we can recover the metric. The computation can be well handled.

Now, the $[4-1]$-soliton solution is cast in the coordinates $(u,v)$ and parameterised by $c$, $d$, $\varkappa$, and $C_{1,2,3}$. The form is still rather complicated and will not be presented explicitly. Recall that once there is a phantom soliton, there is a one-parameter redundancy. We now show that $C_3$ is redundant. The strategy we use is similar to that used in proving that the parameter $C_3$ is redundant in the 3-soliton solution. So we first set $C_3=0$ and reparameterise $d$ and $C_{1,2}$ by doing the simultaneous substitutions as follows:
\begin{align}
 C_1&\rightarrow\frac{(1+c)C_1C_3-(c+d)C_1+(d-1)C_3}{(d-1)C_1C_3+(1+c)C_3-c-d}\,, \quad d\rightarrow \frac{(1+C_3)d-2C_3}{1-C_3}\,,\nonumber\\
C_2&\rightarrow\frac{(1-c)C_2C_3+(c-d)C_2+(d-1)C_3}{(d-1)C_2C_3+(1-c)C_3+c-d}\,.
\end{align}
Now, if we perform a linear coordinate transformation by doing the substitution as
\begin{align}
\tau&\rightarrow \tau+\frac{2\varkappa^2(d-1)C_3}{1-C_3}\phi\,,
\end{align}
and choose $k_0$ as
\begin{align}
k_0&\rightarrow k_0(1-C_3)^2[(1+c)C_3-(1-d)C_1C_3-c-d]^2[(1-c)C_3-(1-d)C_2C_3+c-d]^2\nonumber\\
&\times [((c+d-2)C_3-c+d)((c-d+2)C_3-c-d)]^{-2}\,,
\end{align}
the metric of the $[4-1]$-soliton solution (with $C_3=0$) is brought to its original form, with $C_3$ an arbitrary parameter. So the parameter $C_3$ is redundant, and we set $C_3=0$ in the rest of this subsection.

Now, the $[4-1]$-soliton solution, cast in the coordinates $(u,v)$, contains five non-trivial parameters $c$, $d$, $\varkappa$ and  $C_{1,2}$. We observe that $g_{uu}$ contains a factor $R$ linear in terms of both $u$ and $v$ in the form
\begin{align}
R=\gamma_1+\gamma_2 u+ v+\gamma_3 uv\,,
\label{R-factor}
\end{align}
for some constants $\gamma_{1,2,3}$ which are functions of $d$ and $C_{1,2}$. All metric components are simplified if we use $\gamma_{1,2,3}$, instead of $d$ and $C_{1,2}$, as fundamental parameters. This is achieved by equating
\begin{align} C_1=\frac{(1+c)(1-\gamma_3)}{c(\gamma_1+\gamma_2)-1-\gamma_3}\,,\quad C_2=\frac{c(\gamma_1-\gamma_2)-1+\gamma_3}{(1-c)(1+\gamma_3)}\,,\quad d=\frac{2-c(\gamma_1+\gamma_3)}{1-\gamma_2}\,.
\end{align}
The $[4-1]$-soliton solution parameterised by $c$, $\varkappa$ and $\gamma_{1,2,3}$ has a much simpler form, which will not be presented, however.

Inspired by the derivation of the Pleba\'nski--Demia\'nski solution (\ref{metric_PD}) from the 3-soliton solution, shown in Sec.~\ref{sec_3-soliton}, we are led to perform a M\"obius transformation on the coordinates $(u,v)$ and attempt to absorb as many parameters as possible into the structure function. We thus define
\begin{align} u=\frac{\delta_2x+\delta_3}{x+\delta_1}\,,\quad v=\frac{\delta_2y+\delta_3}{y+\delta_1}\,,
\end{align}
for some constants $\delta_{1,2,3}$ to be determined. The M\"obius transformation preserves the form of $R$ in (\ref{R-factor}), in the sense that now $g_{xx}$ contains a factor in the form $R'=\gamma'_1+\gamma'_2 x+ y+\gamma'_3 xy$. It is possible to adjust the parameters $\delta_{1,2,3}$ of the M\"obius transformation such that
\begin{align}
R'=y+\nu x\,,\quad {\text{i.e.,}}\quad \gamma'_1=\gamma'_3=0\,,\quad \gamma'_2=\nu\,.
\end{align}
This can be achieved by equating
\begin{align} \gamma_1=-\frac{(1+\nu)\delta_2\delta_3}{\delta_1\delta_2+\nu\delta_3}\,,\quad \gamma_2=\frac{\delta_3+\nu\delta_1\delta_2}{\delta_1\delta_2+\nu\delta_3}\,,\quad \gamma_3=-\frac{(1+\nu)\delta_1}{\delta_1\delta_2+\nu\delta_3}\,.
\end{align}
These equations reparameterise $\gamma_{1,2,3}$ in terms of $\nu$ and $\delta_{1,2}$; it is obvious that $\delta_3/\delta_1$, and thus $\delta_3$, is an irrelevant parameter. After the coordinate transformations and parameter redefinitions, the $[4-1]$-soliton solution is now cast in the coordinates $(x,y)$ and parameterised by $c$, $\varkappa$, $\delta_{1,2}$, $(\delta_3)$ and $\nu$.

The structure function in the coordinates $(x,y)$ is now quartic, with four roots given by
\begin{align}
x_1=-\delta_1\,,\quad x_2=\frac{\delta_1-\delta_3}{\delta_2-1}\,,\quad x_3=-\frac{\delta_1+\delta_3}{1+\delta_2}\,,\quad x_4=-\frac{\delta_1+c\delta_3}{1+c\delta_2}\,.
\end{align}
We can now solve $\delta_{1,2,3}$ and $c$ in terms of the roots $x_{1,...,4}$ and substitute them into the metric components. We identify these four roots as those of a quartic polynomial
\begin{align}
X=a_0+a_1x+a_2x^2+a_3x^3+a_4x^4\,,
\end{align}
by equating
\begin{align}
a_0/a_4&=x_1x_2x_3x_4\,,\quad a_1/a_4=-(x_1x_2x_3+x_1x_2x_4+x_1x_3x_4+x_2x_3x_4)\,, \nonumber\\
a_2/a_4&=x_1x_2+x_1x_3+x_1x_4+x_2x_3+x_2x_4+x_3x_4\,,\quad a_3/a_4=-(x_1+x_2+x_3+x_4)\,.
\label{structure_function_coefficients}
\end{align}
This redefines the parameters $x_{1,...,4}$ in terms of $a_{0,...,4}$, but the explicit expressions cannot be written down. $X$ appears as a factor in the denominator of $g_{xx}$ and is the structure function of the solution in the coordinates $(x,y)$.

The last task is to write the metric components in terms of the coefficients, namely, the parameters $a_{0,...,4}$ rather than the roots, namely, $x_{1,...,4}$ of the structure function $X$. At this point, $g_{\tau\tau}$ contains $x_{1,...,4}$ symmetrically through combinations that appear on the right-hand sides of the equations in Eq.~(\ref{structure_function_coefficients}), so it can be directly written in terms of $a_{0,...,4}$. This is not the case for $g_{\tau\phi}$ and $g_{\phi\phi}$, since they contain the parameters $x_{1,...,4}$ in a non-symmetric way. We recall that, with $g_{\tau\tau}$ fixed, we have the freedom to perform a linear coordinate transformation in the form $\tau\rightarrow \tau+b\phi$ for some constant $b$. We hope that the parameters $x_{1,...,4}$ enter the transformed metric components $g_{\tau\phi}$ and $g_{\phi\phi}$ symmetrically, so the latter can be reparameterised explicitly in terms of $a_{0,...,4}$. Indeed, we find that, if $b$ and $\varkappa^2$ are appropriately chosen as follows:
\begin{align}
b&= a_4[\nu^2x_2x_3-x_1x_4-\nu(x_1x_2+x_1x_3+2x_1x_4+x_2x_4+x_3x_4)]/(1-\nu^2)\,,\nonumber\\
\varkappa^2&= a_4(x_1x_2+x_1x_3-2x_1x_4-2x_2x_3+x_2x_4+x_3x_4)/4\,,
\end{align}
all metric components $g_{ab}$ can be explicitly written in terms of $a_{0,...,4}$, and so will be $g_{xx}$ and $g_{yy}$ if $k_0$ is chosen as
\begin{align}
k_0&= 4k(1-\nu)^3(x_1+\nu x_3)^2(x_4+\nu x_3)^2(1+\nu)^{-1}\Big[a_4^3(x_1-x_3)(x_1-x_4)(x_2-x_3)\nonumber\\
&\times (x_3-x_4)(2x_1x_4-2\nu x_2x_3-(1-\nu)(x_1x_3+x_2x_4))\nonumber\\
&\times (2x_1x_4-2\nu x_2x_3-(1-\nu)(x_1x_2+x_3x_4))\Big]^{-2}\,.
\end{align}
We remark that now $k_0$ and so $k$, instead of $\varkappa$, is chosen as the scale parameter. The metric is slightly simplified by performing the following linear coordinate transformations:
\begin{align}
\tau&\rightarrow \tau/\sqrt{1-\nu^2}\,,\qquad \phi\rightarrow\sqrt{1-\nu^2}\phi\,.
\end{align}
The $[4-1]$-soliton solution is then brought to the following form:
\begin{align}
\label{metric_new_ricci}
\dif s^2&=\frac{\left(F\dif \tau+G\dif \phi\right)^2}{(x-y)HF}+\frac{kH}{(x-y)^3}\left(\frac{\dif x^2}{X}-\frac{\dif y^2}{Y}-\frac{XY}{kF}\,\dif \phi^2\right),\nonumber\\
H&=(\nu x+ y)[(\nu x-y)(a_1-a_3xy)-2(1-\nu)(a_0-a_4x^2y^2)]\,,\quad F=y^2X-x^2Y\,,\nonumber\\
G&=(\nu^2a_0+2\nu a_3y^3+2\nu a_4y^4-a_4y^4)X+(a_0-2\nu a_0-2\nu a_1x-\nu^2 a_4x^4)Y\,,\nonumber\\
X&=a_0+a_1x+a_2x^2+a_3x^3+a_4x^4\,,\quad Y=a_0+a_1y+a_2y^2+a_3y^3+a_4y^4\,,
\end{align}
which is precisely the solution given in \cite{Chen:2015vva}.

\subsubsection{The $[4-2-0]$-soliton solution}

There are two separate $[4-2]$-soliton solutions, namely, the $[4-2-0]$-soliton solution and the $[4-1-1]$-soliton solution. They are both ALF and have two turning points. The only known solutions with these properties are the 2-centred Taub-NUT and the Kerr-NUT solutions. We now show how they are exactly recovered from the two $[4-2]$-soliton solutions. The reader is reminded that to preserve symmetries, we do not set $C_3=0$.

We first show that the $[4-2-0]$-soliton solution is equivalent to the 2-centred Taub-NUT solution. We choose to eliminate $\mu_2$ and $\mu_4$ by setting
\begin{align}
C_2=1\,,\qquad C_4=1\,.
\label{C24e1}
\end{align}
Since there are two free solitons, we prefer to use the prolate spheroidal coordinates $(p,q)$ (see Appendix~\ref{sec_prolate_sph}). So we define
\begin{align}
z_1=z_{\text{I}}\,,\quad z_2=d_1\iota\,,\quad z_3=z_{\text{II}}\,,\quad z_4=d_2\iota\,,
\end{align}
for some constants $\iota$ and $d_{1,2}$. This identifies the two free solitons $\mu_{1,3}$ as
\begin{align}
\mu_1=\mu_{\text{I}}\,,\quad \mu_3=\mu_{\text{II}}\,,
\end{align}
and the two phantom solitons $\mu_{2,4}$ as
\begin{align} \mu_{2,4}=\iota(W_{1,2}-pq+d_{1,2})\,,
\end{align}
where $W_{i}\equiv W(d_{i})$. The expression $W$ as a function of $d$ is given by (\ref{W}). The joining-up conditions (\ref{C24e1}) ensure that \emph{both} $W_1$ and $W_2$ cancel from all metric components. The $[4-2-0]$-soliton solution is then cast in the coordinates $(p,q)$ with purely algebraic metric components and is parameterised by $\iota$, $d_{1,2}$ and $C_{1,3}$, whose explicit form will not be presented.

We perform linear coordinate transformations by doing the simultaneous substitutions as follows:
\begin{align} 
\tau\rightarrow \tau/\sqrt{\epsilon}+\sqrt{\epsilon}(z_{12}+z_{34})\phi\,,\quad \phi\rightarrow -\sqrt{\epsilon}\phi\,,
\end{align}
and choose $k_0=\epsilon[4z_{13} z_{24}(1-C_1)(1-C_3)]^{-2}$, where $\epsilon$ is an arbitrary constant. The $[4-2-0]$-soliton solution is then brought to the 2-centred Taub-NUT solution in the familiar form, namely, the metric (\ref{multi_TN}) with $s=2$, with the coordinates $(r,\theta)$ identified as (\ref{rtheta2topq}) or (\ref{weyltortheta}) and the parameters $a_{1,2}$ and $n_{1,2}$ identified as $a_{1,2}=z_{\text{I,II}}$ and 
\begin{align}
n_1=-\frac{\epsilon z_{12}z_{14}(1+C_1)}{2z_{13}(1-C_1)}\,,\quad n_2=-\frac{\epsilon z_{23}z_{34}(1+C_3)}{2z_{13}(1-C_3)}\,.
\end{align}

\subsubsection{The $[4-1-1]$-soliton solution}

Now, we show that the $[4-1-1]$-soliton solution is equivalent to the Kerr-NUT solution. In this case, we eliminate $\mu_2$ and $\mu_4$ by setting
\begin{align}
C_2=1\,,\qquad C_4= -1\,.
\end{align}
The $[4-1-1]$-soliton solution is then cast in the coordinates $(p,q)$ as done in the previous subsection, and it is now parameterised by $\iota$, $d_{1,2}$ and $C_{1,3}$.

The $[4-1-1]$-soliton solution is equivalent to the 2-soliton solution. Recall that the latter solution is cast in the coordinates $(p,q)$ in Sec.~\ref{sec_2-soliton} and is parameterised by $\iota$ and $C_{1,2}$. Starting from this solution, if we redefine its parameters $C_{1,2}$ as follows:
\begin{align}
C_{1}\rightarrow \frac{(d_1-d_2)+(d_1+d_2+ 2)C_1}{(d_1-d_2)C_1+(d_1+d_2+ 2)}\,,\quad C_{2}\rightarrow \frac{(d_1-d_2)+(d_1+d_2- 2)C_3}{(d_1-d_2)C_3+(d_1+d_2- 2)}\,,
\end{align}
perform a linear coordinate transformation by doing the substitution as
\begin{align}
\tau\rightarrow \tau-\iota(d_1-d_2)\phi\,,
\end{align}
and choose its $k_0$ as
\begin{align}
k_0\rightarrow 4\iota^4 k_0[(d_1-d_2)C_1+(d_1+d_2+ 2)]^2[(d_1-d_2)C_3+(d_1+d_2- 2)]^2\,,
\end{align}
we then obtain the present form of the $[4-1-1]$-soliton solution. So the $[4-1-1]$-soliton solution is equivalent to the 2-soliton solution and, thus, the Kerr-NUT solution.

For both of the $[4-2]$-soliton solutions, we see that the parameters $C_{1,3}$ are redundant and can be set to zero without loss of generality, consistent with the observation that each phantom soliton introduces a one-parameter redundancy.

\subsection{The $5$-soliton solution}

The static limit of the 5-soliton solution can be obtained by setting $k_0=(C_2C_4)^{-2}$ and taking
\begin{align}
C_{1,3,5}\rightarrow 0\,,\quad C_{2,4}\rightarrow \infty\,,
\end{align}
with a resulting metric given by
\begin{align}
g_{{\text{acc 2-Sch}}}&=\left[ \begin {array}{cc} {\frac{\mu_1\mu_3\mu_5}{\mu_2\mu_4} }&0\\0&{\frac{\rho^2\mu_2\mu_4}{\mu_1\mu_3\mu_5}}\end {array} \right],\nonumber\\
f_{{\text{acc 2-Sch}}}&=\frac{\mu_{13}^2 \mu_{15}^2\mu_{35}^2\mu_{24}^2 R_{12}^2 R_{23}^2 R_{34}^2R_{45}^2 R_{14}^2 R_{25}^2 }{\mu_1^3\mu_2^3\mu_3^3\mu_4^3\mu_5^3 R_{11} R_{22} R_{33} R_{44} R_{55}}\,.
\end{align}
We recognise this solution as the accelerating double-Schwarzschild solution.

The 5-soliton solution can be naturally interpreted as describing an accelerating system consisting of two Kerr-NUT black holes in the presence of an acceleration horizon. It has a rather complicated form and will not be explicitly presented. In the following, we focus on the special case with three free solitons, so we study $[5-2]$-soliton solutions. There are two separate classes, namely, the $[5-2-0]$-soliton solution and the $[5-1-1]$-soliton solution. They are ALE with three turning points. The only known solutions with these properties are the 3-centred Gibbons--Hawking solution and the Pleba\'nski--Demia\'nski solution. We now show how they are exactly recovered from the two $[5-2]$-soliton solutions.

\subsubsection{The $[5-2-0]$-soliton solution}

We first show that the $[5-2-0]$-soliton solution is equivalent to the 3-centred Gibbons--Hawking solution. We choose to eliminate $\mu_2$ and $\mu_4$ by setting
\begin{align}
C_2=1\,,\qquad C_4=1\,.
\label{join_up_5m2m0}
\end{align}
Since there are three free solitons, we prefer to cast the solution in the coordinates $(u,v)$ (see Appendix~\ref{sec_C-metric-coordinates}). So we define
\begin{align}
z_1=z_{\text{A}}\,,\quad z_2=d_1\varkappa^2\,,\quad z_3=z_{\text{B}}\,,\quad z_4=d_2\varkappa^2\,,\quad z_5=z_{\text{C}}\,,
\end{align}
for some constants $c$, $\varkappa$ and $d_{1,2}$. This identifies the three free solitons $\mu_{1,3,5}$ as
\begin{align}
\mu_1=\mu_{\text{A}}\,,\quad \mu_3=\mu_{\text{B}}\,,\quad \mu_5=\mu_{\text{C}}\,,
\end{align}
and the two phantom solitons $\mu_{2,4}$ as
\begin{align} \mu_{2,4}=\frac{\varkappa^2Q_{1,2}}{u-v}-\left[\frac{\varkappa^2(1-uv)(2+cu+cv)}{(u-v)^2}-d_{1,2}\varkappa^2\right],
\end{align}
where $Q_i\equiv Q(d_i)$. The expression $Q$, as a function of $d$, is given by (\ref{Q}). The joining-up conditions (\ref{join_up_5m2m0}) ensure that both $Q_1$ and $Q_2$ cancel from all metric components. The $[5-2-0]$-soliton solution is then cast in the coordinates $(u,v)$ with purely algebraic metric components.

We perform linear coordinate transformations by doing the simultaneous substitutions (\ref{linear_transformation_1-soliton}) and choose $k_0=[16z_{24}z_{13}z_{15}z_{35}(1-C_1)(1-C_3)(1-C_5)]^{-2}$. The $[5-2-0]$-soliton solution is then brought to the 3-centred Gibbons--Hawking solution in the familiar form, namely, the metric (\ref{multi_GH}) with $s=3$, with the coordinates $(r,\theta)$ identified as (\ref{weyltortheta}) or equivalently
\begin{align}
\label{rthetatouv}
r\sin\theta=\frac{2\varkappa^2\sqrt{-G(u)G(v)}}{(u-v)^2}\,,\quad r\cos\theta=\frac{\varkappa^2(1-uv)(2+cu+cv)}{(u-v)^2}\,,
\end{align}
and the parameters $a_{1,2,3}$ and $n_{1,2,3}$ identified as $a_{1,2,3}=z_{\text{A,B,C}}$ and
\begin{align}
n_1=\frac{z_{12}z_{14}(1+C_1)}{4z_{13}z_{15}(1-C_1)}\,,\quad n_2=\frac{z_{23}z_{34}(1+C_3)}{4z_{13}z_{35}(1-C_3)}\,,\quad n_3=\frac{z_{25}z_{45}(1+C_5)}{4z_{15}z_{35}(1-C_5)}\,.
\end{align}

\subsubsection{The $[5-1-1]$-soliton solution}

\label{sec_5m1m1}

Now, we show that the $[5-1-1]$-soliton solution is equivalent to the Pleba\'nski--Demia\'nski solution. In this case, we eliminate $\mu_2$ and $\mu_4$ by setting
\begin{align}
C_2=1\,,\qquad C_4=-1\,.
\end{align}
The $[5-1-1]$-soliton solution is then cast in the coordinates $(u,v)$ as done in the previous subsection, and it is now parameterised by $c$, $\varkappa$, $d_{1,2}$ and $C_{1,3,5}$.

The $[5-1-1]$-soliton solution is equivalent to the 3-soliton solution. Recall that the latter solution was cast in the coordinates $(u,v)$ in Sec.~\ref{sec_3-soliton} and is parameterised by $c$, $\varkappa$, and $C_{1,2,3}$. Starting from this solution, if we redefine its parameters $C_{1,2,3}$ as follows:
\begin{align}
\label{parameters_5m1m1e3}
C_{1}&\rightarrow \frac{(d_1-d_2)+(d_1+d_2+ 2c)C_1}{(d_1-d_2)C_1+(d_1+d_2+ 2c)}\,,\quad 
C_{2}\rightarrow\frac{(d_1-d_2)+(d_1+d_2- 2c)C_3}{(d_1-d_2)C_3+(d_1+d_2- 2c)}\,,\nonumber\\ C_{3}&\rightarrow\frac{(d_1-d_2)+(d_1+d_2-2)C_5}{(d_1-d_2)C_5+(d_1+d_2-2)}\,,
\end{align}
and choose its $k_0$ as
\begin{align}
k_0\rightarrow& 4\varkappa^{12}k_0[(d_1-d_2)C_1+(d_1+d_2+ 2c)]^2[(d_1-d_2)C_3+(d_1+d_2- 2c)]^2\nonumber\\
&\times [(d_1-d_2)C_5+(d_1+d_2-2)]^2\,,
\end{align}
we then obtain the present form of the $[5-1-1]$-soliton solution. Note that the redundant parameter $C_3$ in the 3-soliton solution has been set to a particular value in (\ref{parameters_5m1m1e3}). So the $[5-1-1]$-soliton solution is equivalent to the 3-soliton solution and, thus, the Pleba\'nski--Demia\'nski solution.

For both of the $[5-2]$-soliton solutions, we see that the parameters $C_{1,3,5}$ are redundant and can be set to zero without loss of generality.

\subsection{The $6$-soliton solution}

The static limit of the 6-soliton solution can be obtained by setting $k_0=(C_2C_4C_6)^{-2}$ and taking
\begin{align}
C_{1,3,5}\rightarrow 0\,,\quad C_{2,4,6}\rightarrow \infty\,,
\end{align}
with a resulting metric given by
\begin{align}
g_{{\text{3-Sch}}}&=\left[ \begin {array}{cc} {\frac{\mu_1\mu_3\mu_5}{\mu_2\mu_4\mu_6} }&0\\0&{\frac{\rho^2\mu_2\mu_4\mu_6}{\mu_1\mu_3\mu_5}}\end {array} \right],\nonumber\\
f_{{\text{3-Sch}}}&=\frac{\mu_{13}^2\mu_{15}^2\mu_{35}^2\mu_{24}^2\mu_{26}^2\mu_{46}^2R_{12}^2R_{23}^2R_{34}^2R_{45}^2R_{56}^2 R_{14}^2  R_{25}^2R_{36}^2R_{16}^2  }{\mu_1^5\mu_2^3 \mu_3^5\mu_4^3\mu_5^5\mu_6^3 R_{11} R_{22} R_{33} R_{44} R_{55}R_{66}}\,.
\end{align}
We recognise this solution as the triple-Schwarzschild solution.

The 6-soliton solution can be naturally interpreted as describing a system consisting of three Kerr-NUT black holes. It has a rather complicated form and will not be explicitly presented. In the following, we focus on the special case with three free solitons, so we study $[6-3]$-soliton solutions. There are two separate classes, namely, the $[6-3-0]$-soliton solution and the $[6-2-1]$-soliton solution. They are ALF with three turning points. The only known solutions with these properties are the 3-centred Taub-NUT solution and the solution (\ref{metric_new_ricci}). We now show how they are exactly recovered from the two $[6-3]$-soliton solutions.

\subsubsection{The $[6-3-0]$-soliton solution}

We first show that the $[6-3-0]$-soliton solution is equivalent to the 3-centred Taub-NUT solution. We choose to eliminate $\mu_{2}$, $\mu_4$ and $\mu_6$ by setting
\begin{align}
C_2=1\,,\quad C_4=1\,,\quad C_6=1\,.
\label{join_up_6m3m0}
\end{align}
Since there are three free solitons, we prefer to cast the solution in the coordinates $(u,v)$ (see Appendix~\ref{sec_C-metric-coordinates}). So we define
\begin{align}
z_1=z_{\text{A}}\,,\quad z_2=d_1\varkappa^2\,,\quad z_3=z_{\text{B}}\,,\quad z_4=d_2\varkappa^2\,,\quad z_5=z_{\text{C}}\,,\quad z_6=d_3\varkappa^2\,,
\end{align}
for some constants $c$, $\varkappa$ and $d_{1,2,3}$. This identifies the three free solitons $\mu_{1,3,5}$ as
\begin{align}
\mu_1=\mu_{\text{A}}\,,\quad \mu_3=\mu_{\text{B}}\,,\quad \mu_5=\mu_{\text{C}}\,,
\end{align}
and the three phantom solitons $\mu_{2,4,6}$ as
\begin{align} \mu_{2,4,6}=\frac{\varkappa^2Q_{1,2,3}}{u-v}-\left[\frac{\varkappa^2(1-uv)(2+cu+cv)}{(u-v)^2}-d_{1,2,3}\varkappa^2\right].
\end{align}
The joining-up conditions (\ref{join_up_6m3m0}) ensure that $Q_{1,2,3}$ cancel from all metric components. The $[6-3-0]$-soliton solution is then cast in the coordinates $(u,v)$ with purely algebraic metric components.

We perform linear coordinate transformations by doing the simultaneous substitutions as follows:
\begin{align}
\tau\rightarrow \tau/\sqrt{\epsilon}+\sqrt{\epsilon}(z_{12}+z_{34}+z_{56})\phi\,,\quad \phi\rightarrow -\sqrt{\epsilon}\phi\,,
\end{align}
and choose $k_0=\epsilon[64z_{13}z_{15}z_{35}z_{24}z_{26}z_{46}(1-C_1)(1-C_3)(1-C_5)]^{-2}$, where $\epsilon$ is an arbitrary constant. The $[6-3-0]$-soliton solution is then brought to the 3-centred Taub-NUT solution in the familiar form, namely, the metric (\ref{multi_TN}) with $s=3$, with the coordinates $(r,\theta)$ identified as (\ref{weyltortheta}) or equivalently (\ref{rthetatouv}) and the parameters $a_{1,2,3}$ and $n_{1,2,3}$ identified as $a_{1,2,3}=z_{\text{A,B,C}}$ and
\begin{align}
n_{1}=-\frac{\epsilon z_{12}z_{14}z_{16}(1+C_1)}{2z_{13}z_{15}(1-C_1)}\,,\quad n_{2}=-\frac{\epsilon z_{23}z_{34}z_{36}(1+C_3)}{2z_{13}z_{35}(1-C_3)}\,,\quad  n_3=-\frac{\epsilon z_{25}z_{45}z_{56}(1+C_5)}{2z_{15}z_{35}(1-C_5)}\,.
\end{align}

\subsubsection{The $[6-2-1]$-soliton solution}

\label{sec_6m2m1}

Now, we show that the $[6-2-1]$-soliton solution is equivalent to the solution (\ref{metric_new_ricci}). In this case, we eliminate $\mu_{2}$, $\mu_4$ and $\mu_6$ by setting
\begin{align}
C_2=1\,,\quad C_4=1\,,\quad C_6=-1\,.
\end{align}
The $[6-2-1]$-soliton solution is then cast in the coordinates $(u,v)$ as done in the previous subsection, and it is now parameterised by $c$, $\varkappa$ , $d_{1,2,3}$ and $C_{1,2,3}$.

The $[6-2-1]$-soliton solution is equivalent to the $[4-1]$-soliton solution. Recall that the latter solution was cast in the coordinates $(u,v)$ in Sec.~\ref{sec_4m1} and is parameterised by $c$, $d$, $\varkappa$, and $C_{1,2,3}$. Starting from this solution, if we redefine its parameters $d$ and $C_{1,2,3}$ as follows:
\begin{align}
\label{parameters_6m2m1e4m1}
C_{1}&\rightarrow \frac{(d_1-d_3)(d_2-d_3)+[d_1d_2-d_3^2+2c^2+(d_3+ 2c)(d_1+d_2)]C_1}{(d_1-d_3)(d_2-d_3)C_1+d_1d_2-d_3^2+2c^2+(d_3+ 2c)(d_1+d_2)}\,,\nonumber\\
C_{2}&\rightarrow\frac{(d_1-d_3)(d_2-d_3)+[d_1d_2-d_3^2+2c^2+(d_3- 2c)(d_1+d_2)]C_3}{(d_1-d_3)(d_2-d_3)C_3+d_1d_2-d_3^2+2c^2+(d_3- 2c)(d_1+d_2)}\,,\nonumber\\ C_3&\rightarrow\frac{(d_1-d_3)(d_2-d_3)+[d_1d_2-d_3^2+2+(d_3-2)(d_1+d_2)]C_5}{(d_1-d_3)(d_2-d_3)C_5+d_1d_2-d_3^2+2+(d_3-2)(d_1+d_2)}\,,\nonumber \\
d&\rightarrow d_1+d_2-d_3\,,
\end{align}
and choose its $k_0$ as
\begin{align}
k_0\rightarrow &16 \varkappa^{16}k_0(d_1+d_2-d_3-1)^{-2}(d_1+d_2-d_3+c)^{-2}(d_1+d_2-d_3-c)^{-2}(d_1-d_2)^2\nonumber\\
&
\times [(d_1-d_3)(d_2-d_3)C_1+d_1d_2-d_3^2+2c^2+(d_3+ 2c)(d_1+d_2)]^2\nonumber\\
&\times [(d_1-d_3)(d_2-d_3)C_3+d_1d_2-d_3^2+2c^2+(d_3- 2c)(d_1+d_2)]^2
\nonumber\\
&\times [(d_1-d_3)(d_2-d_3)C_5+d_1d_2-d_3^2+2+(d_3-2)(d_1+d_2)]^2
\,,
\end{align}
we then obtain the present form of the $[6-2-1]$-soliton solution. Note that the redundant parameter $C_3$ in the $[4-1]$-soliton solution has been set to a particular value in (\ref{parameters_6m2m1e4m1}). So the $[6-2-1]$-soliton solution is equivalent to the $[4-1]$-soliton solution and, thus, the solution (\ref{metric_new_ricci}).

For both of the $[6-3]$-soliton solutions, we see that the parameters $C_{1,3,5}$ are redundant and can be set to zero without loss of generality.

\section{Classification of multi-soliton solutions on flat space}

\label{sec_classification}

\subsection{Classification of multi-soliton solutions on flat space}

\begin{table}[!t]
	\begin{center}
		\begin{tabular}{|c|l|l|l|l|}
			\hline
			\multicolumn{2}{|l|}{The $n$-soliton solution}  & \multicolumn{3}{|l|}{$[n-m]$-soliton solutions} \\
			\hline
			$n$ & Interpretations & $[n-m]$&$[n-m_{1}-m_{2}]$ &Interpretations\\
			\hline
			1 & Accelerating nothing & \multicolumn{2}{|l|}{$[1-0]$} &1-centred Gibbons--Hawking \\
			\hline
			2 & Kerr-NUT & \multicolumn{2}{|l|}{$[2-1]$} &1-centred Taub-NUT\\
			\hline
			3 & Pleba\'nski--Demia\'nski &\multicolumn{2}{|l|}{$[3-1]$}&2-centred Gibbons--Hawking  \\
			\hline
			\multirow{3}{*}{4}  &\multirow{3}{*}{Double Kerr-NUT} &\multicolumn{2}{|l|}{$[4-1]$}&Solution (\ref{metric_new_ricci})\\
			\cline{3-5}
			&&\multirow{2}{*}{$[4-2]$}& {$[4-2-0]$}&2-centred Taub-NUT\\
			\cline{4-5}
			&&&$[4-1-1]$ &{Kerr-NUT}\\
			\hline
			\multirow{2}{*}{5}&\multirow{2}{*}{\parbox[t]{3.5cm}{Accelerating \\double Kerr-NUT}} &\multirow{2}{*}{$[5-2]$}& {$[5-2-0]$}&3-centred Gibbons--Hawking \\
			\cline{4-5}
			&&&$[5-1-1]$ &Pleba\'nski--Demia\'nski\\
			\hline
			\multirow{2}{*}{6} &\multirow{2}{*}{Triple Kerr-NUT} &\multirow{2}{*}{$[6-3]$}&{$[6-3-0]$}& 3-centred Taub-NUT\\
			\cline{4-5}
			&&&$[6-2-1]$ & Solution (\ref{metric_new_ricci})\\
			\hline
		\end{tabular}
		\caption{A summary of the $n$- and $[n-m]$-soliton solutions on flat space for $n\le 6$ and $n-m\leq 3$. The physical interpretations of the $n$-soliton solution are indicated in the second column. The last three columns contain $[n-m]$-soliton solutions and their interpretations. Details of these results can be found in Sec.~\ref{sec_examples}.}
		\label{summary}
		\medskip
		\medskip
		\begin{tabular}{|c|l|l|c|c|}
			\hline
			$s$ & Classes of solutions & Solitonic representations&ALE/ALF&$p$ \\
			\hline
			\multirow{2}{*}{1} & 1-centred Gibbons--Hawking & $[1-0]$ &ALE&0\\
			\cline{2-5}
			& 1-centred Taub-NUT & $[2-1]$ &ALF&1\\
			\hline
			\multirow{3}{*}{2} & 2-centred Gibbons--Hawking & $[3-1]$&ALE&2\\
			\cline{2-5}
			& Kerr-NUT & $[2-0-0]\sim[4-1-1]$&\multirow{2}{*}{ALF}&	\multirow{2}{*}{3}\\
			\cline{2-3}
			& 2-centred Taub-NUT & $[4-2-0]$& &\\
			\hline
			\multirow{4}{*}{3}&Pleba\'nski--Demia\'nski&$[3-0-0]\sim [5-1-1]$&\multirow{2}{*}{ALE}&	\multirow{2}{*}{4}\\
			\cline{2-3}
			&3-centred Gibbons--Hawking&$[5-2-0]$&&\\
			\cline{2-5}
			&Solution (\ref{metric_new_ricci})&$[4-1-0]\sim [6-2-1]$&\multirow{2}{*}{ALF}&	\multirow{2}{*}{5}\\
			\cline{2-3}
			&3-centred Taub-NUT&$[6-3-0]$&&\\
			\hline
		\end{tabular}
		\caption{The results in Table~\ref{summary} are rearranged according to the free-soliton number $s=n-m$. The asymptotic structure (being either ALE or ALF) and the number of independent parameters $p$ of the solutions in this table are also indicated. Two solutions connected by the symbol $\sim$ are equivalent.}
		\label{s-soliton}
	\end{center}
\end{table}

In the previous section, we studied in detail the $n$- and $[n-m]$-soliton solutions on flat space with $n\leq 6$ and $n-m\leq 3$. Various known instanton solutions have been obtained. The results are summarised in Table~\ref{summary}. We notice that in this table, several solutions appear more than once. A concise rearrangement of these multi-soliton solutions can be based on the free-soliton number $s$---the number of free solitons $s=n-m$. The rearranged results are shown in Table~\ref{s-soliton}.

We make a few remarks on the results of these tables. Firstly, we see from Table~\ref{s-soliton} that, for each fixed $s$, there are several separate classes of solutions, some of which are ALE and some of which are ALF. ALE and ALF solutions are obtainable from $[n-m]$-soliton solutions for odd $n$ and even $n$, respectively. So $[n-m]$-soliton solutions have the same asymptotic structure as the corresponding $n$-soliton solution. This means that, not surprisingly, eliminating a soliton of the $n$-soliton solution in the way we have described does not change the asymptotic structure.

Secondly, let us count the number of independent parameters of the $[n-m]$-soliton solutions in these two tables. As we mentioned, the $n$-soliton solution for even $n$ is a $(2n-1)$-parameter class of solutions, while for odd $n$, it is a $(2n-2)$-parameter class. When we eliminate $m$ solitons from the $n$-soliton solution, the corresponding $m$ BZ parameters are fixed according to (\ref{condition_joinup}). From our explicit examples in Sec.~\ref{sec_examples}, we see that in this process, additional $m$ parameters become redundant. We expect that this is a general feature of $[n-m]$-soliton solutions on flat space. Hence, the number of independent parameters of $[n-m]$-soliton solutions is the number of independent parameters of the $n$-soliton solution minus $2m$, namely, $2n-2m-1$ for even $n$ and $2n-2m-2$ for odd $n$. In terms of the free-soliton number $s$, this means that ALE solutions have $2s-2$ independent parameters, while ALF solutions have $2s-1$ independent parameters, consistent with the results in Table~\ref{s-soliton}.

One may naively think that we can add free parameters to a solution by adding pairs of solitons on it and then eliminating them. The above counting of parameters indicates that this is not the case, however. By adding pairs of solitons and then eliminating them, we may obtain new classes of solutions that are cousins of the original solutions; however, in this process, the number of free solitons and the number of independent parameters are not changed. As an example, we can add two solitons on the Pleba\'nski--Demia\'nski solution to obtain the $5$-soliton solution (solitonically, this is $5\sim 3+2$) and then eliminate them so that they become two positive phantom solitons. The resulting solution is the $[5-2-0]$-soliton solution, which is equivalent to the $3$-centred Gibbons--Hawking solution. Both the Pleba\'nski--Demia\'nski and the 3-centred Gibbons--Hawking solutions have three free solitons and four independent parameters, but they are certainly inequivalent.

Another remark is that the same solution can have multiple equivalent solitonic representations. In Table~\ref{s-soliton}, we see that the Kerr-NUT solution can be represented as $[2-0-0]$ or equivalently as $[4-1-1]$. We will shortly prove that there is a general equivalence relation
\begin{align}
[(n+2)-(m_1+1)-(m_2+1)]\sim[n-m_{1}-m_{2}]\,.\label{annilation}
\end{align}
Other examples of this relation can be found in Table~\ref{s-soliton}: $[3-0-0]\sim[5-1-1]$ and $[4-1-0]\sim[6-2-1]$.

We have seen that various known solutions have been reproduced from multi-soliton solutions on flat space. Now, we show how all multi-soliton solutions on flat space can be classified. The first observation is that solutions with different free-soliton number $s$ are obviously different. We denote the set of multi-soliton solutions on flat space with free-soliton number $s$ as $\mathfrak{S}(s)\equiv\{[n-m]:n-m=s\}$, i.e.,
\begin{align}
\mathfrak{S}(s)\equiv\Big\{[n-m_{1}-m_{2}]:n-m_{1}-m_{2}=s\text{ and }\lfloor\frac{n}{2}\big\rfloor\geq m_{1}\geq m_{2}\Big\}\,,\label{Ss}
\end{align}
two solutions being equivalent if they are related by coordinate transformations. The set $\mathfrak{S}$ of multi-soliton solutions on flat space is then
\begin{align}
\mathfrak{S}\equiv\Big\{[n-m_1-m_2]\Big\}=\bigcup_{s=0}^{\infty} \mathfrak{S}(s)\,.
\label{S}
\end{align}
Since two solutions with different free-soliton numbers are inequivalent, we have the relation $\mathfrak{S}(s_1)\bigcap \mathfrak{S}(s_2)=\emptyset$ if $s_1\neq s_2$.\footnote{Note that the solution $(\ref{metric_seed_even_Euclidean})\sim [0-0-0]\in \mathfrak{S}(0)$ can be mapped to $(\ref{1-centred_GH})\sim [1-0-0]\in \mathfrak{S}(1)$ by coordinate transformations since both are flat-space metrics. This is, however, a coincidence and should be taken as an exception.}

At first sight, from (\ref{Ss}), $\mathfrak{S}(s)$ contains an infinite number of elements indexed by two integers $m_{1}$ and $m_{2}$. This is not true, since (\ref{annilation}) tells us that for the solutions in $\mathfrak{S}$, one positive phantom soliton and one negative phantom soliton annihilate each other completely. For a given free-soliton number $s$, multi-soliton solutions on flat space only depend on the number $s_z= m_{1}-m_{2}$. Here, $s_z$ will be called the phantom-soliton number of the solution: it is the number of positive phantom solitons minus the number of negative phantom solitons. The reader is reminded that $s_z$ is by definition different from the total number of phantom solitons $m=m_1+m_2$. We thus use the pair of integers $(s,s_z)$ consisting of the free-soliton number $s=n-m$ and the phantom-soliton number $s_z= m_{1}-m_{2}$ to denote an element in $\mathfrak{S}(s)$:
\begin{align}
(s,s_z)\equiv\Big\{[n-m_1-m_2]:n-m_1-m_2=s,m_1-m_2=s_z\Big\}/\sim\,,
\end{align}
where the equivalence relation $\sim$ is given by (\ref{annilation}). The condition (\ref{nmmplus}) reads $s_z\geq 0$ and $\lfloor\frac{n}{2}\rfloor=\lfloor\frac{s+s_z+2m_2}{2}\rfloor\geq m_1=s_z+m_2$, so we have
\begin{align}
0\leq s_z\leq s\,.
\end{align}
This reduces the number of elements of $\mathfrak{S}(s)$ to $s+1$.

It is instructive to give a concrete representative to the equivalence class of solutions $(s,s_z)$. An obvious choice is $[(s+s_z)-s_z-0]$, and we simply write $(s,s_z)=[(s+s_z)-s_z-0]$. This representative has $m_2=0$ and thus the least possible value of $n=s+s_z$. Another possible choice is given as follows:
 \begin{center}
 	\begin{tabular}{|c|c|c|c|c|c|}
 		\hline
 		$(s,s)$&$(s,s-1)$&$(s,s-2)$&$(s,s-3)$&...&$(s,0)$     	 \\
 		\hline
 		\scriptsize$[2s-s-0]$&\scriptsize$[(2s-1)-(s-1)-0]$&\scriptsize$[2s-(s-1)-1]$&\scriptsize  $[(2s-1)-(s-2)-1]$&... &\scriptsize  $[\left(s+2\lfloor\frac{s}{2}\rfloor\right)-\lfloor\frac{s}{2}\rfloor-\lfloor\frac{s}{2}\rfloor]$\\
 		\hline
 	\end{tabular}
 \end{center}
It is not difficult to see that this representation establishes a one-to-one correspondence between elements of $\mathfrak{S}(s)$ and $[2s-s]$- and $[(2s-1)-(s-1)]$-soliton solutions. So in order to find all multi-soliton solutions with, say, three free solitons, namely, $\mathfrak{S}(3)$, we only need to consider all $[5-2]$- and $[6-3]$-soliton solutions.

To summarise, the set of multi-soliton solutions on flat space has a partition as (\ref{S}). Each cell $\mathfrak{S}(s)$ of the partition contains $s+1$ elements and can be written as
\begin{align}
\label{spin_labeling}
\mathfrak{S}(s)=\Big\{(s,s_z): \quad 0\leq s_z\leq s\Big\}\,.
\end{align}
Representatives of these elements are given in the table of the previous paragraph. Here, $s$ and $s_z$ are the free-soliton number and the phantom-soliton number of the solution $(s,s_z)$, respectively. Their symbols are chosen by the analogy of this classification scheme to the quantum states of a particle with spin $s$. The only differences are that here $s$ cannot take half-integer values and that $s_z$ takes only non-negative values. This classification of multi-soliton solutions on flat space in terms of the free-soliton number and the phantom-soliton number is shown in Table~\ref{spin-classification}, where only solutions for the first few $s$ are presented. 

\begin{table}[!t]
\begin{center}
\begin{tabular}{|c|c|c|c|c|c|c|c|}
\cline{1-1}
$s$&\multicolumn{1}{r}{}&\multicolumn{1}{r}{}&\multicolumn{1}{r}{} &\multicolumn{1}{r}{}&  \multicolumn{1}{r}{}&  \multicolumn{1}{r}{} &  \multicolumn{1}{r}{} \\
\cline{1-2}
\multirow{2}{*}{0}& \scriptsize $[0-0-0]$&\multicolumn{1}{r}{} &\multicolumn{1}{r}{}&\multicolumn{1}{r}{}&\multicolumn{1}{r}{}  &  \multicolumn{1}{r}{} &  \multicolumn{1}{r}{}\\
& \large$\text{\bf FS(0TN)}\atop \uparrow$&\multicolumn{1}{r}{} &\multicolumn{1}{r}{}&\multicolumn{1}{r}{}&\multicolumn{1}{r}{}  &  \multicolumn{1}{r}{} &  \multicolumn{1}{r}{}\\		
\cline{1-3}
\multirow{2}{*}{1}&\scriptsize $[1-0-0]$&\scriptsize$[2-1-0]$&\multicolumn{1}{r}{} &\multicolumn{1}{r}{}&\multicolumn{1}{r}{}   &  \multicolumn{1}{r}{} &  \multicolumn{1}{r}{}\\	
& \large $\hphantom{\rightharpoondown }\text{1GH}\leftharpoonup  \atop \uparrow$ &\large$\text{\bf 1TN}\atop \uparrow$ &\multicolumn{1}{r}{}&\multicolumn{1}{r}{}&\multicolumn{1}{r}{}  &  \multicolumn{1}{r}{} &  \multicolumn{1}{r}{} \\
\cline{1-4}
\multirow{2}{*}{2}	& \scriptsize$[4-1-1]$& \scriptsize$[3-1-0]$&\scriptsize$[4-2-0]$ &\multicolumn{1}{r}{}&\multicolumn{1}{r}{}   &  \multicolumn{1}{r}{} &  \multicolumn{1}{r}{}\\
&\large $\text{\bf KN}\atop \uparrow$&\large $\rightharpoondown  \text{2GH} \leftharpoonup \atop \uparrow$&\large$\text{\bf 2TN}\atop \uparrow$ & \multicolumn{1}{r}{}&\multicolumn{1}{r}{}  &  \multicolumn{1}{r}{} &  \multicolumn{1}{r}{}\\
\cline{1-5}
\multirow{2}{*}{3}& \scriptsize$[5-1-1]$& \scriptsize$[6-2-1]$&\scriptsize$[5-2-0]$&\scriptsize$[6-3-0]$&\multicolumn{1}{r}{}  &  \multicolumn{1}{r}{}  &  \multicolumn{1}{r}{}\\
&\large $\hphantom{\rightharpoondown }\text{PD}\leftharpoonup \atop \uparrow$&\large $\text{\bf (\ref{metric_new_ricci})}\atop \uparrow$&\large$\rightharpoondown \text{3GH}\leftharpoonup \atop \uparrow$&\large$\text{\bf 3TN}\atop \uparrow$  &\multicolumn{1}{}{}&  \multicolumn{1}{r}{} &  \multicolumn{1}{r}{}\\
\cline{1-6}
\multirow{2}{*}{4}& \scriptsize$[8-2-2]$& \scriptsize$[7-2-1]$&\scriptsize$[8-3-1]$&\scriptsize$[7-3-0]$& \scriptsize$[8-4-0]$ &  \multicolumn{1}{}{}  &  \multicolumn{1}{r}{}\\
&\large $\text{\bf dKN}\atop \uparrow$ &\large$\rightharpoondown \diamondsuit\leftharpoonup \atop \uparrow$ &\large$\clubsuit\atop \uparrow$&\large $\rightharpoondown \text{4GH}\leftharpoonup \atop \uparrow$ &\large$\text{\bf 4TN}\atop \uparrow$&  \multicolumn{1}{r}{} &  \multicolumn{1}{r}{}\\
\cline{1-7}
\multicolumn{1}{r}{}&\multicolumn{6}{|c|}{$\vdots$}     	 \\
\cline{2-8}
\multicolumn{1}{r|}{}&0&1&2&  3&4 &  \multicolumn{1}{r|}{\hphantom{$\twoheadrightarrow$}$...$\hphantom{$\twoheadrightarrow$}}& {\hphantom{$\twoheadrightarrow$}}$s_z$\hphantom{$\twoheadrightarrow$} \\
\cline{2-8}			
\end{tabular}
\caption{Classification of multi-soliton solutions on flat space in terms of the free-soliton number $s$ and the phantom-soliton number $s_z$. Each solution is thus determined by the pair $(s,s_z)$, and a solitonic representation of it is given. GH stands for Gibbons--Hawking and TN stands for Taub-NUT; their number of centres is placed in front. So $2$TN stands for the 2-centred Taub-NUT solution. FS stands for flat space (\ref{metric_seed_even_Euclidean}), KN stands for Kerr-NUT, PD stands for Pleba\'nski--Demia\'nski, and dKN stands for double Kerr-NUT. The symbols $\diamondsuit$ and $\clubsuit$ denote solutions whose solitonic representations are as indicated. Solutions in boldface, including \text{\bf (\ref{metric_new_ricci})} and $\clubsuit$, are ALF; the rest are ALE. An arrow represents a limit in which one solution is reduced to the other following the direction of the arrow. Different types of arrows represent different limits, whose details are described in the main text.}
\label{spin-classification}
\end{center}
\end{table}

Since $n=s+s_z+2m_2= s+s_z\text{ mod }2$, we see that the solution $(s,s_z)$ is ALE if $s+s_z$ is odd and ALF if $s+s_z$ is even. Hence, in Table~\ref{spin-classification}, ALE and ALF solutions appear alternatively in both the horizontal and the vertical directions. In this table, each arrow represents a limit in which one solution reduces to the other following the direction of the arrow. The three types of arrows represent three types of limits. One can freely follow arrows of this table to go from one solution to another by taking the corresponding limits. For example, we can obtain flat space (\ref{metric_seed_even_Euclidean}) from (\ref{metric_new_ricci}) by following the arrows in the sequence $\leftharpoonup$, $\uparrow$, $\rightharpoondown$, $\uparrow$, $\leftharpoonup$ and $\uparrow$. The corresponding limits represented by the three types of arrows will be explained below.

\subsection{Solitonic calculus on flat space}

\label{sec_soliton_calculus}

In this subsection, we develop a calculation formalism to manipulate multi-soliton solutions on flat space without referring to their explicit forms. We call this formalism solitonic calculus on flat space and use it to prove (\ref{annilation}) and to understand the limits in Table~\ref{spin-classification}.

Our major tool is the decomposition formula (\ref{decomposition_solitons}) of the ISM, which holds when the same seed is used on both its left- and right-hand sides. Since in this paper even-soliton and odd-soliton solutions on flat space are constructed using different seeds, we need to be very careful to apply the formula (\ref{decomposition_solitons}). From this point onwards, we define the solitonic summation $n_1+n_2+...+n_p$ for $p\ge 1$, as the solution obtained from the $n_1$-soliton solution on flat space by applying an $n_2$-soliton transformation, followed by an $n_3$-soliton transformation, and so on. With this definition, Eq.~(\ref{decomposition_solitons}) holds only when $n$ and $n_1$ are both even or both odd, so the decomposition formula now becomes
\begin{align}
n\sim n_1+n_2+...+n_p, \text{  if   }\sum_{i=1}^p n_i=n \text{ and }n= n_1\text{ mod }{2}.
\label{decomposition_solitons_flat}
\end{align}
As an example of this formula, we have
\begin{align}
3\sim 1+2\nsim 2+1.
\end{align}
The following solitonic equivalence will also be used
\begin{align}
1+2k\sim\text{1GH}+2k\,,
\label{decomposition_odd}
\end{align}
which follows from the result that the 1-soliton solution is equivalent to the 1-centred Gibbons--Hawking solution. As discussed in Sec.~\ref{sec_transition_limit}, even-soliton and odd-soliton solutions on flat space are related by the transition limit, which we denote by the arrow $\rightarrow$, namely,
\begin{align}
n\rightarrow n-1\,.
\label{transition_limit}
\end{align}
A number of solitons can be eliminated at one time, or one by one, and in the latter case the order does not matter. So if we want to eliminate $m$ solitons, we can first eliminate $k$ solitons and then eliminate $m-k$ solitons. Symbolically, we have
\begin{align}
n\ominus m\sim n\ominus k\ominus (m-k)\,.
\end{align}
This is true regardless of whether the phantom solitons are positive or negative, as long as the total number of phantom solitons of each type is fixed. 

With the above basic solitonic formulas, we are now ready to prove (\ref{annilation}). The proof proceeds as follows: 
\begin{align}
[(n+2)-(m_{1}+1)-(m_{2}+1)]&\sim(n+2)\ominus (m_{1}+1)^+\ominus (m_{2}+1)^-\nonumber\\
&\sim \Big\{\begin{array}{ll}
2\ominus 1^{+}\ominus 1^{-}+n\ominus m_{1}^+\ominus m_{2}^-&\text{even $n$}\\\nonumber3\ominus1^+\ominus1^-+(n-1)\ominus m_{1}^+\ominus m_{2}^-&\text{odd $n$}
\end{array}\\\nonumber
&\sim \Big\{\begin{array}{ll}
[2-1-1]+n\ominus m_{1}^+\ominus m_{2}^-&\text{even $n$}\\\nonumber[3-1-1]+(n-1)\ominus m_{1}^+\ominus m_{2}^-&\text{odd $n$}
\end{array}\\\nonumber
&\sim \Big\{\begin{array}{ll}
n\ominus m_{1}^+\ominus m_{2}^-&\text{even $n$}\\\nonumber
\text{1GH}+(n-1)\ominus m_{1}^+\ominus m_{2}^-&\text{odd $n$}
\end{array}\\
&\sim n\ominus m_{1}^+\ominus m_{2}^-\sim [n-m_{1}-m_{2}]\,.
\end{align}
The decomposition formula (\ref{decomposition_solitons_flat}) has been used in various steps. For the fourth equivalence relation, we have used the results $[2-1-1]\sim (\ref{metric_seed_even_Euclidean})$ and $[3-1-1]\sim \text{1GH}$, proved in Secs.~\ref{sec_2m1} and \ref{sec_3m1m1}, respectively. For the fifth equivalence, we have used (\ref{decomposition_odd}).

Now, we turn our attention to understanding the limits in Table~\ref{spin-classification}. Solutions in each column in this table have the same phantom-soliton number $s_z$ and increasing free-soliton number $s$ downwards. Each solution $(s,s_z)$ reduces to its upper floor $(s-1,s_z)$ in the (vertical) limit denoted as $\uparrow$. These limits are just the transition limit (\ref{transition_limit}):
\begin{align}
(s,s_z)\sim& [n-m_1-m_2]\sim n\ominus m_1^+\ominus m_2^-\rightarrow\nonumber\\
& (n-1)\ominus m_1^+\ominus m_2^-\sim [(n-1)-m_1-m_2]\sim (s-1,s_z)\,.
\end{align}
The first column in Table~\ref{spin-classification} has zero phantom-soliton number, and its entries are nothing but the $s$-soliton solution. The key feature of the transition limit is that the location of one free soliton is taken to (plus or minus) infinity, and the soliton thus disappears completely.  

We mentioned that the ALE limit $\epsilon\rightarrow 0$ of the Taub-NUT solution (\ref{self-dual Taub-NUT}) is the Gibbons--Hawking solution (\ref{1-centred_GH}). How can this be understood from a solitonic point of view? Recall that in Sec.~\ref{sec_2m1} it is shown that the 1-centred Taub-NUT solution can be solitonically represented as $2\ominus 1^+$, with the NUT charge $n$ identified as (\ref{nut_charge_1TN}). To take its ALE limit $\epsilon\rightarrow 0$ with fixed $n$, one must send $z_2\rightarrow \pm \infty$ at the same time. This means that in this limit, the location of the positive phantom soliton ($1^+$) in the 1-centred Taub-NUT solution $2\ominus 1^+$ should be sent to (plus or minus) infinity. We denote this limit as
\begin{align}
2\ominus 1^+\sim \text{1TN}  {\rightharpoonup} \text{1GH}\sim 1\,.
\label{decom_limit1}
\end{align}
Similarly, one can take the ALE limit of the 1-centred Taub-NUT solution represented by $2\ominus 1^-$ by sending the location of the negative phantom soliton to (plus or minus) infinity; we denote this limit as
\begin{align}
2\ominus 1^-\sim \text{1TN}{\rightharpoondown} \text{1GH}\sim 1\,.
\label{decom_limit2}
\end{align}
In the above two limits $\rightharpoonup$ and $\rightharpoondown$ of the 1-centred Taub-NUT solution, the compact direction $\partial_\tau$ blows up. So we call these two limits collectively as the decompactification limit. Though similar to the transition limit, it is fundamentally different from the latter in one aspect. In the transition limit, the location of a free soliton is sent to infinity while its BZ parameter is either fixed as $0$ or sent to infinity. In the decompactification limit, however, the location of a phantom soliton is sent to infinity; the BZ parameter of the phantom soliton is of course fixed as either $1$ or $-1$.

It is clear that the transition limit reduces $s$ by 1 and keeps $s_z$ fixed; on the other hand, the decompactification limit reduces or increases $s_z$ by 1 and keeps $s$ fixed. So the transition limit works in the vertical direction in Table~\ref{spin-classification}, while the decompactification limit works in the horizontal direction. We now demonstrate that the limits $\leftharpoonup$ and $\rightharpoondown$ in Table~\ref{spin-classification} (note that we do not distinguish $\leftharpoonup$ and $\rightharpoonup$, as well as $\leftharpoondown$ and $\rightharpoondown$) are indeed the decompactification limit; they are essentially the same as (\ref{decom_limit1}) and (\ref{decom_limit2}). We take the limits $\text{PD}\leftharpoonup(\ref{metric_new_ricci})\rightharpoondown\text{3GH}$ as an example. The solution (\ref{metric_new_ricci}) has a solitonic representation $[6-2-1]$. Its decompactification limit to the Pleba\'nski--Demia\'nski solution goes as follows:
\begin{align}
(\ref{metric_new_ricci})&\sim [6-2-1]\sim (2+4)\ominus2^+\ominus1^-\sim (2\ominus1^+)+(4\ominus1^+\ominus1^-)\nonumber\\
&\sim\text{1TN}+(4\ominus1^+\ominus1^-)\rightharpoonup\text{1GH}+(4\ominus1^+\ominus1^-)\sim[5-1-1]\sim\text{PD}\,,
\label{ALE_limit1}
\end{align}
which is essentially the same as (\ref{decom_limit1}). Its decompactification limit to the 3-centred Gibbons--Hawking solution goes as
\begin{align}
(\ref{metric_new_ricci})&\sim[6-2-1]\sim(2+4)\ominus2^+\ominus1^-\sim(2\ominus1^-)+(4\ominus2^+)\nonumber\\
&\sim\text{1TN}+(4\ominus2^+)\rightharpoondown \text{1GH}+(4\ominus2^+)\sim [5-2-0]\sim \text{3GH}\,,
\label{ALE_limit2}
\end{align}
which is essentially the same as (\ref{decom_limit2}). In these two limits, a positive or negative phantom soliton is eliminated, resulting in a decrease or increase of $s_z$ by one unit.

Along similar lines, it is not difficult to understand all the horizontal arrows in Table~\ref{spin-classification}. So for any given free-soliton number, an ALF solution reduces to its two neighbouring ALE solutions in two different decompactification limits represented by the arrows $\leftharpoonup$ and $\rightharpoondown$, and an ALE solution has two neighbouring ALF generalisations. This of course excludes the first and last solutions in each row of Table~\ref{spin-classification}, which have only one ALE neighbour in the decompactification limit, or one ALF generalisation. All these relations have simple interpretations in terms of the solitonic calculus.

\subsection{Even-soliton solutions on Gibbons--Hawking and Taub-NUT}

\label{sec_even_soliton_TN}

In Table~\ref{s-soliton}, we see that the $s$-centred Gibbons--Hawking solution is the $[(2s-1)-(s-1)-0]$-soliton solution or the solution $(s,s-1)$. It is ALE and has $2s-2$ independent parameters. The $s$-centred Taub-NUT solution is the $[2s-s-0]$-soliton solution or the solution $(s,s)$. It is ALF and contains $2s-1$ independent parameters. For $s\leq 3$, these results have been explicitly proven in Sec.~\ref{sec_examples}. We believe that the above solitonic representations of the multi-centred Gibbons--Hawking and Taub-NUT solutions are valid for arbitrary $s$:
\begin{align}
\label{TN_soliton_representations}
(s,s)&\sim \text{$s$-centred Taub-NUT};\nonumber\\ (s,s-1)&\sim \text{$s$-centred Gibbons--Hawking}.
\end{align}
The coordinate transformations and parameter identifications to establish the above equivalences can be obtained by identifying the rod structures of both sides of (\ref{TN_soliton_representations}); the results can be explicitly written down and are very similar to those in the simple cases $s=1,2,3$ studied in Sec.~\ref{sec_examples}. We have not yet achieved a complete proof of (\ref{TN_soliton_representations}) for general $s$, but we believe that what remains is of a technical nature. According to (\ref{TN_soliton_representations}), the first two elements in each column of our classification in Table~\ref{spin-classification} are the $s_z$-centred Taub-NUT and $(s_z+1)$-centred Gibbons--Hawking solutions.

Using the solitonic calculus developed in the previous subsection, it is not difficult to prove the solitonic formula
\begin{align}
(s+2k,s_z)\sim (s,s_z)+2k\,,
\label{column_relation}
\end{align}
for any positive integer $k$. This means that the solution $(s+2k,s_z)$ is the $2k$-soliton solution on the seed $(s,s_z)$. Together with (\ref{TN_soliton_representations}), this allows us to interpret solutions in each column in Table~\ref{spin-classification} as
\begin{align}
(s_z+2k,s_z)&\sim \text{$2k$-soliton solution on $s_z$-centred TN};\nonumber\\
(s_z+2k+1,s_z)&\sim \text{$2k$-soliton solution on $(s_z+1)$-centred GH}.
\label{spin_column}
\end{align}
For example, the solution (\ref{metric_new_ricci}) is the 2-soliton solution on the 1-centred Taub-NUT, while the Pleba\'nski--Demia\'nski solution is the 2-soliton solution on the 1-centred Gibbons--Hawking. Among the multi-centred Taub-NUT and Gibbons--Hawking solutions, only the 0-centred Taub-NUT and the 1-centred Gibbons--Hawking are known to admit Lorentzian sections and they are in the first column. So only solutions in this column admit Lorentzian sections.

In Table~\ref{spin-classification}, rows are indexed by the free-soliton number $s$. Consider the solution $(s,s-2k)$ in the $s$ row. It is the $2k$-soliton solution on the seed $(s-2k,s-2k)\sim$ the $(s-2k)$-centred Taub-NUT. The latter (seed) solution can be obtained from the $s$-centred Taub-NUT solution by removing its $2k$ centres by setting the corresponding NUT charges to zero. This allows us to interpret the solution $(s,s-2k)$ as replacing $k$ pairs of centres of the $s$-centred Taub-NUT by $k$ pairs of BZ solitons. Similarly, we can interpret the solution in the same row $(s,s-1-2k)$ as replacing $k$ pairs of centres of the $s$-centred Gibbons--Hawking by $k$ pairs of BZ solitons. Symbolically, we have
\begin{align}
(s,s-2k)&\sim \text{$s$-centred TN with $2k$ centres replaced by $2k$ BZ solitons;}\nonumber\\ 
(s,s-1-2k)&\sim \text{$s$-centred GH with $2k$ centres replaced by $2k$ BZ solitons.}
\label{TN_GH_cousins}
\end{align}
In this sense, solutions in the $s$ row are either ALF cousins of the $s$-centred Taub-NUT solution or ALE cousins of the $s$-centred Gibbons--Hawking solution. So the $3$-centred Taub-NUT has an ALF cousin (\ref{metric_new_ricci}), while the $3$-centred Gibbons--Hawking has an ALE cousin which is the Pleba\'nski--Demia\'nski solution. For a given free-soliton number $s$, the number of all ALF cousins is $1+\lfloor\frac{s}{2}\rfloor$, and the number for all ALE cousins is $\lfloor\frac{1+s}{2}\rfloor$. The reader is reminded that these numbers are the numbers of separate classes of ALF and ALE solutions, respectively, with $s$ turning points.

In view of (\ref{TN_GH_cousins}), the phantom-soliton number $s_z$ could be roughly understood as a measure of self-duality of the solutions in each row of Table~\ref{spin-classification}: the solutions with $s_z=s$ or $s-1$ have the maximal self-duality, while the solution with $s_z=0$ has the minimal self-duality. The larger the phantom-soliton number a solution has, the closer it is to its self-dual cousin.

To end this section, we remark that the set of multi-soliton solutions on flat space is equal to the set of even-soliton solutions on the seeds of the multi-centred Taub-NUT and Gibbons--Hawking solutions. Here, we need to define the seeds to include the $0$- and $1$-centred Taub-NUT and the $1$-centred Gibbons--Hawking solutions, in addition to the more-centred ones. Symbolically, we can write
\begin{align}
\mathfrak{S}=\Big\{\text{Even-soliton solutions on Gibbons--Hawking and Taub-NUT}\Big\}\,.
\end{align}
 This can be seen from the interpretation (\ref{spin_column}).

\section{Discussion}

\label{sec_discussion}

We have defined and systematically studied gravitational multi-soliton solutions on flat space, namely, the $n$- and $[n-m]$-soliton solutions. These solutions are locally regular as defined in this paper. In the Euclidean regime, they are gravitational-instanton solutions with two axial symmetries; in the Lorentzian regime (if it exists at all), they are stationary and axisymmetric solutions. We have shown that multi-soliton solutions on flat space are classified as in Table~\ref{spin-classification} in terms of the free-soliton number $s$ and the phantom-soliton number $s_z$. It is also shown that they give rise to even-soliton solutions on the seeds of the multi-centred Gibbons--Hawking and Taub-NUT solutions.

There are a few potential applications and possible extensions of the results in this paper. The rod structure of the Lorentzian $n$-soliton solution obtained in Sec.~\ref{subsec_rod_structure_Lorentzian} can be used to identify the subclass of solutions that describes interacting multiple Kerr solutions, with or without acceleration. In the Lorentzian section, if we adopt the choice of a positive $k_0$ for $n=0,1\,\, {\text{mod}}\,\, 4$ and a negative $k_0$ otherwise,\footnote{The static limit of the Lorentzian $n$-soliton solution on flat space can be obtained by setting $k_0=i^{-2\lfloor\frac{n}{2}\rfloor}(\mathcal{C}_2\mathcal{C}_4...\mathcal{C}_{2\lfloor\frac{n}{2}\rfloor})^{-2}$ and then taking $\mathcal{C}_{1,3,...,2\lfloor\frac{n+1}{2}\rfloor-1}\rightarrow 0$ followed by $\mathcal{C}_{2,4,...,2\lfloor\frac{n}{2}\rfloor}\rightarrow \infty$. For this limit to have the Lorentzian signature $(-,+,+,+)$, it is clear that $k_0$ is positive for $n=0,1\,\, {\text{mod}}\,\, 4$ and negative otherwise. This is consistent with the sign choice of $k_0$ for the general Lorentzian $n$-soliton solution in the main text.} it is not difficult to see that rods with odd labels (i.e., Rods 1, 3, ...) are space-like, and those with even labels (i.e., Rods 2, 4, ...) are time-like. An axis can be consistently defined in the space-time described by the $n$-soliton solution if the NUT charge carried by each black hole is zero. This requires that, in terms of the rod structure, all space-like rods have parallel directions:
\begin{align} \frac{a_1}{b_1}=\frac{a_3}{b_3}=...=\frac{a_{2\lfloor \frac{n}{2}\rfloor+1}}{b_{2\lfloor \frac{n}{2}\rfloor+1}}\,.
\label{absence_NUT_charge}
\end{align}
If, further, the following conditions were satisfied:
\begin{align}
|b_1|=|b_3|=...=|b_{2\lfloor \frac{n}{2}\rfloor+1}|\,,
\label{absence_conical_singularity}
\end{align}
the multiple Kerr black holes would be in equilibrium (notice that, for even $n$, $b_1=b_{n+1}$ is automatically satisfied as proven in (\ref{even_equalb})). We should, however, mention an additional and practically more severe constraint that does not manifest itself in the rod structure, namely, that $f$ should not change sign and is positive for all $\rho>0$. Efforts may be made to see if (\ref{absence_NUT_charge}), (\ref{absence_conical_singularity}) and the additional constraint can be satisfied simultaneously for general $n\ge 4$.

The well-known Tomimatsu--Sato solution \cite{Tomimatsu:1972zz} with a distortion parameter $\delta$ can also be obtained from multi-soliton solutions on flat space. More specifically, it can be obtained as a special limit from the $2\delta$-soliton solution by dividing the solitons into two groups with equal elements and fusing the solitons in each group to a single one \cite{Tomimatsu:1981td}. Now, the accelerating generalisation of the $2\delta$-soliton solution is the $(2\delta+1)$-soliton solution. It should then be possible to take a similar limit of the $(2\delta+1)$-soliton solution to obtain the accelerating Tomimatsu--Sato solution. This solution should essentially involve only three solitons and can be cast in the C-metric-like coordinates $(u,v)$ (see Appendix~\ref{sec_C-metric-coordinates}).

In Table~\ref{spin-classification}, we have classified multi-soliton solutions on flat space in terms of the pair $(s,s_z)$. The free-soliton number $s$ of a solution equals the number of turning points in its rod structure; it also has the interpretation as the Euler characteristic if the solution can be made completely regular \cite{Chen:2010zu}. The phantom-soliton number $s_z$ is, however, not captured by the rod structure of the solution, and it would be interesting to find a geometrical or topological or even algebraic interpretation of it. One may further ask whether all locally regular stationary and axisymmetric solutions, and all locally regular instanton solutions with two axial symmetries, have solitonic representations on flat space so that they are all classified according to Table~\ref{spin-classification}. (Non-)uniqueness properties of these solutions may be proved to help resolve related issues.

An interesting application of multi-soliton solutions on flat space is to extract from them completely regular gravitational instantons with two axial symmetries. In fact, the new completely regular AF instanton discovered in \cite{Chen:2011tc} can be extracted from the solution (\ref{metric_new_ricci}) \cite{Chen:2015vva}, and we have shown here that the latter solution is equivalent to the $[4-1]$-soliton solution. The rod structure studied in this paper will be particularly helpful in this direction, since for the two axial symmetries to be globally well defined for a solution, it is required that all pairs of adjacent rod directions are related by GL$(2,\mathbb{Z})$ transformations \cite{Chen:2010zu}. This condition ensures that the coordinate identifications (\ref{rod_identifications}) and (\ref{turning_point_identifications}) needed to make all rods and their turning points regular are consistent with each other. Among the most promising new completely regular instantons are multi-Taub-bolt instantons, and AF instantons with four or more turning points.

We showed in Sec.~\ref{sec_even_soliton_TN} that multi-soliton solutions on flat space contain solutions that are non-self-dual ALF cousins of the multi-centred Taub-NUT solution, obtainable from the latter by replacing pairs of its centres by pairs of BZ solitons. One can study these solutions, when embedded in type IIA string theory, as systems describing multiple D6-branes. The ALF cousin solutions can also be viewed as even-soliton perturbations of the multi-centred Taub-NUT solution. These perturbations are thus continuous (recall that an odd-soliton perturbation of a solution is finite, while an even-soliton perturbation is continuous). They may have interesting interpretations in terms of D6-brane dynamics.

Assuming the presence of two commuting Killing vectors, the Einstein--Maxwell equations were shown to be integrable by Alekseev \cite{Alekseev:1980mba,Alekseev:1988mba}. Soliton solutions in supergravity theories can also be constructed \cite{Belinski:2015nha}. A generalisation of multi-soliton solutions on flat space considered in this paper to these two contexts will be worth pursuing.

Another direction of research is to study locally regular stationary and axisymmetric multi-soliton solutions in $D$ dimensions with $D-2$ linearly independent and mutually commuting Killing vectors. Various such classes of solutions have already been constructed in the literature using the ISM, on certain specially designed seeds derived using a technique due to Pomeransky \cite{Pomeransky:2005sj}. In view of the results of this paper, our question to ask is whether locally regular higher-dimensional stationary and axisymmetric solutions have solitonic representations on higher-dimensional flat space. We note that soliton solutions on the higher-dimensional Minkowski seed have been studied in detail in the literature (see, e.g., \cite{Belinsky:1980ae,Verdaguer:1993um}). Other forms of flat space in higher dimensions have been classified using the generalised Weyl formalism in \cite{Emparan:2001wk}, suitable to apply the ISM. A systematic study of soliton solutions on these forms of flat space may be worth pursuing. The generalisation of (\ref{general_g0}) to $D$ dimensions
\begin{align}
\label{generalised_LeviC}
g_0={\text{diag}\{\epsilon_1\rho^{s_1},\epsilon_2\rho^{s_2},,...,\epsilon_{D-2}\rho^{s_{D-2}}\}}\,,
\end{align}
is also interesting to consider, where $s_1=0$ can be set without loss of generality.

\section*{Acknowledgements}

The author would like to thank Kenneth Hong for discussions and Edward Teo for comments on the manuscript and for discussions. This work was supported by the Academic Research Fund (WBS No.: R-144-000-333-112) from the National University of Singapore.

\appendix

\section{Prolate spheroidal coordinates and C-metric-like coordinates}

\label{sec_coordinates}

We define two types of coordinates that are particularly useful to study solutions with two and three turning points or free solitons, respectively. In these coordinates, the free solitons present in the solutions can be expressed algebraically, with all square roots eliminated simultaneously. These coordinates are frequently used in Sec.~\ref{sec_examples}. Here, we follow Appendixes G and H of \cite{Harmark:2004rm}.

\subsection{Prolate spheroidal coordinates}

\label{sec_prolate_sph}

We define the prolate spheroidal coordinates $(p,q)$ by their relations to the Weyl--Papapetrou coordinates $(\rho,z)$ as
\begin{align} \rho=\iota\sqrt{(1-p^2)(q^2-1)}\,,\quad z=\iota pq\,,
\label{definition_pq}
\end{align}
where $\iota$ is a positive constant and
\begin{align}
\label{range_prolate}
-1\leq p\leq 1\,,\quad q\geq 1\,.
\end{align}
This range of the coordinates $(p,q)$ precisely covers the upper-half complex plane parameterised in the Weyl--Papapetrou coordinates as $(\rho\geq 0,-\infty<z<\infty)$. The flat metric on this upper-half plane becomes
\begin{align}
\dif \rho^2+\dif z^2=\iota^2(q^2-p^2)\left(\frac{\dif p^2}{1-p^2}-\frac{\dif q^2}{1-q^2}\right),
\label{flat_metric_pq}
\end{align}
in the coordinates $(p,q)$.

\begin{figure}[!t]
\begin{center}
\includegraphics[]{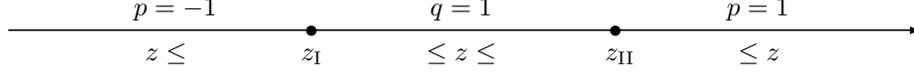}
\caption{The locations of the turning points and rods in the prolate spheroidal coordinates $(p,q)$ and the Weyl--Papapetrou coordinates $(\rho,z)$ for solutions with two free solitons $\mu_{\text{I,II}}$.}
\label{fig_ps}
\end{center}
\end{figure}

For a solution with two free solitons, we can appropriately shift the coordinate $z$ such that these solitons are located, respectively, at
\begin{align}
z_{\text{I}}=-\iota\,,\quad z_{\text{II}}=\iota\,,
\end{align}
for some value of $\iota$. With such definitions, the two free solitons can be expressed algebraically as
\begin{align}
\mu_{\text{I}}=\iota (1-p)(q-1)\,,\quad \mu_{\text{II}}=\iota(1-p)(q+1)\,.
\label{mu_III}
\end{align}
The equalities in (\ref{range_prolate}) correspond to the three rods of the solution in its rod structure. The locations of the turning points and rods of the solution in the coordinates $(p,q)$ and the Weyl--Papapetrou coordinates $(\rho,z)$ are shown in Fig.~\ref{fig_ps}. We remark that the signs on the right-hand sides of the equations in (\ref{mu_III}) are fixed by insisting on the range (\ref{range_prolate}).

The inverse relations to (\ref{definition_pq}) are
\begin{align}
\label{pq_inverse}
p=\frac{\rho^2-\mu_{\text{I}}\mu_{\text{II}}}{\rho^2+\mu_{\text{I}}\mu_{\text{II}}}\,,\quad q=\frac{\mu_{\text{II}}+\mu_{\text{I}}}{\mu_{\text{II}}-\mu_{\text{I}}}\,.
\end{align}
An alternative expression for $p$ is $p=1-\frac{\mu_{\text{II}}-\mu_{\text{I}}}{2\iota}$. The relations (\ref{pq_inverse}), together with (\ref{flat_metric_pq}), tell us that any solution written in the coordinates $(p,q)$ with algebraic metric components in the form
\begin{align}
\dif s^2=g_{ab}(p,q)\,\dif x^a\dif x^b+\xi(p,q)\left(\frac{\dif p^2}{1-p^2}-\frac{\dif q^2}{1-q^2}\right),
\label{pq_metrics}
\end{align}
satisfying the condition (\ref{rho}), i.e., $|\det g|=\iota^2{(1-p^2)(q^2-1)}$, can be cast in the Weyl--Papapetrou coordinates with two solitons $\mu_{\text{I}}$ and $\mu_{\text{II}}$.

\subsection{C-metric-like coordinates}

\label{sec_C-metric-coordinates}

We define the C-metric-like coordinates $(u,v)$ by their relations to the Weyl--Papapetrou coordinates $(\rho,z)$ as
\begin{align}
\label{definition_C} \rho=\frac{2\varkappa^2\sqrt{-G(u)G(v)}}{(u-v)^2}\,,\quad z=\frac{\varkappa^2(1-uv)(2+cu+cv)}{(u-v)^2}\,,
\end{align}
where $c$ and $\varkappa$ are two positive constants with $0< c< 1$, the function $G(u)= (1+cu)(1-u^2)$, and
\begin{align}
\label{range_C-metric}
-1\leq u\leq 1\,,\quad -1/c\leq v\leq -1\,.
\end{align}
This range of the coordinates $(u,v)$ precisely covers the upper-half complex plane parameterised in the Weyl--Papapetrou coordinates as $(\rho\geq 0,-\infty<z<\infty)$. The flat metric on this upper-half plane becomes
\begin{align}
\label{flat_metric_C}
\dif \rho^2+\dif z^2=\frac{\varkappa^4(c+u+v+cuv)[c^2(1-uv)^2-(2+cu+cv)^2]}{(u-v)^3}\left(\frac{\dif u^2}{G(u)}-\frac{\dif v^2}{G(v)}\right),
\end{align}
in the coordinates $(u,v)$.

\begin{figure}[!t]
\begin{center}
\includegraphics[]{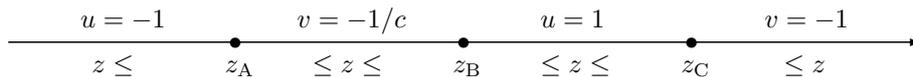}
\caption{The locations of the turning points and rods in the C-metric-like coordinates $(u,v)$ and the Weyl--Papapetrou coordinates $(\rho,z)$ for solutions with three free solitons $\mu_{\text{A,B,C}}$.}
\label{fig_C}
\end{center}
\end{figure}

For a solution with three free solitons, we can appropriately shift the coordinate $z$ such that these solitons are located, respectively, at
\begin{align}
z_{\text{A}}=-c\varkappa^2\,,\quad z_{\text{B}}=c\varkappa^2\,,\quad z_{\text{C}}=\varkappa^2\,,
\end{align}
for some values of $c$ and $\varkappa$. With such definitions, the three free solitons can be expressed algebraically as
\begin{align}
\mu_{\text{A}}&= \frac{2\varkappa^2(1-u)(-1-v)(1+cv)}{(u-v)^2}\,,\quad \mu_{\text{B}}=\frac{2\varkappa^2(1-u)(-1-v)(1+cu)}{(u-v)^2}\,,\nonumber\\
\mu_{\text{C}}&=\frac{2\varkappa^2(v^2-1)(1+cu)}{(u-v)^2}\,.
\label{mu_ABC}
\end{align}
The equalities in (\ref{range_C-metric}) correspond to the four rods of the solution in its rod structure. The locations of the turning points and rods of the solution in the coordinates $(u,v)$ and the Weyl--Papapetrou coordinates $(\rho,z)$ are shown in Fig.~\ref{fig_C}. We remark that the signs on the right-hand sides of the equations in (\ref{mu_ABC}) are fixed by insisting on the range (\ref{range_C-metric}).

The inverse relations to (\ref{definition_C}) are
\begin{align}
\label{uv_inverse}
u=1-\frac{2\mu_{\text{B}}R_{\text{AC}}}{\mu_{\text{C}}R_{\text{AB}}}\,,\quad v=1-\frac{2R_{\text{AC}}}{R_{\text{AB}}}\,,
\end{align}
or, equivalently,
\begin{align}
u=1-\frac{(1+c)\mu_{\text{AB}}}{c\mu_{\text{AC}}}\,,\quad v=1-\frac{(1+c)\mu_{\text{C}}\mu_{\text{AB}}}{c\mu_{\text{B}}\mu_{\text{AC}}}\,,
\end{align}
where $R_{\text{AB}}\equiv \rho^2+\mu_{\text{A}}\mu_{\text{B}}$ and likewise for $R_{\text{AC}}$, $\mu_{\text{AB}}\equiv \mu_{\text{A}}-\mu_{\text{B}}$ and likewise for $\mu_{\text{AC}}$. The relations (\ref{uv_inverse}), together with (\ref{flat_metric_C}), tell us that any solution written in the coordinates $(u,v)$ with algebraic metric components in the following form:
\begin{align}
\label{C-metric_metrics}
\dif s^2=g_{ab}(u,v)\,\dif x^a\dif x^b+\zeta(u,v)\left(\frac{\dif u^2}{G(u)}-\frac{\dif v^2}{G(v)}\right),
\end{align}
satisfying the condition (\ref{rho}), i.e., $|\det g|=-\frac{4\varkappa^4{G(u)G(v)}}{(u-v)^4}$, can be cast in the Weyl--Papapetrou coordinates with three solitons $\mu_{\text{A}}$, $\mu_{\text{B}}$ and $\mu_{\text{C}}$. This statement still holds if $G(u)$ is a quartic polynomial \cite{Harmark:2004rm}. This can be proved by observing that the form (\ref{C-metric_metrics}) is preserved under a M\"obius transformation, simultaneously applied on the coordinates $u$ and $v$.  The function $G(u)$ is called the structure function of a solution in the form (\ref{C-metric_metrics}). Under the M\"obius transformation, a cubic structure function becomes a quartic one, while a quartic structure function remains quartic. This is not surprising since a cubic structure function is really a special case of a quartic one with one infinite root.

\bigskip\bigskip

{\renewcommand{\Large}{\large}
}

\end{document}